\documentclass[longauth]{aa}  

\usepackage{graphicx}
\usepackage{lscape, float}
\usepackage{txfonts}
\usepackage{subcaption}
\usepackage{lscape}
\usepackage{hyperref}%
\usepackage{arydshln}
\usepackage{color, soul} 
\usepackage[rightcaption]{sidecap}
\usepackage{afterpage}

\newcommand{\mearth}{M_\oplus}
\newcommand{\rearth}{R_{\rm \oplus}}

\def\ms{\hbox{\,m\,s$^{-1}$}}         
  
\def\kms{\hbox{\,km\,s$^{-1}$}}       

\begin{document}

	\title{A quarter century of spectroscopic monitoring of the nearby M dwarf Gl~514}
	
	\subtitle{A super-Earth on an eccentric orbit moving in and out of the habitable zone \thanks{Based on HIRES, CARMENES and HARPS observations, the latter carried out during programs 072.C-0488, 183.C-0437, and 191.C-0873.} }
	\titlerunning{An eccentric low-mass planet moving through the habitable zone of Gl~514 }
	
	\author{M.~Damasso \inst{1}
		\and M.~Perger \inst{2,3}
		\and J.~M.~Almenara \inst{4}   
		\and D.~Nardiello \inst{5,6}
		\and M.~P\'erez-Torres \inst{7}
		\and A.~Sozzetti \inst{1}
		\and N.~C.~Hara \inst{8}
		\and A.~Quirrenbach \inst{9}
		\and X.~Bonfils \inst{4}
		\and M.~R.~Zapatero Osorio \inst{10}
		\and N.~Astudillo-Defru \inst{11}
		\and J.~I.~Gonz\'alez Hern\'andez \inst{12,13}
		\and A.~Su\'arez Mascare\~{n}o \inst{13,12}
		\and P.~J.~Amado \inst{7}
		\and T.~Forveille \inst{4}
		\and J.~Lillo-Box \inst{14}
		\and Y.~Alibert \inst{15}
		\and J.~A.~Caballero \inst{14}
		\and C. Cifuentes \inst{14}
		\and X.~Delfosse \inst{4}
		\and P.~Figueira \inst{16,17}
		\and D.~Galad\'{\i}-Enr\'{\i}quez \inst{18}
		\and A.~P.~Hatzes \inst{19}
		\and Th.~Henning \inst{20}
		\and A.~Kaminski \inst{9}
		\and M.~Mayor \inst{8}
		\and F.~Murgas \inst{13,12}
		\and D.~Montes \inst{21}
		\and M.~Pinamonti \inst{1}
		\and A.~Reiners \inst{22}
		\and I.~Ribas \inst{2,3} 
		\and V.~J.~S.~B\'ejar \inst{13,12}
		\and A.~Schweitzer \inst{23}
		\and M.~Zechmeister \inst{22}
	} 
	
	\institute{INAF - Osservatorio Astrofisico di Torino, Via Osservatorio 20, I-10025 Pino Torinese, Italy\\
		\email{mario.damasso@inaf.it}
		\and Institut de Ciències de l’Espai (ICE, CSIC), Campus UAB, Carrer de Can Magrans s/n, 08193 Bellaterra, Spain
		\and Institut d’Estudis Espacials de Catalunya (IEEC), 08034 Barcelona, Spain
		\and Universit\'e Grenoble Alpes, CNRS, IPAG, F-38000 Grenoble, France
		\and Aix Marseille Univ, CNRS, CNES, LAM, Marseille, France
		\and INAF - Osservatorio Astronomico di Padova, Vicolo dell'Osservatorio 5, IT-35122, Padova, Italy
		\and Instituto de Astrof\'isica de Andaluc\'ia (IAA-CSIC), Glorieta de la Astronom\'ia s/n, E-18008 Granada, Spain 
		\and Observatoire Astronomique de l’Universit\'e de Gen\'eve, 51 Chemin des Pegasi, 1290 Versoix, Switzerland
		\and Landessternwarte, Zentrum f\"ur Astronomie der Universit\"at Heidelberg, K\"onigstuhl 12, 69117 Heidelberg, Germany
		\and Centro de Astrobiolog\'ia (CSIC-INTA), Carretera de Ajalvir km 4, E-28850 Torrej\'on de Ardoz, Madrid, Spain
		\and Departamento de Matem\'atica y F\'isica Aplicadas, Universidad Cat\'olica de la Sant\'is\'ima Concepci\'on, Concepción, Chile
		\and Departamento de Astrof\'isica, Universidad de La Laguna, 38206 La Laguna, Tenerife, Spain
		\and Instituto de Astrof\'isica de Canarias, 38205 La Laguna, Tenerife, Spain 
		\and Centro de Astrobiolog\'ia (CAB, CSIC-INTA), Depto. de Astrof\'isica,
		ESAC campus, 28692 Villanueva de la Ca\~nada (Madrid), Spain
		\and Physikalisches Institut, University of Bern, Gesellsschaftstrasse 6, CH-3012
		Bern, Switzerland
		\and European Southern Observatory, Alonso de Cordova 3107, Vitacura, Santiago, Chile
		\and Instituto de Astrof\'isica e Ci\^encias do Espaço, Universidade do Porto, CAUP, Rua das Estrelas, 4150-762 Porto, Portugal
		\and Observatorio de Calar Alto, Sierra de los Filabres, ES-04550-G\'ergal, Almer\'ia, Spain
		\and Th\"uringer Landessternwarte Tautenburg, Sternwarte 5, D-07778 Tautenburg, Germany
		\and Max-Planck-Institut f\"ur Astronomie, K\"onigstuhl 17, D-69117 Heidelberg, Germany
		\and Departamento de F\'isica de la Tierra y Astrof\'isica $\&$ IPARCOS-UCM (Instituto de F\'isica de Part\'iculas y del Cosmos de la UCM), Facultad de Ciencias F\'isicas,
		Universidad Complutense de Madrid, E-28040 Madrid, Spain
		\and Institut f\"ur Astrophysik, Georg-August-Universit\"at, Friedrich-Hund-Platz 1, D-37077 G\"ottingen, Germany
		\and Hamburger Sternwarte, Gojenbergsweg 112, D-21029 Hamburg, Germany
	}
	
	
	
	\abstract
	{Statistical analyses based on \textit{Kepler} data show that most of the early-type M dwarfs host multi-planet systems consisting of Earth to sub-Neptune sized planets with orbital periods up to $\sim$250 days, and that at least one such planet is likely located within the habitable zone. M dwarfs are therefore primary targets to search for potentially habitable planets in the solar neighbourhood.}
	{We investigated the presence of planetary companions around the nearby (7.6 pc) and bright ($V=9$ mag) early-type M dwarf Gl~514, analysing 540 radial velocities collected over nearly 25 years with the HIRES, HARPS, and CARMENES spectrographs.}
	{The data are affected by time-correlated signals at the level of 2--3 \ms due to stellar activity, that we filtered out testing three different models based on Gaussian process regression. As a sanity cross-check, we repeated the analyses using HARPS radial velocities extracted with three different algorithms. We used HIRES radial velocities and {\sl Hipparcos-Gaia} astrometry to put constraints on the presence of long-period companions, and we analysed \textit{TESS} photometric data.}
	{We found strong evidence that Gl~514 hosts a super-Earth on a likely eccentric orbit, residing in the conservative habitable zone for nearly $34\%$ of its orbital period. The planet Gl~514\,b has minimum mass $m_b\sin i_b=5.2\pm0.9$ $\mearth$, orbital period $P_b=140.43\pm0.41$ days, and eccentricity $e_b=0.45^{+0.15}_{-0.14}$. No evidence for transits is found in the \textit{TESS} light curve. There is no evidence for a longer period companion in the radial velocities and, based on astrometry, we can rule out a $\sim0.2$ $M_{\rm Jup}$ planet at a distance of $\sim3-10$ au, and massive giant planets/brown dwarfs out to several tens of au. We discuss the possible presence of a second low-mass companion at a shorter distance from the host than Gl~514\,b.  }
	{Gl~514\,b represents an interesting science case to study the habitability of planets on eccentric orbits. We advocate for additional spectroscopic follow-up to get more accurate and precise planetary parameters. Further follow-up is also needed to investigate sub \ms and shorter period signals. }
	
	
	\maketitle
	%
	
	\section{Introduction}
	\label{sec:intro}
	Thanks to the most rewarding detection techniques, nowadays the search for and characterisation of low-mass planets orbiting within the habitable zone (HZ) of M dwarfs are hot topics, and very promising prospects for further investigation are offered especially by exoplanets detected around the nearest and brightest stars. Some of the earlier radial velocity (RV) surveys have focused on low-mass stars (e.g. \citealt{ endl03},\citealt{zech2009,bonfils2013}), and they are among the main targets of new spectrographs that have been developed and started working in the last decade (e.g. \citealt{mahade2012,artigau2014,affer2016,quirrenbach18,pepe2021}). Following the milestone\textit{ Kepler/K2} mission, in the realm of transit photometry the Transiting Exoplanet Survey Satellite (\textit{TESS}; \citealt{ricker2016}) is presently the major source of planet candidates around the brightest nearby stars, and M dwarfs are main targets to detect small-size planets in the HZ. By far, mostly RV and transit surveys allowed for the first assessments of the occurrence rate of HZ planets around M dwarfs (e.g. \citealt{bonfils2013,dressing2015,hsu2020,sabotta2021,Pinamonti22}, and stunning detections have been made in the last few years, especially around mid-to-late type M dwarfs (e.g. \citealt{anglada2016,astudillo2017,cloutier2017,gillon2017,bonfils2018,zech2019}). 
	
	So far, the detection and characterisation of HZ planets has proven to be more complicated for early-type M dwarfs (M0V--M2V), as testified by the relatively lower number of discoveries compared to mid-to-late type M dwarfs. We used the NASA exoplanet archive\footnote{\url{https://exoplanetarchive.ipac.caltech.edu/}. Queried on 7 April 2022.}, complemented by the archive of confirmed potentially habitable low-mass planets maintained by the Planetary Habitability Laboratory (PHL)\footnote{\url{http://phl.upr.edu/projects/habitable-exoplanets-catalog} (updated to 6 Dec 2021)}, and by the most recent papers on specific planetary systems with no data reported in both catalogues, to compile a list of planets with radius $R<2.5$ $\rearth$ and/or mass $m<10$ $\mearth$, and an insolation flux $S$ in the range 0.25--1.65 $S_{\oplus}$, that have been detected around M-dwarfs with $T_{\rm eff}$ in the range 3650--3900 K.
	We found three transiting planets detected by \textit{Kepler}, namely Kepler-186\,f \citep{Quintana2014}, Kepler-705\,b and Kepler-1229\,b \citep{Morton_2016} (even the bona-fide KOI-4427.01 planet could be added to this group, see \citealt{Torres_2015}). Due to the faintness of their host stars, all these planets do not have yet a measured mass through radial velocities. The list of transiting planets also includes K2-3\,d \citep{crossfield15}, a 1.6 $\rearth$ planet orbiting near the inner edge of the HZ ($P_d=44.57$ d;$S_d=1.6 S_\oplus$), detected by \textsc{Kepler/K2}. Its brightness ($V=12.2$) compared to that of the previous systems makes K2-3 an excellent target to determine the planetary masses via high-precision Doppler spectroscopy. Actually, this task proved to be much more challenging than expected due to the difficulty in filtering out the signal induced by stellar activity. Only a mass upper limit has been presently determined for K2-3\,d, despite the precise orbital ephemeris provided by transits and more than 400 RVs available \citep{damasso2018,kosiarek2019}.
	In the list compiled for M0V--M2V dwarfs, as described above, we found only one non-transiting planet detected through the radial velocity technique, the $\sim$7 $\mearth$ super-Earth Gl~229~A~c \citep{feng2020} in a red-brown dwarf binary system at a distance of 5.75 pc.
	
	In this paper, we report the detection of a super-Earth that moves on an eccentric orbit through the HZ of Gl~514 (BD+11~2576), an M0.5V--M1.0V star ($V=9$ mag) located at a distance of 7.6 pc. The detection is based on nearly 25 years of RV monitoring with the HIRES, HARPS, and CARMENES spectrographs. The manuscript is organised as follows. In Sect. \ref{sec:starparam} we summarise the main astrophysical properties of the host Gl~514. In Sect. \ref{sec:overviewdata} general information about the spectroscopic and photometric dataset we analysed in this work is provided. We discuss in Sect. \ref{sec:activitycharacterization} the properties of the stellar magnetic activity of Gl~514 as derived from the analysis of spectroscopic activity diagnostics, and we use the results to interpret the frequency content observed in the RV dataset (explored in Sect. \ref{sec:RVperiodogram}). The core of the paper is represented by Sect. \ref{sec:rvmodelling}, where we present and discuss the detection of the super-Earth Gl~514~b based on an extensive analysis of 540 RVs. In Sect.  \ref{sec:LCanalysis} we present the analysis of \textit{TESS} photometric data, and in Sect. \ref{sec:astrometry} we discuss the astrometric sensitivity to wide-separation companions. Conclusions and future perspectives are discussed in Sect. \ref{sec:conclusions}. Additional results supporting our finding are shown in the Appendix. 
	
	
	\section{Stellar fundamental parameters}
	\label{sec:starparam}
	
	\begin{table}
		\caption[]{Astrometric, photometric, and spectroscopically derived stellar properties of Gl~514.}
		\label{tab:starparam}
		\centering
		\tiny
		\begin{tabular}{l c c}
			\hline
			\hline
			\noalign{\smallskip}
			\multicolumn{3}{c}{{\it Gl~514, BD+11 2576, HIP~65859}}  \\
			\noalign{\smallskip}
			Parameter     &  Value & Refs. \\
			\noalign{\smallskip}
			\hline
			\noalign{\smallskip}
			\noalign{\smallskip}
			\textit{Astrometry:} \\
			\noalign{\smallskip}
			$\alpha$ (J2000) & 13h 29m 59.79s & [1,2,3]  \\
			\noalign{\smallskip}
			$\delta$ (J2000) & $+$10$^\circ$ 22' 37.8'' & [1,2,3]  \\
			\noalign{\smallskip}
			$\mu_\alpha \cdot \cos \delta$ [mas yr$^{-1}$] & $1127.341\pm0.027$ & [1,2,3]  \\
			\noalign{\smallskip}
			$\mu_\delta$ [mas yr$^{-1}$] & $-1073.888\pm0.014$ & [1,2,3]  \\
			\noalign{\smallskip}
			$\varpi$ [mas] & $131.101\pm0.027$ & [1,2,3]   \\
			\noalign{\smallskip}
			$d$ [pc] & $7.618\pm0.003$ & [4]  \\
			\noalign{\smallskip}
			\noalign{\smallskip}
			\textit{Photometry:} \\
			\noalign{\smallskip}
			$U$ &  $11.7$ & [5] \\
			\noalign{\smallskip} 
			$B$ &  $10.5$ & [5]\\
			\noalign{\smallskip}
			$V$ &  $9.0$ & [5]\\
			\noalign{\smallskip}
			$G$ &  $8.2053\pm0.0027$ & [1,2,3] \\
			\noalign{\smallskip}
			$J$  & $5.902\pm0.018$ & [6] \\
			\noalign{\smallskip}
			$H$   & $5.300\pm0.033$ & [6] \\
			\noalign{\smallskip}
			$K_s$ &  $5.036\pm0.027$ &  [6]\\
			\noalign{\smallskip}
			\noalign{\smallskip}
			\textit{Stellar Parameters:} \\
			\noalign{\smallskip}
			$T_\mathrm{eff}$ [K] & $3728\pm68$ &  [8] \\
			& $3714\pm51$ &  [9] \\
			\noalign{\smallskip}
			$\log g$ [dex] & $4.76\pm0.04$ &  [8] \\
			& $4.81\pm0.07$ &  [9] \\
			\noalign{\smallskip}
			[Fe/H] [dex] & $-0.14\pm0.09$ &  [8] \\
			& $-0.15\pm0.16$ &  [9] \\
			\noalign{\smallskip}
			$M_\star$ [M$_\odot$] & $0.510\pm0.051$ &  [8] \\
			\noalign{\smallskip}
			$R_\star$ [R$_\odot$] & $0.500\pm0.047$  & [8]  \\
			\noalign{\smallskip}
			$L_\star$ [L$_\odot$] & $0.043\pm0.009$ & [8]  \\
			\noalign{\smallskip}
			$P_{\star,\, \rm rot}$ [days] & $28.0\pm2.9$ & [10]  \\
			& $30.8\pm0.3$ (H$\alpha$) & [11] \\
			& $30.3\pm0.2$ (Ca\textsc{II} infrared triplet) & [11] \\ 
			& 30 (Ca\textsc{II} $H\&K$) & [12] \\
			& $30.6\pm0.3$ (GP fit of RVs) & [7] \\
			\noalign{\smallskip}
			$<\log R^\prime_{HK}>$  & $-5.10\pm0.06$  &  [10] \\
			\noalign{\smallskip}
			($U$,$V$,$W$) [km s$^{-1}$] & ($57.42\pm0.03$,$-8.11\pm0.02$, & [7]  \\
			& $-3.54\pm0.09$) & \\
			\noalign{\smallskip}
			Kinematical age [Gyr] & $>0.8$ & [7]  \\
			\noalign{\smallskip}
			Conservative HZ\tablefootmark{a} [au] & [0.207,0.411] & [13]\\
			\noalign{\smallskip}
			\hline
		\end{tabular}
		\tablefoot{
			\tablefoottext{a}{For planet mass $m_p=5 \mearth$}}
		\tablebib{[1] \citealt{gaiacoll2016}; [2] \citealt{gaiacoll2020}; [3] \citealt{Lindegren21}; [4] \citealt{BailerJones2018b}; [5] \citealt{Koen2010}; [6] \citealt{Cutri2003}; [7] This work; [8] \citealt{Maldonado2015}; [9] \citealt{Schweitzer19}; [10] \citealt{Mascaren02015}; [11] \citealt{Fuhrmeister19}; [12] \citealt{astudillo17}; [13] \citealt{kopparapu13,kopparapu14}.}
	\end{table}
	
	
	Gl~514 is a M0.5--M1 dwarf located at a distance of 7.6 pc from us. Its main astrophysical properties are summarised in Table \ref{tab:starparam}. An earlier estimate of the stellar rotation period was provided by \cite{Mascaren02015} from the analysis of spectroscopic activity indexes based on part of the same HARPS spectra analysed in this work. A following and independent result by \cite{Fuhrmeister19}, and based on CARMENES spectra (also part of our dataset), shows that $P_{\star,\, \rm rot}=30.8\pm0.3$ d and $30.3\pm0.2$ d, derived from the analysis of the H$\alpha$ and Ca{\sc II} infrared triplet lines, respectively. As we will discuss in Sect. \ref{sec:activitycharacterization} and \ref{sec:harpscarmervanalysis}, our derived values for $P_{\star,\, \rm rot}$ are in agreement with the previous determinations.
	
	The Galactic space velocities $U$, $V$, and $W$ of Gl\,514 were derived using the {\sl Gaia} coordinates and proper motions. We also employed the {\sl Gaia} radial velocity 
	to calculate the $U$ , $V$ , and $W$ heliocentric velocity components in the directions of the Galactic center, Galactic rotation, and north Galactic pole, respectively, with the formulation developed by \cite{johnson1987}. Note that the right-handed system is used and that we did not subtract the solar motion from our calculations. The uncertainties associated with each space velocity component were obtained from the observational quantities and their error bars after the prescription of \cite{johnson1987}. According to the resulting space velocities, Gl\,514 has kinematics that deviate from that of known young stellar moving groups in the $UV$ vs $WV$ planes (e.g. \citealt{gagne2018}). Therefore, the kinematical age of Gl\,514 is likely $>$ 0.8 Gyr. Given its positive systemic RV, Gl~514 is moving away from us and, based on data from {\sl Gaia} DR2, \cite{bailerjones2018a} calculated that the closest approach to the Sun occurred nearly 30 Myr ago, at a distance of 7.4 pc.
	
	\begin{table*}[ht]
		\caption{General properties of the radial velocity time series analysed in this work.}          
		\label{table:rvproperties}      
		\centering   
		\begin{tabular}{lccccc}       
			\hline\hline              
			\textbf{Instrument} & \textbf{Time span} & \textbf{No. RVs} & \textbf{Pipeline} & \textbf{RMS} & \textbf{$\overline{\sigma}_{\rm RV}$} \\ 
			& [days] & & & [$\ms$] & [$\ms$] \\ 
			\hline              
			\noalign{\smallskip}
			HIRES & 6095 & 104 & \cite{TalOr2019} & 3.8 & 1.6 \\
			\noalign{\smallskip}
			HARPS$_{pre-2015}$ & 3963 & 142 & TERRA & 2.8 & 0.7  \\
			\noalign{\smallskip}
			& & & NAIRA & 2.9 & 0.6 \\
			\noalign{\smallskip}
			& & & \cite{trifonov20}, SERVAL extraction & 3.0 & 0.9 \\
			\noalign{\smallskip}
			HARPS$_{post-2015}$ & 266 & 20 & TERRA & 2.5 & 0.7 \\
			\noalign{\smallskip}
			& & & NAIRA & 2.8 & 0.6 \\
			\noalign{\smallskip} 
			& & & \cite{trifonov20}, SERVAL extraction & 2.6 & 0.9 \\
			\noalign{\smallskip}
			CARMENES-VIS & 1909 & 274 & SERVAL & 2.6 & 1.6   \\
			\noalign{\smallskip}
			\hline
		\end{tabular}
		\tablefoot{Total time spans -- HARPS: 4352 days; HARPS+CARMENES: 6154 days; HIRES+HARPS+CARMENES: 8761 days.}  
	\end{table*}

	\section{Overview of the spectroscopic and photometric dataset}
	\label{sec:overviewdata}
	
	Gl\,514 was monitored with the HIRES, HARPS, and CARMENES spectrographs, collecting a total of 540 (104+162+274) RV measurements over 24 years. HIRES observations cover the time span between 7 April 1997 and 14 December 2014. We used the publicly available data derived by \cite{TalOr2019}.\footnote{https://cdsarc.unistra.fr/viz-bin/cat/J/MNRAS/484/L8}
	
	With HARPS, mounted at the ESO La Silla (Chile) 3.6m telescope, the star was observed between 27 May 2004 and 26 April 2016. The spectra are publicly available through the ESO Archive\footnote{\url{http://archive.eso.org/wdb/wdb/adp/phase3_main/form}}. The majority of the spectra (142 out of 162) were collected before the fiber upgrade intervention occurred in May 2015 (we label that as the \textit{pre-2015} dataset), and they have mean signal-to-noise S/N=82 measured at a reference wavelength of 5500 $\AA$. The remaining 20 spectra collected after May 2015 (the \textit{post-2015} dataset) have mean S/N=94.  
	We extracted the HARPS RVs using different pipelines, all based on template-matching. One is \textsc{TERRA} (\textsc{v9.0}; \citealt{AngladaEscud2012}), an algorithm which has been proven to be particularly efficient with M dwarfs (e.g. \citealt{perger2017}). We used the RVs extracted from orders $\geq22$ (corresponding to wavelengths $\lambda\geq4400$ \AA, according to the notation adopted by \textsc{TERRA}), excluding the bluest and lower S/N orders from the RV computation (in this case, we did not use in the computation the spectral region between $\sim$3800--4000 $\AA$). This selection minimises the RMS and uncertainties of the RVs. Pre- and post-2015 spectra were treated as coming from different instruments, and two template spectra were calculated for each group. To account for the reported offset introduced by the fiber upgrade, when modelling the RVs we considered a zero-point ($\gamma$) and an uncorrelated jitter term ($\sigma_{jit}$) as free parameters for each dataset separately. We also exploited two other RV dataset extracted with alternative template-matching pipelines, as a sanity cross check. One dataset was extracted with the NAIRA algorithm \citep{astudillo2015}. NAIRA recipe is based on a stellar template built from the median of the whole set of spectra, with uncertainties $\sigma_{\rm RV}$ computed following \cite{bouchy2001}, and the perspective acceleration subtracted using proper motion and parallax from {\sl Gaia} DR2 \citep{gaiacoll2016,gaia2018}. The third dataset is represented by the nightly zero-point (NZP) corrected RVs published by \cite{trifonov20}\footnote{Also available at \url{https://www2.mpia-hd.mpg.de/homes/trifonov/HARPS_RVBank.html}}, which were derived using the template-matching code SERVAL \citep{2018A&A...609A..12Z}.
	The RVs extracted with TERRA and NAIRA are listed in Tables \ref{table:rvterradata} and \ref{table:rvnairadata}. 
	
	The CARMENES spectrograph at the 3.5m telescope of the Calar Alto Observatory in Spain \citep{quirrenbach18} observed Gl~514 from 10 January 2016 to 2 April 2021, with only a small overlap with the last HARPS observations. In this work, we use the RVs extracted from the VIS channel, covering the wavelength range 5200--9600 \AA. The data of the near-infrared channel are characterised by a S/N which is by far lower, and a mean $\sigma_{\rm RV}$ uncertainty of 8.7 $\ms$, therefore they are not particularly useful to detect and characterise RV signals with a semi-amplitude of few \ms, which are those we are interested in. Originally, CARMENES collected 321 spectra of Gl\,514, which were reduced using the \textsc{caracal} pipeline to obtain calibrated 1D spectra \citep{zechmeister2014,caballero2016}, from which we extracted the RVs using the template-matching code \textsc{SERVAL} \citep{2018A&A...609A..12Z}. During a few nights, CARMENES observed the star more than once. We binned the corresponding RVs on a nightly basis\footnote{As it will be shown later in the paper, we discuss signals with periods of several days, including activity and planets, therefore the binning does not affect the analysis.}, then discarded 14 measurements with $\sigma_{\rm RV}>3 \ms$ (nearly twice the mean value of $\sigma_{\rm RV}$), resulting in a total of 274 RVs. The RVs are listed in Table \ref{table:rvcarmedata}. 
	
\begin{figure}
		\centering
		\includegraphics[width=\hsize]{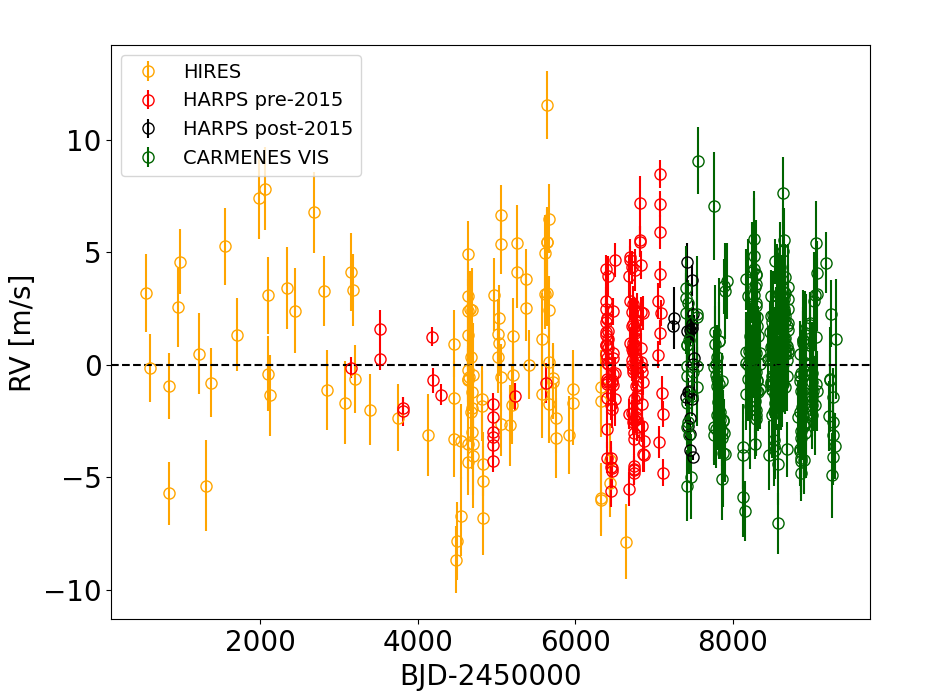}
		\caption{Time series of the HIRES (orange dots), HARPS$_{\rm TERRA}$ (red and black dots), and CARMENES-VIS (green dots) radial velocities analysed in this work. The average has been subtracted from each dataset.}
		\label{fig:rv_harps_hires_terra}
	\end{figure}

	A summary of the main properties of each RV datset is provided in Table \ref{table:rvproperties}. The whole RV time series is shown in Fig. \ref{fig:rv_harps_hires_terra} (for clarity, only the HARPS measurements extracted with TERRA are shown). HIRES RVs have the longest time span and a less dense sampling. Due to its properties, we will only use the HIRES dataset to test the presence of long-term signals, since it is not suitable to search for low-amplitude and shorter period signals compared to the other two dataset. 
	CARMENES RVs are characterised by a typical internal error $\sigma_{\rm RV}$ of 1.6 \ms, equal to that of HIRES data, but the time span is nearly three times shorter and the sampling is much denser. The larger RV uncertainties compared with those of the HARPS data are mostly due to the lower average exposure time of $440\pm180$\,s, while for HARPS this is $900\pm40$\,s.
	
	Concerning the photometric observations, \textit{TESS} monitored Gl~514 from 19 March to 15 April 2020 (sector 23). We analysed the long-cadence light curve extracted from the Full Frame Images (FFIs) by using the PATHOS pipeline \citep{Nardiello2019, Nardiello2020a}. Before the measurement of the flux of the target star, the light curve extractor \textsc{IMG2LC} \citep{Nardiello2015, Nardiello2016} subtracts from each FFI all its neighbour stars by adopting empirical Point Spread Functions and information from the {\sl Gaia} DR2 catalog \citep{gaia2018}. The flux of the target star is measured with different photometric apertures (with radii from 1 to 4 pixels), fitting the empirical PSF. The systematic effects that affect the raw light curve are corrected by using the cotrending basis vectors (CBVs) extracted and applied as in \cite{Nardiello2020a}. 
	We also analysed the short-cadence light curve released by the \textit{TESS} team. We did not use the Pre-search Data Conditioning Simple Aperture Photometry (PDCSAP) flux because of some systematic effects due to over-corrections and/or injection of spurious signals. We corrected the Simple Aperture Photometry (SAP) flux by applying CBVs obtained by using the SAP light curves of the stars in the same Camera/CCD in which Gl~514 falls, and following the same procedure used for extracting the CBVs for the long-cadence light curves \citep{Nardiello2020a}. Gl~514 falls on a CCD heavily affected by straylight contamination, which in turn badly affects the overall quality of the data.  
	
	\section{Stellar magnetic activity characterisation}
	\label{sec:activitycharacterization}
	
	We used spectroscopic diagnostics to characterise the time variability induced by the magnetic activity of Gl~514. This is a fundamental analysis to correctly interpret the nature of significant periodic signals possibly found in the RV time series. We analysed the frequency content of the data for each individual instrument using the Generalised Lomb-Scargle algorithm (GLS; \citealt{zech2009}). The main peaks identified in the periodogram are summarised in Table \ref{table:activdiagnpeaks}, together with their false alarm probabilities (FAPs) calculated through a bootstrap with replacement analysis (i.e. an element of the dataset may be drawn multiple times from the original sample).
	
	\begin{table*}
		\caption{Peaks with the highest power observed in the GLS periodograms of the spectroscopic chromosperic activity indicators extracted from HIRES, HARPS, and CARMENES-VIS spectra (see Fig. \ref{fig:hiressindex}, \ref{fig:harpsindex}, and \ref{fig:carmeactind}). }          
		\label{table:activdiagnpeaks}      
		\centering    
		\small
		\begin{tabular}{lccc}       
			\hline\hline              
			\textbf{Activity diagnostic} & \textbf{Frequency} & \textbf{Period} & \textbf{Note}  \\   
			& [days$^{-1}$] & [days] & \\ 
			\hline              
			\noalign{\smallskip}
			\textbf{HIRES}  \\
			\noalign{\smallskip}
			S-index & 0.00136 & 734 & FAP=0.01$\%$ \\
			\noalign{\smallskip}
			& 0.00159 & 629 & FAP=0.01$\%$; alias of the main peak frequency  \\
			&         &     & ($f_{\rm alias}$=0.000229 d$^{-1}$, or $P_{\rm alias}$=4361 days)\\
			\noalign{\smallskip}
			& 0.0325 & 30.7 & FAP$\sim0.1\%$ \\
			\noalign{\smallskip}
			& 0.0673 & 14.9 & FAP$\sim1\%$ \\
			\noalign{\smallskip}
			\textbf{HARPS}  \\
			\noalign{\smallskip}
			CCF FWHM\tablefootmark{a} & 0.00126 & 796 & FAP$=0.01\%$ \\
			\noalign{\smallskip}
			& 0.0304 & 32.9 & FAP$=0.01\%$ \\
			\noalign{\smallskip}
			& 0.00402 & 249 & FAP$=0.01\%$ \\
			\noalign{\smallskip}
			& 0.000705 & 1417 & FAP$=0.01\%$ \\
			\noalign{\smallskip}
			& 0.0332 & 30.1 & FAP$=0.01\%$ \\
			\noalign{\smallskip}
			Ca\textsc{II}\,H$\&$K index & 0.001022 & 978 & FAP$=0.01\%$ \\
			\noalign{\smallskip}
			& 0.03049 & 32.8 & FAP$=0.01\%$\\
			\noalign{\smallskip}
			& 0.0333 & 30.0 & FAP$=0.01\%$; 1-yr alias of the previous frequency \\
			\noalign{\smallskip}
			& 0.00139 & 717 & FAP$\sim0.1\%$ \\
			\noalign{\smallskip}
			H$\alpha$ index & 0.002038 & 491.8 & FAP$=0.01\%$  \\
			\noalign{\smallskip}
			&  & 1024 & FAP$=0.01\%$; alias of the main peak frequency  \\
			&  &      & ($f_{\rm alias}$=0.00303 d$^{-1}$, or $P_{\rm alias}$=329.7 days)\\
			\noalign{\smallskip}
			&  & 207.7 & FAP$=0.1\%$; 1-yr alias of the main peak frequency\\
			\noalign{\smallskip}
			NaI index & 0.00256 & 390.3 & FAP$=0.01\%$ \\
			\noalign{\smallskip}
			& 0.0007 & 1427 & FAP$=0.01\%$ \\
			\noalign{\smallskip}
			& 0.0303 & 33 & FAP$<0.1\%$ \\ 
			\noalign{\smallskip}
			\textbf{CARMENES-VIS}\tablefootmark{b}  \\
			\noalign{\smallskip}
			Ca-IRT index & 0.00157 & 636 & FAP$=0.01\%$ \\
			\noalign{\smallskip}
			&  & 34 & FAP$=0.1\%$ \\ 
			\noalign{\smallskip}
			NaD index & 0.00165 & 606 & $1\%<FAP<10\%$ \\ 
			\noalign{\smallskip}
			& 0.0625 & 16 & $1\%<FAP<10\%$ \\ 
			\noalign{\smallskip}
			dLW & 0.00175  & 570 & FAP$=0.01\%$ \\ 
			\noalign{\smallskip}
			& 0.0294 & 34 & FAP$=1\%$ \\ 
			\noalign{\smallskip}
			CRX & 0.0024 & 410 & FAP$=0.01\%$ \\ 
			\noalign{\smallskip}
			& 0.00565 & 177 & likely related to the data sampling \\ 
			\noalign{\smallskip}
			& 0.0164 & 61 & $1\%<FAP<10\%$ \\
			\hline
		\end{tabular}
		\tablefoot{
			\tablefoottext{a}{The periodogram was calculated on the residuals after removing a long-term trend.}
			\tablefoottext{b}{The peak frequencies are calculated from pre-whitened data, except for the CRX index.}
		}
	\end{table*}
	
	\subsection{HIRES}
\begin{SCfigure*}
		\centering
		\includegraphics[width=1.5\linewidth]{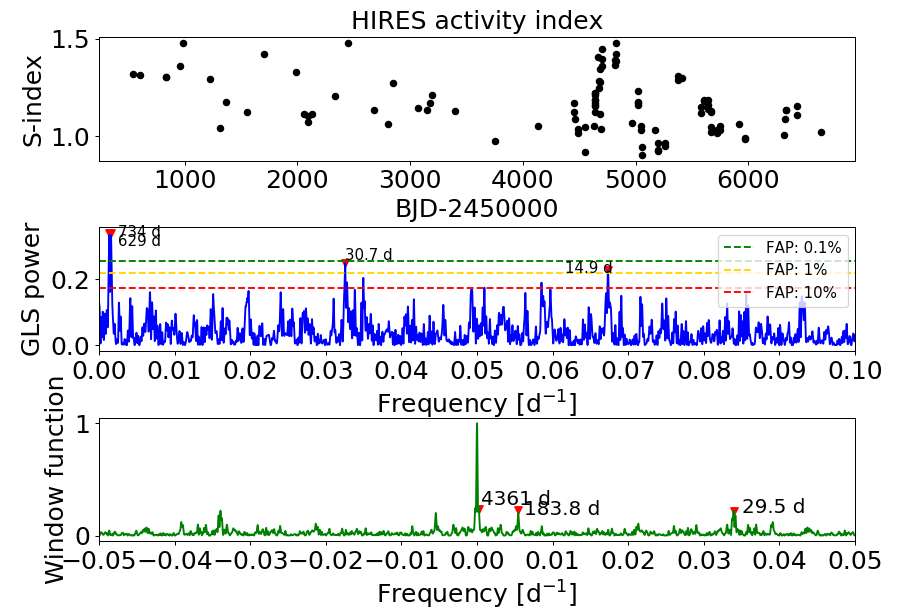}
		\caption{\textit{First and second row:} time series and GLS periodogram of the chromospheric activity S-index derived from HIRES spectra. The levels of FAPs are indicated as horizontal dashed lines (green: FAP 0.1$\%$; yellow: 1$\%$; red: 10$\%$), and have been determined through a bootstrap analysis. \textit{Third row:} window function of the data.}
		\label{fig:hiressindex}
	\end{SCfigure*}

	We analysed the chromospheric $S$-index reported by \cite{TalOr2019}. The time series and GLS periodogram are shown in Fig. \ref{fig:hiressindex}. The main peak occurs at 734 days, with almost the same power of its alias at 629 days (the alias frequency in the window function is 0.000229 d$^{-1}$, corresponding to $P=4361$ days). Due to the low number of points sparsely sampled over a large time span, we can only conclude that this period suggests the presence of a long term modulation, whose accurate properties cannot be determined. The other two significant peaks (FAP $<1\%$) occur at 30.7 and 14.9 days in the original data, and correspond to the stellar rotation period $P_{\star,\, \rm rot}$ and its first harmonic (see Table \ref{tab:starparam}).
	
	\subsection{HARPS}
\begin{SCfigure*}
		\includegraphics[width=1.5\linewidth]{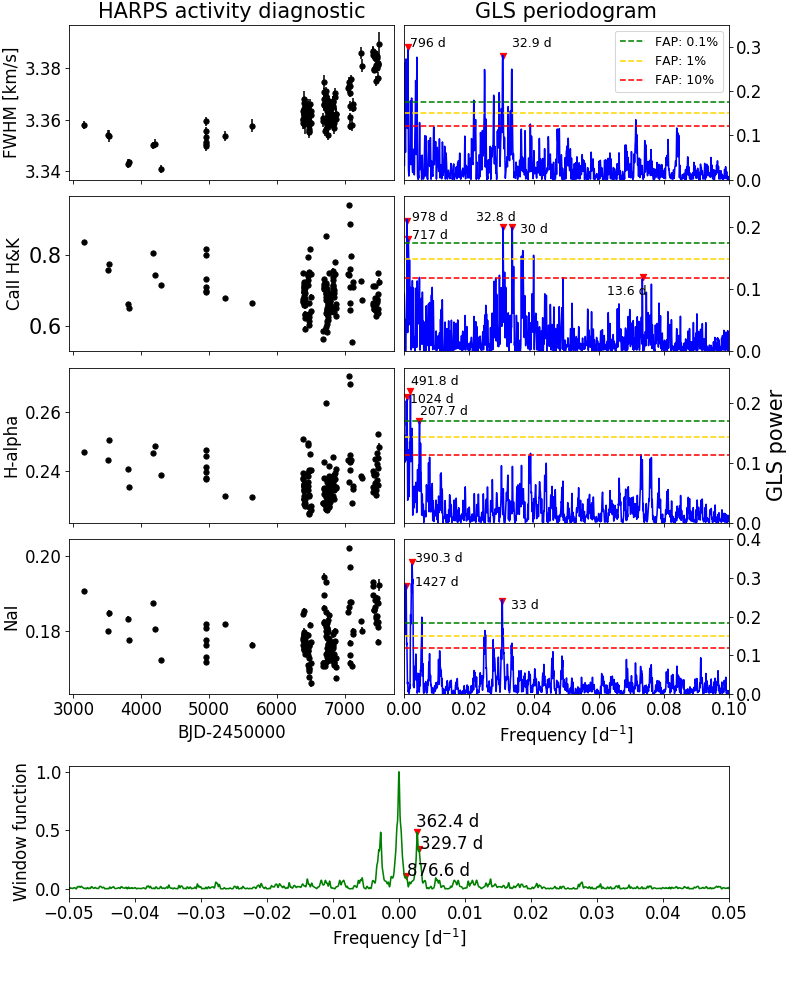}
		\caption{\textit{Rows 1-4:} time series (left panel) and GLS periodograms (right panel) of the CCF FWHM, and activity indexes derived from the spectral lines Ca\textsc{II}\,H$\&$K, H$\alpha$ and Na\textsc{I} of HARPS spectra. For the FWHM, we show the periodogram of the residuals, after removing the long-term trend clearly seen in the time series. The levels of FAPs are indicated as horizontal dashed lines (green: FAP 0.1$\%$; yellow: 1$\%$; red: 10$\%$), and are determined through a bootstrap analysis. \textit{Last row:} window function of the data.}
		\label{fig:harpsindex}
	\end{SCfigure*}

	We derived activity diagnostics based on the Ca\textsc{II}\,H$\&$K, H$\alpha$, and NaI spectral lines using the code \textsc{ACTIN v1.3.6} \citep{gomesdasilva2018}. We also use the FWHM calculated from the cross-correlation function (CCF). We did not use the FWHM calculated by the standard data reduction system (DRS) of HARPS, because it does not perform a color-correction in the case of M dwarfs, which stabilizes the CCF against airmass and seeing variations, as it happens for FGK-type stars. We found that, for M dwarfs, the color-corrected FWHM time series show a reduced RMS which allows for a better characterisation of the variability caused by stellar activity. Based on this improvement, we corrected the CCFs of Gl~514 by re-weighting them against a fixed flux distribution, using files already available within the HARPS DRS. 
	The time series are shown in Fig. \ref{fig:harpsindex} together with the GLS periodograms, and listed in Table \ref{table:actindataharps}. For the FWHM, we show the periodogram after removing a long-term trend. Long-term variability (greater than few hundreds days) is detected for all the indices, even though it is not possible to identify any periodicity (if present) unambiguously. Signals associated to $P_{\star,\, \rm rot}$ appear significant (FAP $<0.1\%$) for the FWHM, Ca\textsc{II}\,H$\&$K, and NaI diagnostics. No significant signals are detected at the first harmonic of $P_{\star,\, \rm rot}$. 
	
	\subsection{CARMENES}
\begin{SCfigure*}
		\centering
		\includegraphics[width=1.5\linewidth]{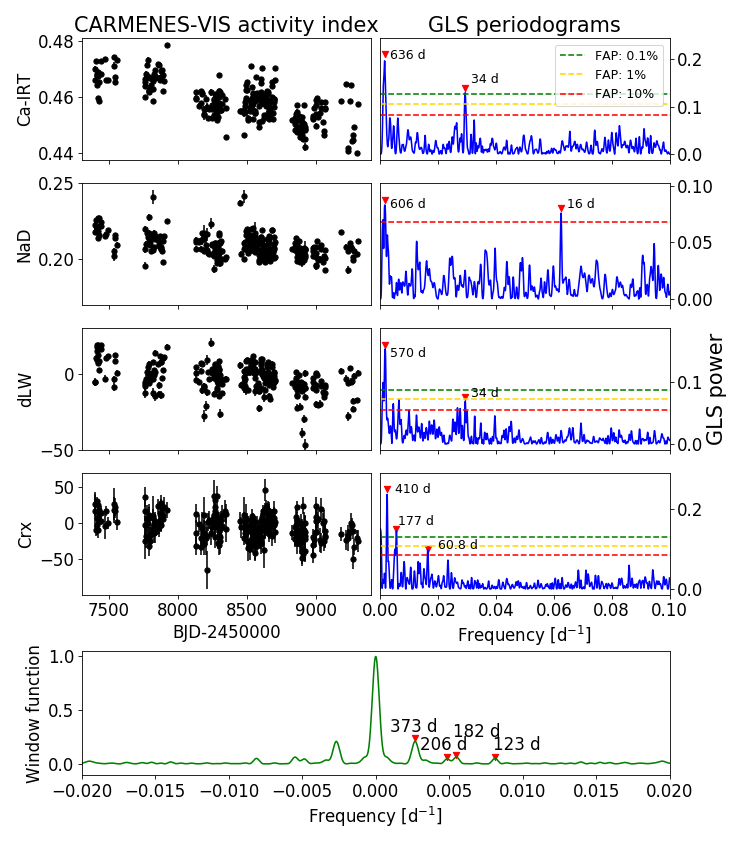}
		\caption{\textit{Left panels:} time-series of activity diagnostics extracted from CARMENES-VIS spectra of Gl~514. \textit{Right panels:} GLS periodograms for each diagnostic. Except for the CRX index, the other periodograms have been calculated after a pre-whitening by removing the clear long-term trend visible in the time-series. The levels of FAPs, determined through a bootstrap analysis, are indicated as horizontal dashed lines (green: FAP 0.1$\%$; yellow: 1$\%$; red: 10$\%$.
			\textit{Last panel:} window function of the CARMENES-VIS data.}
		\label{fig:carmeactind}
	\end{SCfigure*}

	We derived spectroscopic activity diagnostics from the CARMENES-VIS spectra using the code \textsc{SERVAL}. These include the strength of emission lines of molecules sensitive to the chromospheric magnetic fields, such as the calcium infrared triplet (Ca-IRT) and the sodium doublet (NaD). We further derived the time series of the differential line width of the spectral lines (dLW), and the chromatic index (CRX) by measuring the wavelength dependency of the RVs in the different orders of the spectra (see \citealt{2018A&A...609A..12Z} for details). The time series of these activity indicators are shown in the left panels of Fig. \ref{fig:carmeactind}, and the data are listed in Table \ref{table:actindatacarme}. All the time-series show a clear long-term trend, especially the Ca-IRT, NaD, and dLW indexes. For the Ca-IRT, NaD, and dLW indexes, we corrected this trend using a least-square linear fit. The right panels of Fig. \ref{fig:carmeactind} show the GLS periodograms for each activity diagnostic (Ca-IRT, NaD, and dLW: pre-whitened residuals; CRX: original data), together with the FAP levels determined through a bootstrap (with replacement) analysis. Evidence for signals related to the stellar rotation period, or its first harmonic, are present in all the periodograms, except for CRX. The peaks with the highest power occur in the range 410--636 days, suggestive of a mid-term modulation with no immediate interpretation. 
	
	\section{Frequency content analysis of the radial velocities}
	\label{sec:RVperiodogram}
	
	We investigate the frequency content of the different RV dataset and their combinations by calculating the maximum likelihood periodograms (MLP; \citealt{zech2019}), using the implementation of the MLP code included in the \textsc{Exo-Striker} package\footnote{see \url{https://ascl.net/1906.004}}. Differently from the commonly used Generalised Lomb-Scargle periodogram (GLS; \citealt{zech2009}), for each tested frequency the MLP shows the difference $d(ln\,L)$ between the logarithm of the likelihood function corresponding to the best-fit sine function, and that of a constant function. The MLP algorithm includes RV zero points as free parameters in case of dataset from different instruments, as well as instrumental uncorrelated jitter terms. In the following, first we report the results for the HIRES, HARPS, and CARMENES data separately, then we inspect the periodograms for the combined HARPS+CARMENES, and HIRES+HARPS+CARMENES time series. The FAPs indicated in each case are analytical and are calculated by \textsc{Exo-Striker}.
	
	\subsection{HIRES RVs}
	\label{sec:MLPHIRES}
	
	\begin{figure}
		\centering
		\includegraphics[width=\hsize]{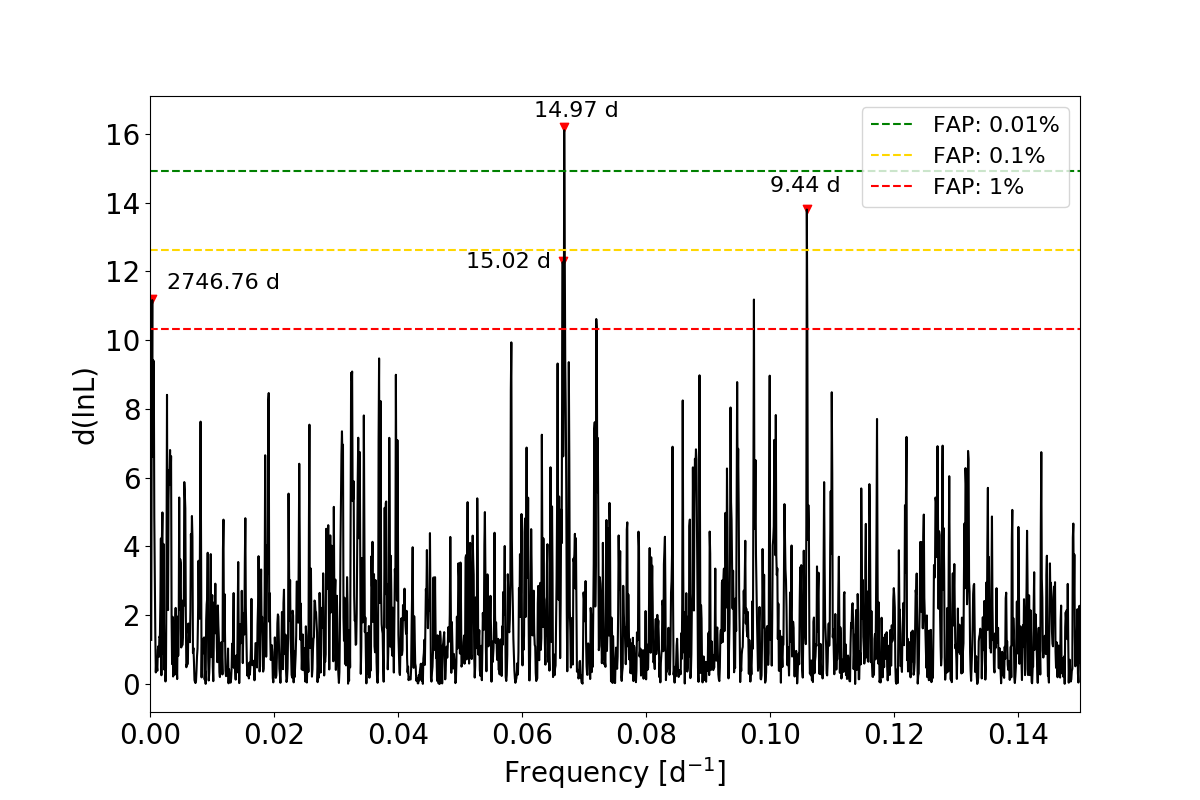}
		\caption{Maximum likelihood periodogram of the HIRES RVs. The periods corresponding to the main peaks are indicated, as well as the analytical false alarm probability (FAP) levels.}
		\label{fig:MLPHIRES}
	\end{figure}
	
	The MLP of the HIRES RVs (Fig. \ref{fig:MLPHIRES}) is dominated by a significant peak at 14.97 d. We note that the peak observed at 15.02 d is likely an alias due to the sampling. We recall that a quite significant peak at 14.9 days is present in the periodogram of the S-index extracted from HIRES spectra, therefore the peaks observed in the RVs are likely related to the first harmonic of $P_{\star,\,rot}$, and should be attributed to stellar activity. Besides a less significant peak at $\sim$9.4 d, which could be related to the second harmonic of $P_{\star,\,rot}$, there is one peak at lower frequency ($P=2746.76$ d) with low significance, which is likely related to the long-term modulation that can be guessed by eye looking at the time series in Fig. \ref{fig:rv_harps_hires_terra}. 
	
	\subsection{HARPS RVs}
	\label{sec:MLPHarps}
	
	\begin{figure}
		\centering
		\includegraphics[width=\hsize]{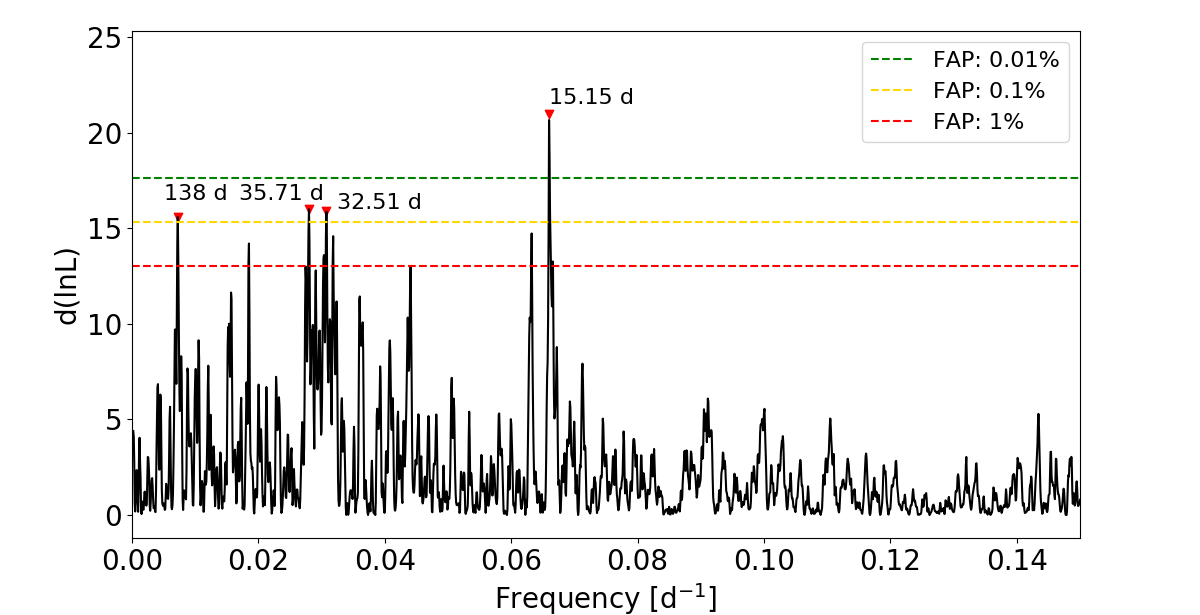}
		\includegraphics[width=\hsize]{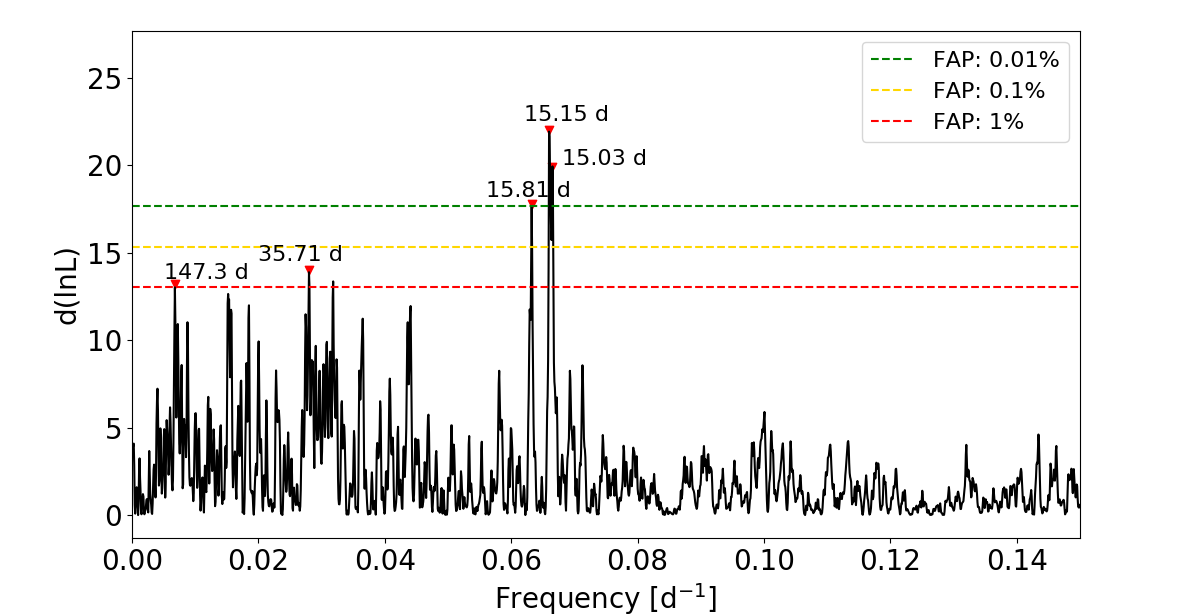}
		\includegraphics[width=\hsize]{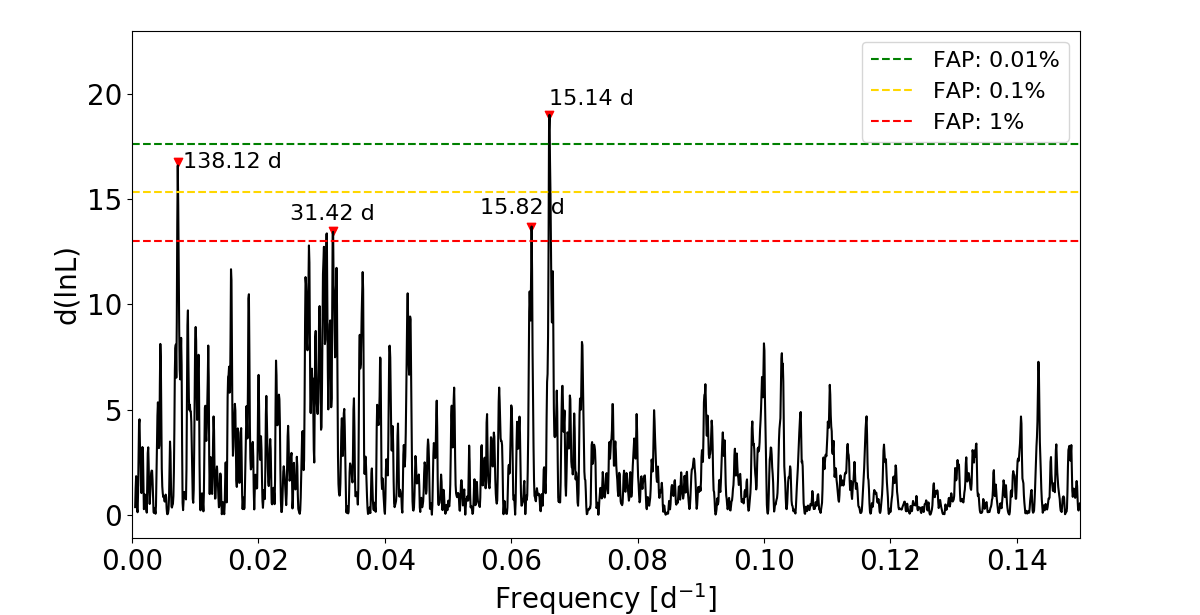}
		\caption{Maximum likelihood periodogram of the HARPS TERRA (first panel), NAIRA (second panel), and Trifonov et al. (2020; third panel) RVs (pre- and post-2015 combined dataset). The periods corresponding to the four highest peaks are indicated, as well as the analytical false alarm probability (FAP) levels.} 
		\label{fig:MLPHARPS}
	\end{figure}
	
	Fig. \ref{fig:MLPHARPS} shows the MLP of the full HARPS RV dataset, for TERRA, NAIRA, and Trifonov et al. (2020) datasets\footnote{Since no significant signals have been detected at high frequencies, we show the periodograms up to 0.1 $d^{-1}$ for more clarity.}. Signals in three different frequency ranges appear significant. As in the case of HIRES data, the highest peak occurs at $\sim$15 d. Peaks with FAP=0.1$\%$ (TERRA) occur at $\sim$33 and 36 days, and they are likely related to $P_{\rm \star,\, rot}$. For the NAIRA dataset, their significance is slightly lower. The MLP of TERRA and Trifonov et al. (2020) data shows another peak with FAP$<0.1\%$ at a period of 138 d, which has a counterpart at $\sim$147 d (FAP$\sim1\%$) in the MLP of the NAIRA RVs. Due to its high significance in two datasets over three, the nature of this signal needs to be examined in more detail using more sophisticated modelling than just a simple and inaccurate pre-whitening. A deeper investigation of this signal is the main focus of our study. We note that there is no evidence for the peak at $\sim$9 d observed in the MLP of HIRES data, and that there are no significant peaks at low frequencies. 
	
	\subsection{CARMENES RVs}
	\label{sec:MLPcarme}
	
\begin{figure}
		\centering
		\includegraphics[width=\hsize]{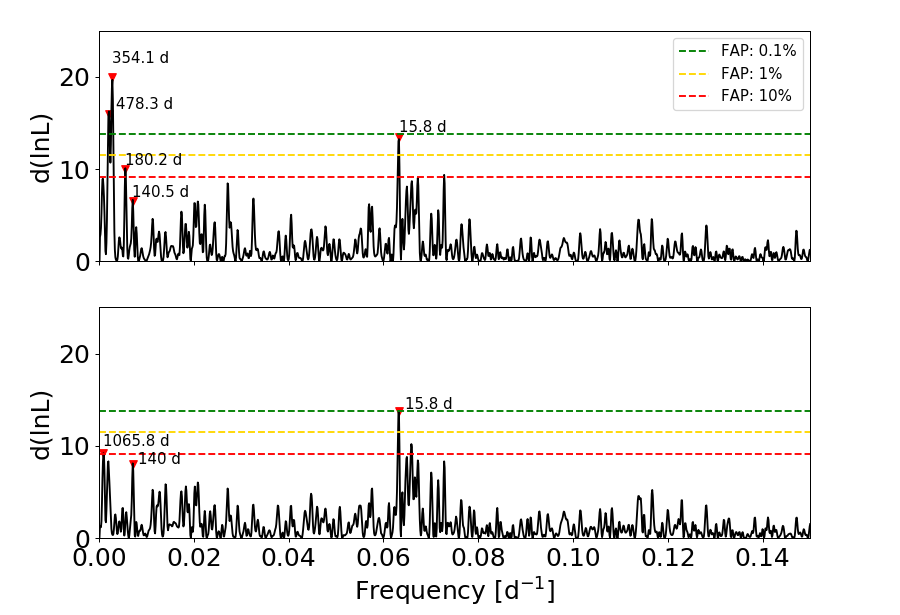}
		\caption{\textit{Upper panel:} Maximum likelihood periodogram of the original CARMENES-VIS RVs. The periods corresponding to the highest peaks are indicated, as well as the analytical false alarm probability (FAP) levels. \textit{Lower panel:} Maximum likelihood periodogram of the pre-whithened RVs, after removing the 354-d signal.} 
		\label{fig:MLPCARMENES}
	\end{figure}

	The MLP of the CARMENES-VIS RVs is  shown in the upper plot of Fig. \ref{fig:MLPCARMENES}. The RVs show very significant signals at 354 and 478\,d and at nearly half the rotational period (15.8 days). We also note the peak at 180 days, with lower FAP. By removing the 354-d signal with a pre-whitening, the 478-d and 180-d signals both disappear, meaning that they are related to each other, while the signal related to the stellar rotation becomes the most significant (second panel of Fig. \ref{fig:MLPCARMENES}). The pre-whitening increases the significance of the 140-d signal. However, its significance in the CARMENES MLP is low, but the MLP itself is dominated by a signal at 15.8 days. We emphasize that the signal at 354 d is detected only in the CARMENES RVs, and that, despite its high significance, we cannot attribute it to a companion of Gl~514, otherwise it would have been detected even in the periodogram of the HARPS RVs. This signal is likely spurious, and it might be due to micro-telluric spectral lines which are not masked by the standard RV extraction pipeline \textsc{SERVAL}, as a few preliminary tests suggest. Given its high significance we took this signal into account when modelling the RVs, and treated it as a sinusoid to fit only the CARMENES RVs. This choice looks reasonable because the MLP provides evidence for the presence of a very significant sinusoidal modulation. 
	
	\subsection{HARPS+CARMENES RVs}
	\label{sec:MLPHarpscarme}
\begin{figure}
		\centering
		\includegraphics[width=\hsize]{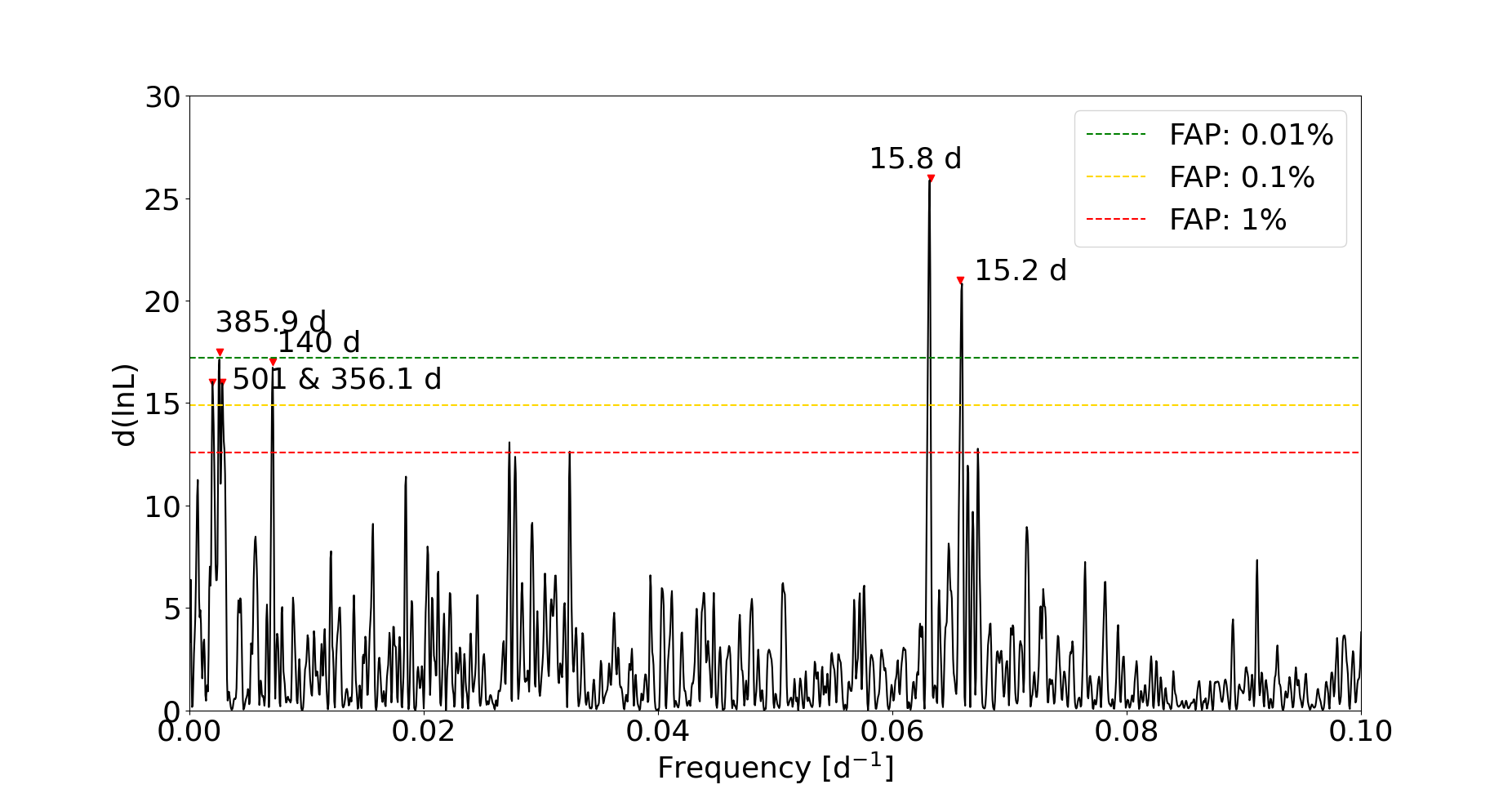}
		\includegraphics[width=\hsize]{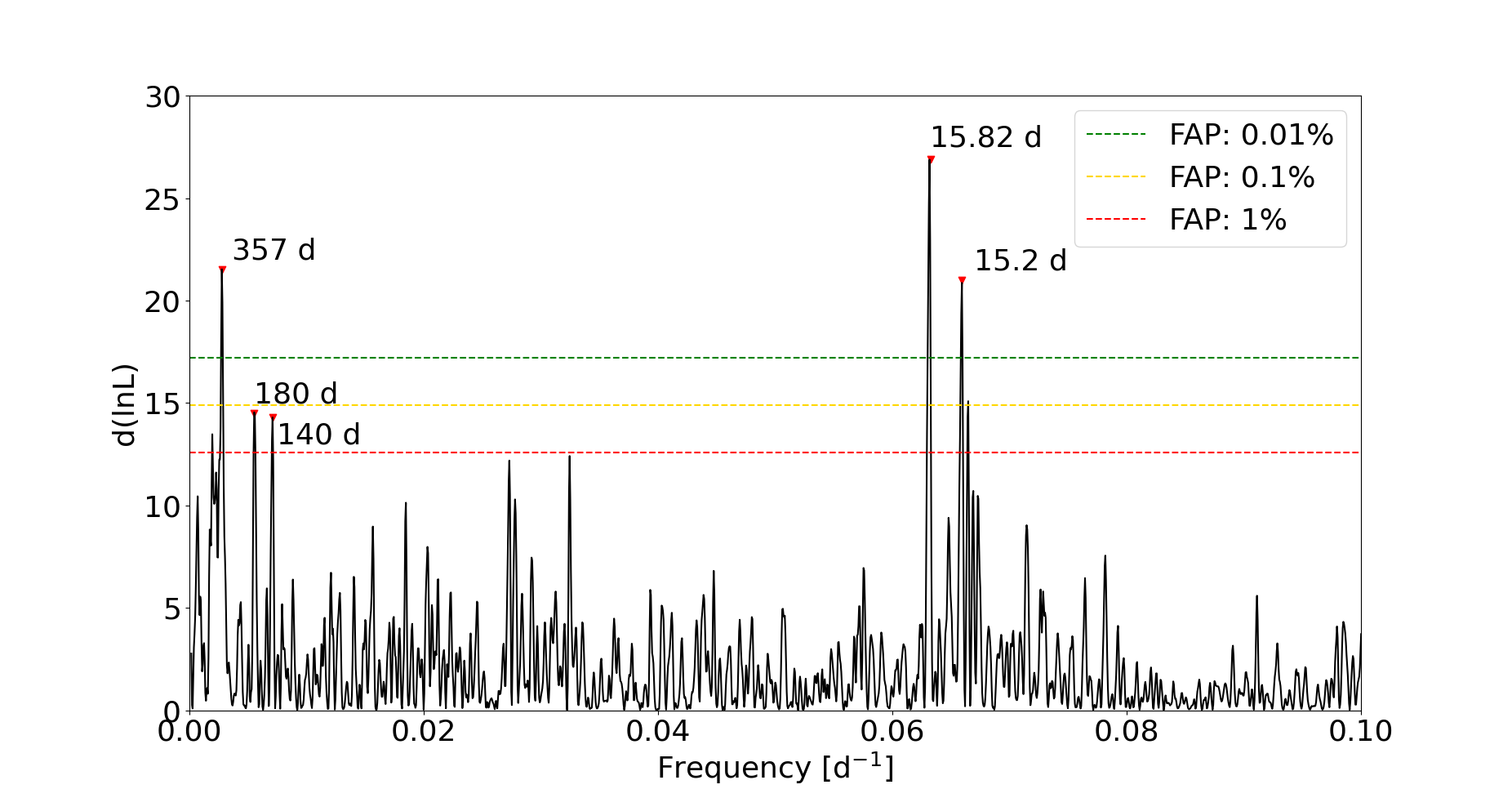}
		\includegraphics[width=\hsize]{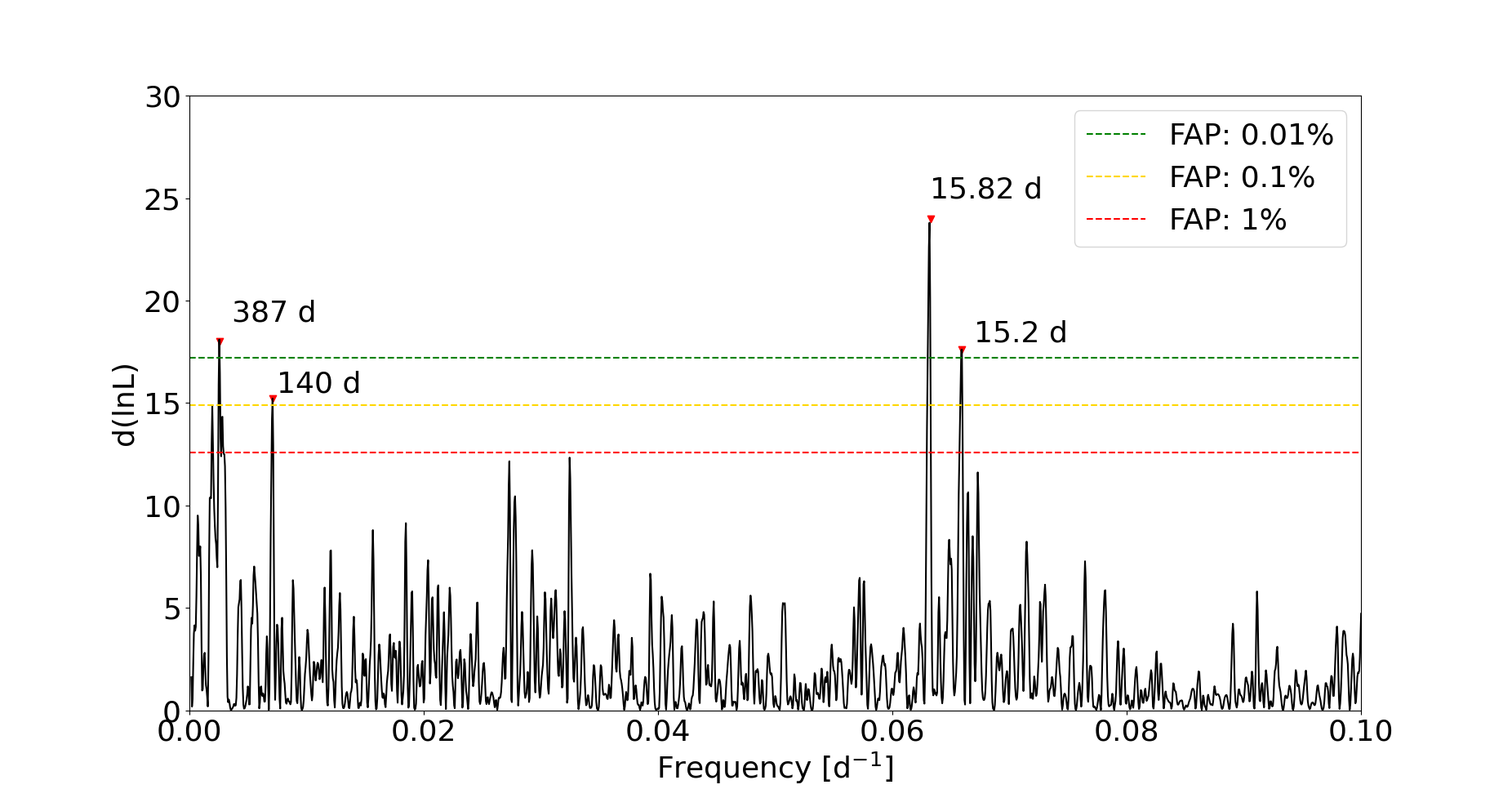}
		\includegraphics[width=\hsize]{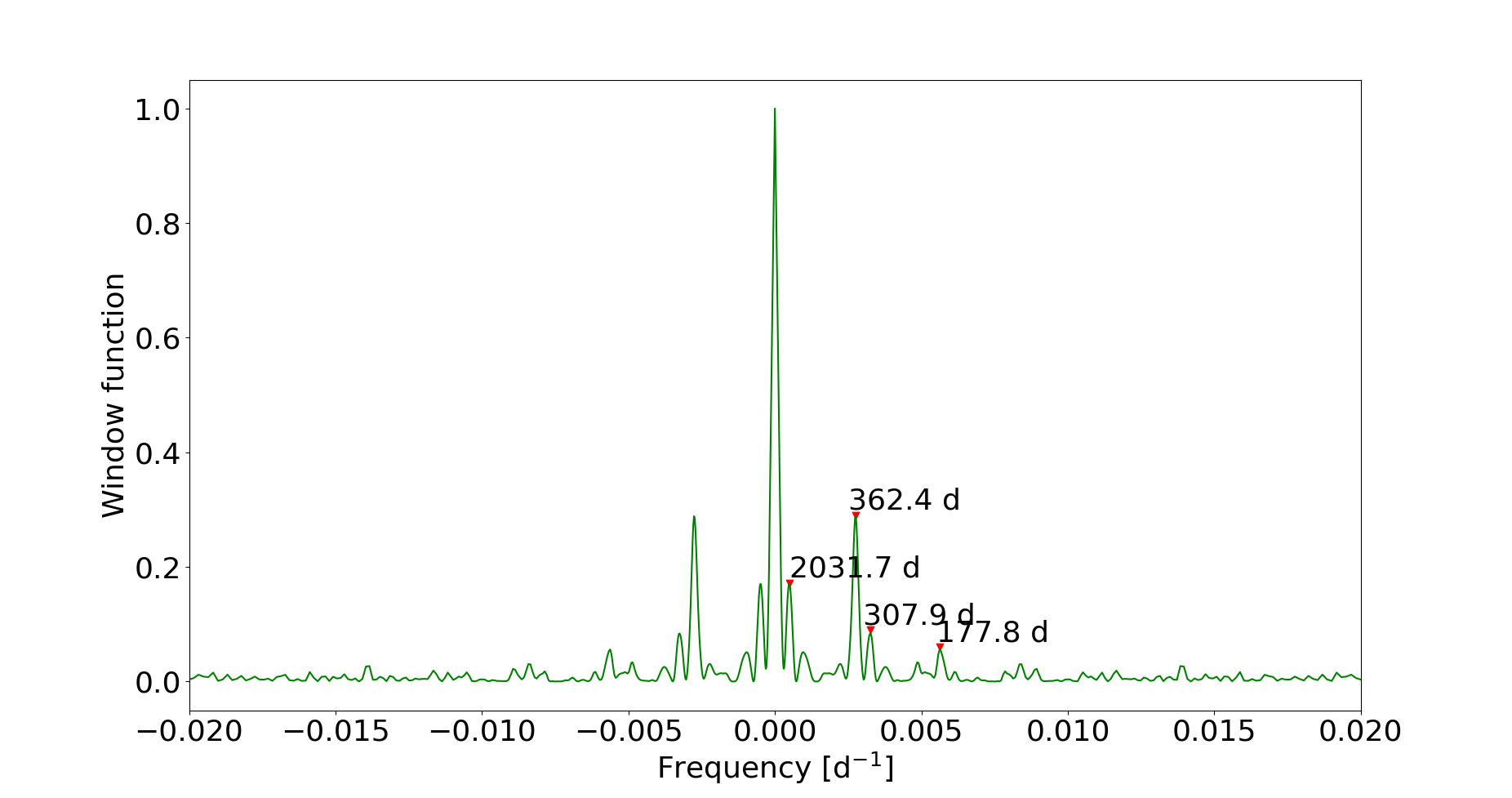}
		\caption{\textit{Panels 1--3:} Maximum likelihood periodograms of the joint HARPS (pre- and post-2015) and CARMENES-VIS RVs (panel 1: HARPS$_{\rm TERRA}$; panel 2: HARPS$_{\rm NAIRA}$; panel 3: HARPS$_{\rm Trifonov\,et\,al.}$). The periods corresponding to the highest peaks are indicated, as well as the analytical false alarm probability (FAP) levels. Offsets have been subtracted from the HARPS and CARMENES RVs, as derived by the MLP procedure. \textit{Lower panel:} window function of the combined dataset. } 
		\label{fig:MLPHARPSCARME}
	\end{figure}

	The MLPs of the combined HARPS and CARMENES-VIS dataset are shown in Fig. \ref{fig:MLPHARPSCARME} for all the different HARPS RV dataset, together with the window function.
	The dominant peak is located at 15.82 days in all the cases, with a best-fit semi-amplitude of the model sinusoid of 1.2--1.3 \ms. The peak at $\sim$140 days reaches a FAP of 0.01$\%$ with the HARPS$_{\rm TERRA}$ RVs, therefore becoming more significant after combining the HARPS and CARMENES RVs together. This period becomes more clear and its significance increases in the case of the HARPS$_{\rm NAIRA}$ RVs, while the FAP increases when using the RVs extracted by Trifonov et al. (2020). In all the cases, the peaks with FAP$\sim1\%$ delimit the range of periods where $P_{\star,\,\rm rot}$ is expected to be located.
	
	\subsection{HIRES+HARPS+CARMENES RVs}
	\label{sec:MLPHarpshirescarme}
	
\begin{figure}
		\centering
		\includegraphics[width=\hsize]{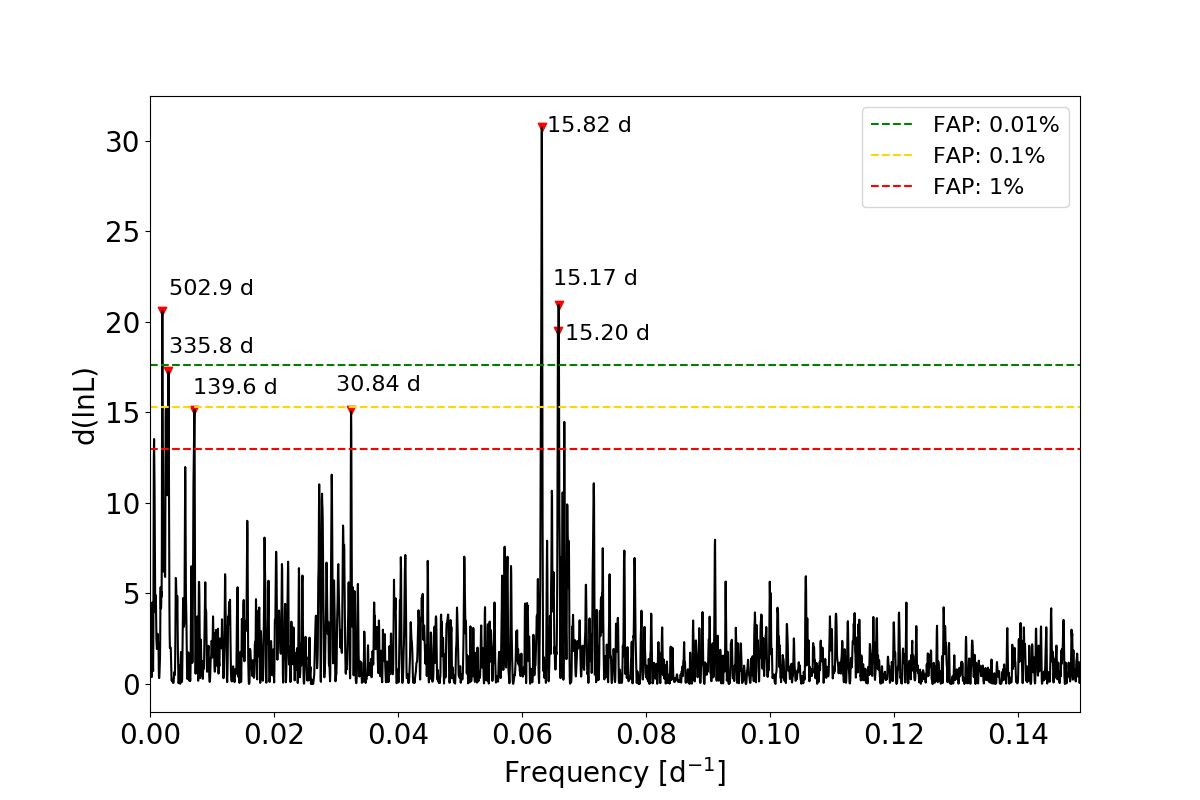}
		\includegraphics[width=\hsize]{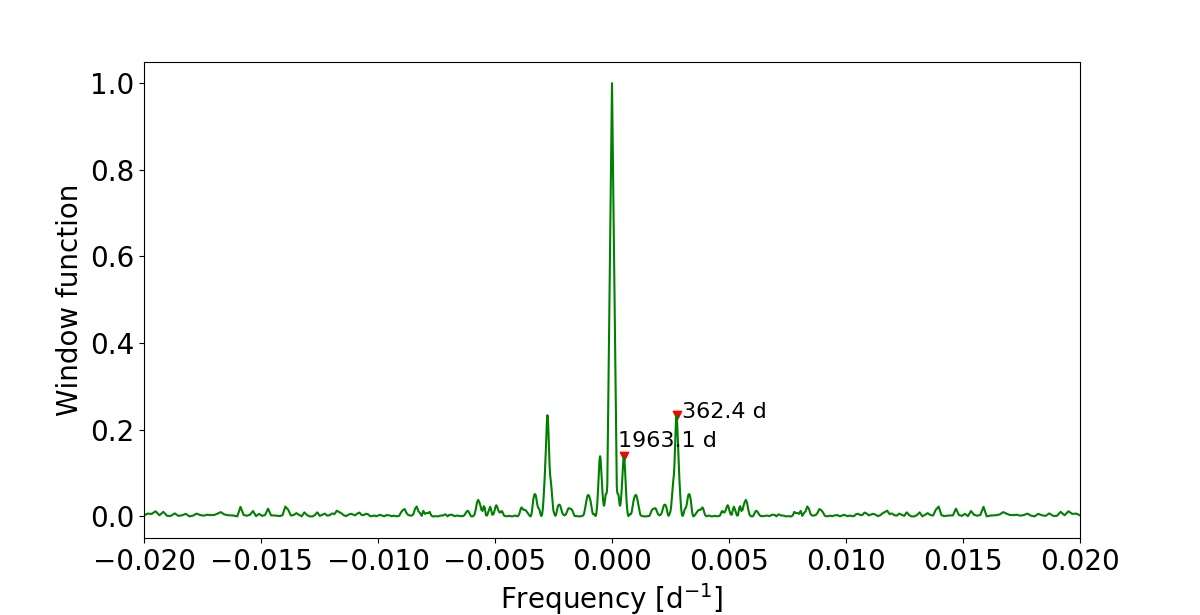}
		\caption{\textit{Upper panel:} Maximum likelihood periodogram of the joint HIRES, HARPS (TERRA; pre- and post-2015) and CARMENES-VIS RVs. The periods corresponding to the highest peaks are indicated, as well as the analytical false alarm probability (FAP) levels. \textit{Lower panel:} Window function of the combined dataset. } 
		\label{fig:MLPHIRESHARPSCARME}
	\end{figure}

	The MLP of the joint HIRES, HARPS$_{\rm TERRA}$, and CARMENES-VIS dataset is shown in Fig. \ref{fig:MLPHIRESHARPSCARME}. It is dominated by activity-related signals, and the inclusion of the HIRES data does not increase the significance of the $\sim$140-d signal, that has a higher FAP close to 0.1$\%$. 
	
	\smallskip
	In summary, the most remarkable results from the frequency content analysis are that the RVs derived from each instrument are dominated by signals related to stellar activity (corresponding to the first harmonic of the 30-d stellar rotation period), and that a peak at $\sim$140 days appears in the periodograms of the HARPS and CARMENES dataset separately, with different levels of significance. This peak does not have a counterpart in the periodograms of the activity diagnostics, and becomes significant (FAP $\sim$0.01$\%$) when the HARPS and CARMENES data are combined together, thanks to the increased time span of the whole dataset. In light of this, we will use more sophisticated models to fit the RVs which take into account the presence of activity-induced variability, and examine the possible planetary nature of the 140-d signal in the combined HARPS+CARMENES dataset. 
	
	
	\section{Radial velocity analysis}
	\label{sec:rvmodelling}
	
	We identified three different GP kernels which, based on the results in the previous Section, we deem all physically justified to be tested on the RVs of Gl~514. They are the standard quasi-periodic (QP) kernel (e.g \citealt{damasso2018}); the recently proposed quasi-periodic with cosine (QPC) kernel \citep{perger2021}; and the so called ``rotational'', or double simple harmonic oscillator (dSHO) kernel. In this study, we use all of them to fit the correlated stellar activity signal, and verify if the detection of the planetary candidate signal depends on the choice of the kernel. A significant detection in all the cases will indeed strongly support the real existence of the 140-d signal. 
	
	Not many cases where the rotational kernel has been applied to RVs are discussed in the literature (see, e.g., \citealt{benatti2021} for an application to a young and very active star), therefore still little is known about the performance of this kernel in recovering planetary Doppler signals, compared to that of the quasi-periodic family. The properties and hyper-parameters of each GP model are described in detail in Appendix \ref{app:GPkernels}. The QPC and rotational kernels contain terms that explicitly depend both on $P_{\star,\,rot}$ and its first harmonic, therefore they look particularly suitable for modelling the activity term in the RVs of Gl~514, assuming that the dominant signal at $\sim$15 d has to be attributed to stellar activity.
	
	We modelled the stellar activity term using the GP regression package \textsc{george} \citep{ambikasaran14}. We tested models with or without planetary signals (testing both circular and eccentric orbits), and explored the full (hyper-)parameter space with the publicly available Monte Carlo (MC) nested sampler and Bayesian inference tool \textsc{MultiNest v3.10} (e.g. \citealt{Feroz2019}), through the \textsc{pyMultiNest} wrapper \citep{Buchner2014}. The priors used in all the analysis described below are summarised in Table \ref{table:priorsrv}. Hereafter, subscripts $b$, $c$, and $d$ refer to the signal with period $\sim$140 days, to a possible innermost planet, and to a possible outermost planetary signal, respectively. The upper limit of the prior on the orbital period $P_b$ was set to 200 days based on the results of the MLP analysis, and in order to guarantee an unbiased analysis. The MC sampler was setup to run with 500 live points and a sampling efficiency of 0.5 in all the cases considered in our study. Model comparison analysis has been performed by calculating the difference $\Delta \ln \mathcal{Z}$ between the natural logarithm of the Bayesian evidences $\mathcal{Z}$ determined by \textsc{MultiNest} for each tested model. The a-priori probability is assumed to be the same for each model, and we follow the scale in \cite{feroz2011} to assess their statistical significance. 
	Hereafter, first we present the results of the analysis of the combined HARPS+CARMENES dataset, which is the most suitable --due to the overall high number of measurements, dense sampling, and precision-- for testing the presence of a low-mass companion in the HZ. Then, we performed an analysis including also the HIRES RV dataset that, due its large time span, is suitable to test the existence of a long-term signal.
	
	\begin{table}
		\caption{Priors used for modelling the HIRES, HARPS, and CARMENES VIS radial velocities. We adopted uninformative $\mathcal{U}$ priors for all the free (hyper-)parameters. }          
		\label{table:priorsrv}      
		\centering      
		\small
		\begin{tabular}{lc}       
			\hline\hline              
			\textbf{Parameter} & \textbf{Prior} \\    
			\hline              
			\noalign{\smallskip}
			\textbf{Stellar Activity - QP kernel} \\
			\noalign{\smallskip}
			$h$ [\ms] & $\mathcal{U}(0,10)$\tablefootmark{a} \\
			\noalign{\smallskip}
			$\theta$ [d] & $\mathcal{U}(20,50)$  \\
			\noalign{\smallskip}
			$\lambda_{\rm QP}$ [d] & $\mathcal{U}(0,1000)$  \\
			\noalign{\smallskip}
			$w$ & $\mathcal{U}(0,1)$  \\
			\noalign{\smallskip}
			\textbf{Stellar Activity - QPC kernel} \\
			\noalign{\smallskip}
			$h_1$ [\ms] & $\mathcal{U}(0,10)$\tablefootmark{a} \\
			\noalign{\smallskip}
			$h_2$ [\ms] & $\mathcal{U}(0,10)$\tablefootmark{a} \\
			\noalign{\smallskip}
			$\theta$ [d] & $\mathcal{U}(20,50)$  \\
			\noalign{\smallskip}
			$\lambda_{\rm QPC}$ [d] & $\mathcal{U}(0,1000)$  \\
			\noalign{\smallskip}
			$w$ & $\mathcal{U}(0,1)$  \\
			\noalign{\smallskip}
			\textbf{Stellar Activity - dSHO kernel} \\
			\noalign{\smallskip}
			$\log A$ & $\mathcal{U}$(0.05,10)\tablefootmark{a} \\
			\noalign{\smallskip}
			$\theta$ [d] & $\mathcal{U}$(20,50)  \\
			\noalign{\smallskip}
			$\log Q_0$ & $\mathcal{U}$(-10,10) \\
			\noalign{\smallskip}
			$\log \Delta Q$ & $\mathcal{U}$(-10,10) \\
			\noalign{\smallskip}
			$f$ & $\mathcal{U}$(0,10) \\
			\noalign{\smallskip}
			\textbf{First Keplerian} \\
			\noalign{\smallskip}
			$K_b$ [\ms] & $\mathcal{U}$(0,5)\\
			\noalign{\smallskip}
			$P_b$ [d] & $\mathcal{U}$(0,200)\tablefootmark{b} \\ 
			\noalign{\smallskip}
			& $\mathcal{U}$(100,200)\tablefootmark{c} \\
			\noalign{\smallskip}
			$T_{conj,\,b}$ [BJD-2\,450\,000] & $\mathcal{U}$(8500,8750) \\
			\noalign{\smallskip}
			$\sqrt{e_{\rm b}}\cos\omega_{\rm \star\:,b}$ & $\mathcal{U}$(-1,1)\\
			\noalign{\smallskip}
			$\sqrt{e_{\rm b}}\sin\omega_{\rm \star\:,b}$ & $\mathcal{U}$(-1,1)\\
			\noalign{\smallskip}
			\textbf{Second Keplerian (innermost)} \\
			\noalign{\smallskip}
			$K_c$ [\ms] & $\mathcal{U}$(0,5)\\
			\noalign{\smallskip}
			$P_c$ [d] & $\mathcal{U}$(0,100)  \\
			\noalign{\smallskip}
			$T_{conj,\,c}$ [BJD-2\,450\,000] & $\mathcal{U}$(8500,8650) \\
			\noalign{\smallskip}
			$\sqrt{e_{\rm c}}\cos\omega_{\rm \star\:,c}$ & $\mathcal{U}$(-1,1)\\
			\noalign{\smallskip}
			$\sqrt{e_{\rm c}}\sin\omega_{\rm \star\:,c}$ & $\mathcal{U}$(-1,1)\\
			\noalign{\smallskip}
			\textbf{Third Keplerian (outermost)\tablefootmark{d}} \\
			\noalign{\smallskip}
			$K_d$ [\ms] & $\mathcal{U}$(0,5)\\
			\noalign{\smallskip}
			$P_d$ [d] & $\mathcal{U}$(200,4350)  \\
			\noalign{\smallskip}
			$T_{conj,\,d}$ [BJD-2\,450\,000] & $\mathcal{U}$(4500,9000) \\
			\noalign{\smallskip}
			$\sqrt{e_{\rm d}}\cos\omega_{\rm \star\:,d}$ & $\mathcal{U}$(-1,1)\\
			\noalign{\smallskip}
			$\sqrt{e_{\rm d}}\sin\omega_{\rm \star\:,d}$ & $\mathcal{U}$(-1,1)\\
			\noalign{\smallskip}
			\textbf{Additional sinusoid}\tablefootmark{e} \\
			\noalign{\smallskip}
			$K_{365-d}$ [\ms] & $\mathcal{U}$(0,10)\\
			\noalign{\smallskip}
			$P_{365-d}$ [d] & $\mathcal{U}$(340,370)  \\
			\noalign{\smallskip}
			$T_{0,\,365-d}$ [BJD-2\,450\,000] & $\mathcal{U}$(8500,8900) \\
			\noalign{\smallskip}
			\textbf{Instrument-related} \\
			\noalign{\smallskip}
			$\gamma$ [\ms] & $\mathcal{U}$(-20,20)\tablefootmark{a} \\
			\noalign{\smallskip}
			$\sigma_{jit}$ [\ms] & $\mathcal{U}$(0,10)\tablefootmark{a} \\
			\noalign{\smallskip}
			\hline
		\end{tabular}
		\tablefoot{
			\tablefoottext{a}{The same prior was used for the corresponding parameter related to HIRES, HARPS (pre- and post-2015), and CARMENES-VIS data.}
			\tablefoottext{b}{Prior used when modelling only one Keplerian.}
			\tablefoottext{c}{Prior used when including a second Keplerian to model planetary signals with $P<200$ days.}
			\tablefoottext{d}{Priors for a planetary signal with period longer than 200 d, that we modelled combining HIRES+HARPS+CARMENES RVs. }
			\tablefoottext{e}{Priors used to model the yearly signal only seen in the CARMENES-VIS RVs.}
		}
	\end{table}
	
	\subsection{HARPS+CARMENES dataset}
	\label{sec:harpscarmervanalysis}
	We tested the three GP kernels on the 378 RVs collected with HARPS and CARMENES-VIS. In addition to the candidate planetary signal, we included a sinusoid to model the 1-year signal only in the CARMENES-VIS data (see Sect. \ref{sec:MLPcarme}).
	Concerning the choice of the prior for the period $P_{\rm 365-d}$ of this signal, initially we have considered a uniform prior between 300 and 600 days, in order to sample also the $\sim$480 days signal seen in the MLP of Fig. \ref{fig:MLPHARPSCARME}. The aim was to check which period would have been selected by the MC sampler. The resulting posterior is not multi-modal, and it looks symmetric around 355 days ($\pm$8 days), as expected since the highest peak in the MLP occurs at this period. This test allowed us to restrict safely the range of the priors for $P_{\rm 365-d}$ and  $T_{\rm0,\,365-d}$, with a considerable decrease in the computing time.   
	We tested both models with the eccentricity $e_b$ of the planet candidate fixed to zero, or treated as a free parameter together with the argument of periastron $\omega_{\star,\, \rm b}$, by using the parametrization $\sqrt{e_{\rm b}}\cos\omega_{\rm \star\:,b}$ and $\sqrt{e_{\rm b}}\sin\omega_{\rm \star\:,b}$. Taking into account the different wavelength range covered by HARPS and CARMENES, the spectrographs could be sensitive to stellar activity at a different level, therefore, even at the cost of increasing the number of free parameters, we adopted different GP amplitudes for each instrument to get a more reliable fit (the same approach is used when including the HIRES RVs, as described below). 
	
	As stated, we repeated the same analyses using HARPS RVs extracted with alternative pipelines. We disclose that we obtained best-fit values of the planetary parameters which are all in agreement within the errors for all the three RV template matching extraction methods. This is indeed an important outcome which supports the detection of Gl~514\,b. For the sake of simplicity, and with no loss of information, hereafter we discuss only the results obtained using \textsc{TERRA}. A summary of the results obtained for the main planetary parameters using \textsc{NAIRA} and \cite{trifonov20} RVs is provided in Appendix \ref{app:nairatrifon}. The main outcomes of the GP analysis are the following: independently from the GP kernel, the model including a Keplerian for a candidate planet is the most significant (strong evidence in all the cases, i.e. $\ln\mathcal{Z}_{1p}$-$\ln\mathcal{Z}_{0p}>5$); the planetary-like signal is detected in all the three cases with $K_b\sim$1.2 \ms (5.5--6$\sigma$ significance), an orbital period $P_b\sim140$ days, determined with an error bar lower than 1 day, and a 3$\sigma$ significant eccentricity $e_b\sim 0.4-0.5$; we get moderate-to-strong evidence ($\Delta\ln\mathcal{Z}=2.5-2.8$) for the model with a Keplerian over a model with a planet on a circular orbit in two cases over three\footnote{This result is generally confirmed when using the two alternative HARPS RV dataset (see Appendix \ref{app:nairatrifon}).}. Overall, the model with the lowest Bayesian evidence is that using the QP kernel, while the highest statistical evidence belongs to the model with the QPC kernel, For that reason, we elect the latter as our reference model. We emphasise that the parameters of the planetary signal are all in agreement for each tested GP kernel.   
	
\begin{figure}
		\centering
		\includegraphics[width=0.5\textwidth]{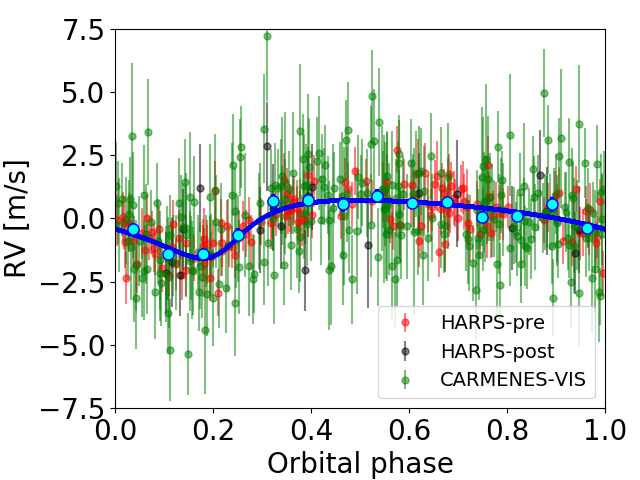}
		\caption{Spectroscopic orbit of Gl~514\,b (blue solid line) based on the best-fit solution tabulated in Table \ref{tab:harpscarmeQPC} (case $e_{b}\neq0$). The error bars of the measurements include uncorrelated jitter terms added in quadrature to the formal RV uncertainties. Cyan dots correspond to RV data averaged over 15 phase bins. }
		\label{fig:foldplanetQPC}
	\end{figure}

	The best-fit values for the parameters of this model are reported in Table \ref{tab:harpscarmeQPC}, while we summarise the results for the QP and dSHO kernel in Tables \ref{tab:harpscarmeQP} and \ref{tab:harpscarmeSHO}. 
	Fig. \ref{fig:foldplanetQPC} shows the spectroscopic orbit of the candidate planet Gl~514\,b based on the best-fit model, and in Fig. \ref{fig:activityQPC} we show the QPC component of the RV time series corresponding to stellar activity. For comparison, we show in Fig. \ref{fig:fold_circ_harpscarmenesQPC} the spectroscopic orbit of Gl~514\,b for the circular case (QPC kernel).
	
	Concerning the results for the hyper-parameters of the correlated activity term QPC, we note that the stellar rotation period $\theta$ is retrieved with high-precision and well in agreement with the literature values, and the characteristic evolutionary time-scale $\lambda_{\rm QPC}$ is nearly four times larger than the value of $\theta$. It is not unusual that, for early-type M dwarfs with rotation periods similar to that of Gl~514, the value of $\lambda_{QPC}$ obtained from the fit of the RVs is of the same order of magnitude of $P_{\star,\, \rm rot}$, as found for instance in Gl~686 ($\theta=37.0^{+5.4}_{-14.4}$ and $\lambda_{\rm QPC}\sim47$ days; \citealt{affer2019}), K2-3 ($\theta=40.4^{+1.1}_{-1.9}$ and $\lambda_{\rm QPC}\sim80$ days; \citealt{damasso2018}), GJ~3998 ($\theta=31.8^{+0.6}_{-0.5}$ and $\lambda_{\rm QPC}\sim69$ days; \citealt{affer2016}), or Gl~15A ($\theta=46.7^{+4.8}_{-4.3}$ and $\lambda_{\rm QPC}\sim72$ days; \citealt{pinamonti18}), where the RVs were modelled with a quasi-periodic GP kernel, and we used the relation $\lambda_{\rm QPC}=2\cdot\lambda_{\rm QP}$ for deriving the time-scales reported in parenthesis. 
	
	\begin{table*}[h]
		\small
		\caption{Best-fit parameters obtained for the model with the QPC kernel applied to HARPS$_{\rm TERRA}$ and CARMENES-VIS RVs. The uncertainties are calculated from the $16^{\rm th}$ and $84^{\rm th}$ percentiles of the posterior distributions. The model with $e_b\neq$0 (values in bold) is our adopted solution}          
		\label{tab:harpscarmeQPC} 
		\centering                 
		\begin{tabular}{lcc}       
			\hline\hline              
			\textbf{Fitted parameter} & \multicolumn{2}{c}{\textbf{Best-fit value}} \\    
			\hline              
			& $e_b$=0 & $e_b\neq$0 \\
			\hline           
			\noalign{\smallskip}
			\noalign{\smallskip}
			$h_{\rm 1,\,HARPS}$ [\ms] & $2.1^{+0.4}_{-0.3}$ & \textbf{2.2$^{+0.4}_{-0.3}$} \\
			\noalign{\smallskip}
			$h_{\rm 2,\,HARPS}$ [\ms] & $1.4\pm0.4$ & \textbf{1.4$\pm0.4$}\\
			\noalign{\smallskip}
			$h_{\rm 1,\,CARMENES}$ [\ms] & $1.1\pm0.3$ & \textbf{1.0$^{+0.3}_{-0.2}$} \\
			\noalign{\smallskip}
			$h_{\rm 2,\,CARMENES}$ [\ms] & $1.3\pm0.3$ & \textbf{1.3$\pm0.3$} \\
			\noalign{\smallskip}
			$\theta$ [d] & $30.7\pm0.3$ & \textbf{30.6$\pm0.3$}\\
			\noalign{\smallskip}
			$\lambda_{\rm QPC}$ [d] & $121^{+30}_{-28}$ & \textbf{122$^{+27}_{-24}$}\\
			\noalign{\smallskip}
			$w$ & $0.34^{+0.05}_{-0.04}$ & \textbf{0.34$^{+0.05}_{-0.04}$} \\
			\noalign{\smallskip}
			$\gamma_{HARPS\,pre-2015}$ [\ms] & $-0.3\pm0.5$ & \textbf{-0.4$\pm0.5$}\\
			\noalign{\smallskip}
			$\gamma_{HARPS\,post-2015}$ [\ms] & $0.0\pm1.1$ & \textbf{0.0$\pm1.1$}\\
			\noalign{\smallskip}
			$\gamma_{CARMENES}$ [\ms] & $0.1\pm0.3$ & \textbf{0.1$\pm0.3$} \\
			\noalign{\smallskip}
			$\sigma_{jit,\,HARPS\,pre-2015}$ [\ms] & $0.8\pm0.1$ & \textbf{0.7$\pm0.1$} \\
			\noalign{\smallskip}
			$\sigma_{jit,\,HARPS\,post-2015}$ [\ms] & $1.5^{+0.5}_{-0.4}$ & \textbf{1.6$^{+0.5}_{-0.4}$} \\
			\noalign{\smallskip}
			$\sigma_{jit,\,CARMENES}$ [\ms] & $0.9\pm0.2$ & \textbf{0.9$\pm0.2$} \\
			\noalign{\smallskip}
			$K_b$ [\ms] & $0.91\pm0.18$ & \textbf{1.15$^{+0.21}_{-0.19}$}\\
			\noalign{\smallskip}
			$P_b$ [d] & $139.90^{+0.68}_{-0.65}$ & \textbf{140.43$\pm0.41$} \\
			\noalign{\smallskip}
			$T_{conj,\,b}$ [BJD-2450000] & $8670.29^{+6.43}_{-6.69}$ & \textbf{8696.21$^{+7.66}_{-13.71}$} \\
			\noalign{\smallskip}
			$\sqrt{e_{\rm b}}\cos\omega_{\rm \star,\:b}$ & - & \textbf{-0.604$^{+0.267}_{-0.143}$}  \\
			\noalign{\smallskip}
			$\sqrt{e_{\rm b}}\sin\omega_{\rm \star,\:b}$ & - & \textbf{-0.209$^{+0.258}_{-0.261}$}  \\
			\noalign{\smallskip}
			$K_{365-d,\,CARMENES}$ [\ms] & $1.2\pm0.3$ & \textbf{1.2$\pm0.3$} \\
			\noalign{\smallskip}
			$P_{365-d,\,CARMENES}$ [d] & $356.98^{+7.92}_{-8.15}$ & \textbf{357.76$^{+7.29}_{-8.15}$} \\
			\noalign{\smallskip}
			$T_{0,365-d \,CARMENES}$ [BJD-2450000] & $8784.4^{+18.7}_{-19.4}$ & \textbf{8785.93$^{+17.56}_{-19.13}$} \\
			\noalign{\smallskip}
			\hline
			\noalign{\smallskip}
			\textbf{Derived parameter} \\ 
			\noalign{\smallskip}
			\hline
			\noalign{\smallskip}
			eccentricity, $e_{\rm b}$ & - & \textbf{0.45$^{+0.15}_{-0.14}$} \\
			\noalign{\smallskip}
			arg. of periapsis, $\omega_{\rm \star,\:b}$ & - & \textbf{-2.49$^{+5.35}_{-0.47}$}  \\
			\noalign{\smallskip}
			min. mass, $m_b$ $\sin i_b$ [$\mearth$] & 4.7$\pm1.0$ & \textbf{5.2$\pm0.9$} \\
			\noalign{\smallskip}
			semi-major axis, $a_b$ [au] & $0.421^{+0.014}_{-0.015}$ & \textbf{0.422$^{+0.014}_{-0.015}$} \\
			\noalign{\smallskip}
			periapsis [au] & - & \textbf{0.231$^{+0.058}_{-0.060}$} \\
			\noalign{\smallskip}
			apoapsis [au] & - & \textbf{0.612$^{+0.064}_{-0.061}$} \\
			\noalign{\smallskip}
			equilibrium temperature, $T_{\rm eq,\,b}$ [K] & $196\pm10$\tablefootmark{a} & Orbit-averaged\tablefootmark{b}: \textbf{202$\pm$11 } \\
			\noalign{\smallskip}
			& & Apoapsis: \textbf{162$^{+12}_{-11}$} \\
			\noalign{\smallskip}
			& & Periapsis: \textbf{264$^{+45}_{-31}$ }\\ 
			\noalign{\smallskip}
			Insolation flux\tablefootmark{c}, $S_b$ [$S_\oplus$] & $0.24^{+0.06}_{-0.05}$ & Orbit-averaged: \textbf{0.28$^{+0.07}_{-0.06}$ } \\
			\noalign{\smallskip}
			& & Apoapsis: \textbf{0.114$^{+0.037}_{-0.030}$ }\\
			\noalign{\smallskip}
			& & Periapsis: \textbf{0.79$^{+0.72}_{-0.31}$} \\ 
			\hline
			\noalign{\smallskip}
			$\ln\mathcal{Z}$ & -958.0 & \textbf{-955.2} \\
			\noalign{\smallskip}
			$\ln\mathcal{Z}_{1p}$-$\ln\mathcal{Z}_{0p}$ & +3.3 & \textbf{+6.1}\\
			\noalign{\smallskip}
			\hline
		\end{tabular}
		\tablefoot{
			\tablefoottext{a}{Derived from the relation $T_{\rm eq}$=$T_{\rm eff}\cdot\sqrt{\frac{R_\star}{2a_b}}\cdot(1-A_B)^{0.25}$, assuming Bond albedo $A_B$=0.\\}
			\tablefoottext{b}{This is the average equilibrium temperature based on the stellar flux received by the planet averaged over the eccentric orbit. This flux-averaged temperature scales with the eccentricity as $(1-e_b^2)^{-\frac{1}{8}}$ with respect to the value for a circular orbit.\\}
			\tablefoottext{c}{For the circular orbit, it is derived from the equation $S_{\rm b}$=$\frac{L_{\star}}{L_{\rm \odot}}\cdot(\frac{au}{a_b})^2$. For the eccentric orbit, the temporal average insolation flux scales with the eccentricity as $\frac{1}{\sqrt{1-e_b^2}}$ with respect to the value for a circular orbit with the same semi-major axis (e.g. \citealt{williams2002})}
		}
	\end{table*}

	\clearpage
	
	\subsubsection{Stability of the 140-day signal over time } 
	\label{sec:stability140d}
\begin{figure}
		\centering
		\includegraphics[width=0.9\linewidth]{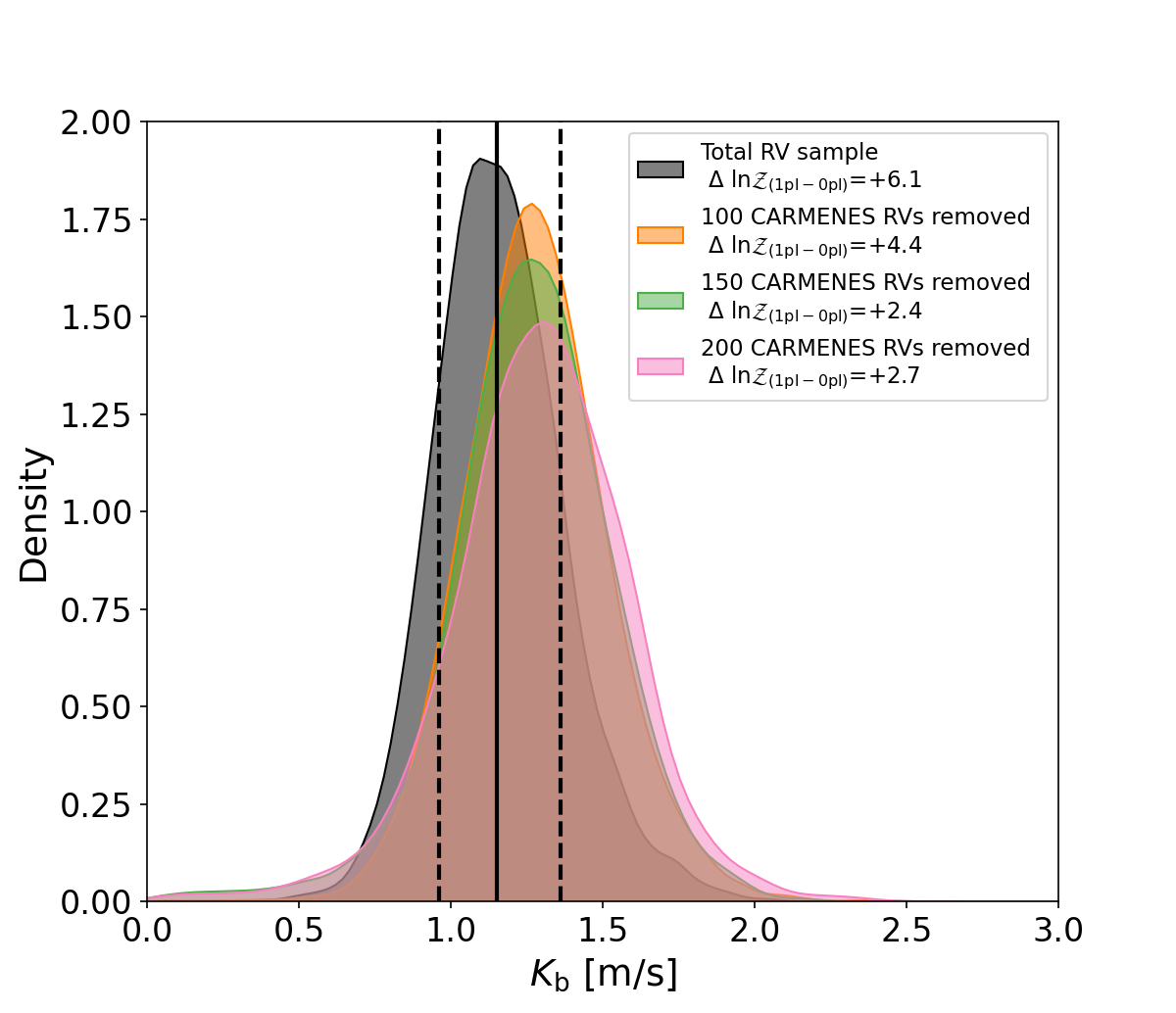}
		\includegraphics[width=0.9\linewidth]{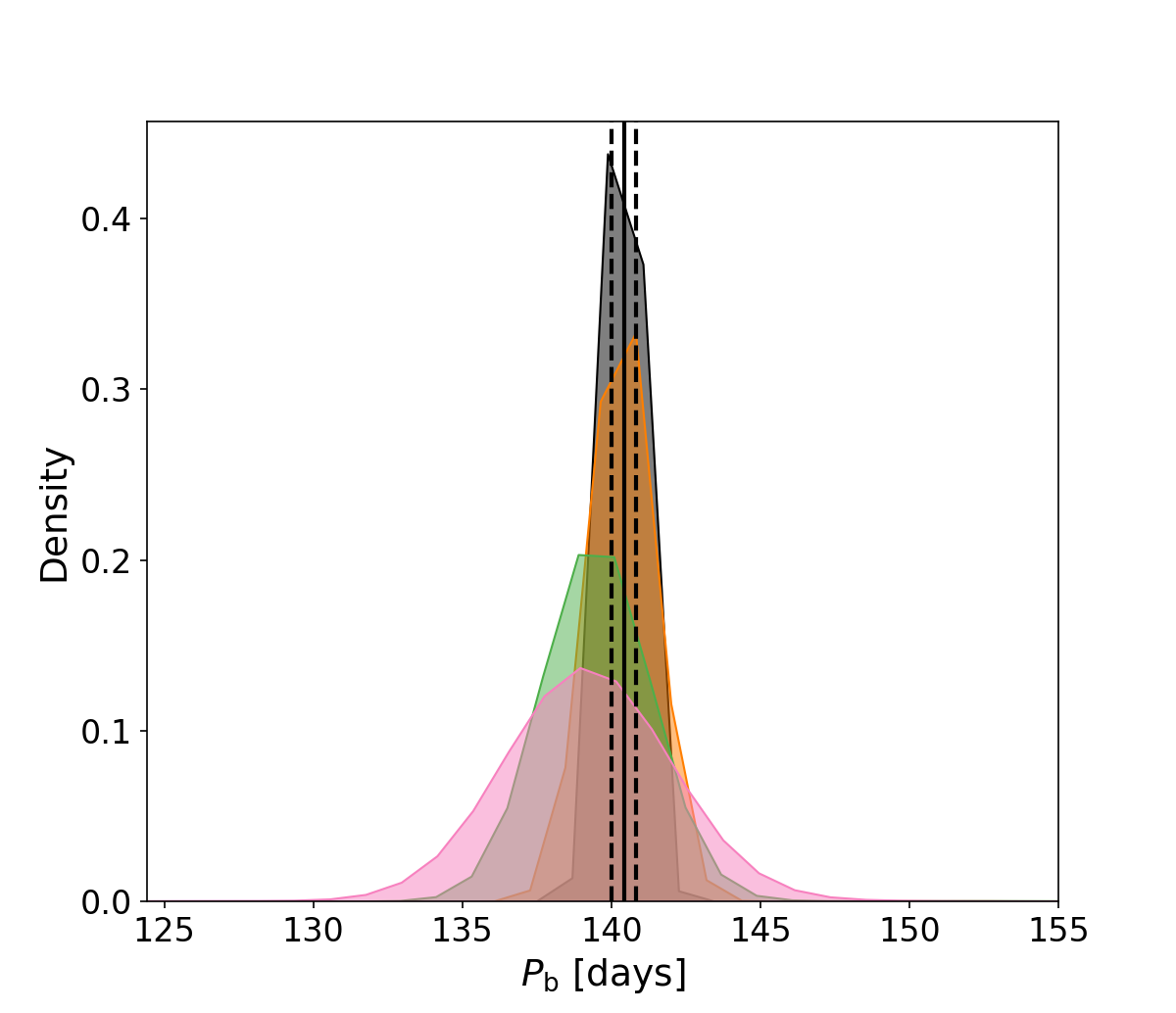}
		\includegraphics[width=0.9\linewidth]{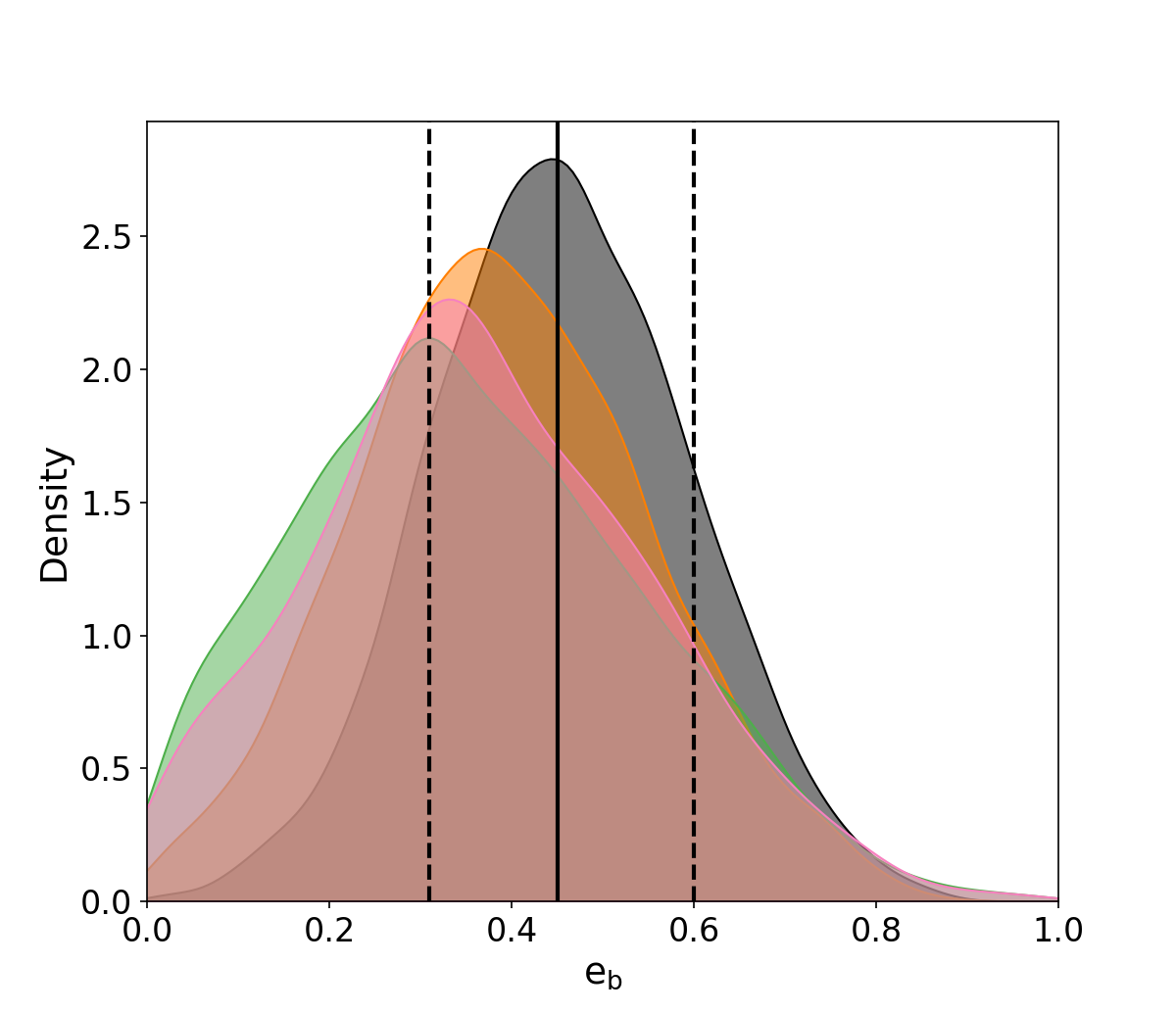}
		\caption{Posterior distributions of the semi-amplitude $K_b$ (top panel), orbital period $P_b$ (middle panel), and eccentricity $e_b$ (bottom panel) of the candidate planet Gl514\,b, derived from a GP QPC analysis of the HARPS$_{\rm TERRA}$+CARMENES VIS radial velocities, after removing an increasing number of CARMENES measurements, beginning from the last epoch. Vertical lines indicate the $16^{\rm th}$ and $84^{\rm th}$ (dashed) and $50^{\rm th}$ (solid) percentiles for the whole HARPS$_{\rm TERRA}$+CARMENES dataset. $\Delta\ln\mathcal{Z}$ denotes the Bayesian evidence difference between a model with and without a Keplerian included. }
		\label{fig:teststabilityQPC}
	\end{figure}

	If the 140-d signal is due to a companion of Gl~514, then its properties, as semi-amplitude and period, must tend to values which are in agreement within the error bars as a function of the progressive increase of the number of RVs. To verify this behaviour, we repeated the GP QPC fit described above on the HARPS$_{\rm TERRA}$+CARMENES RVs (keeping all the priors unchanged) after removing the last 100, 150, and 200 CARMENES measurements. This corresponds to a reduction of 23$\%$, 34.4$\%$, and 45.9$\%$ of the total HARPS$_{\rm TERRA}$+CARMENES data, respectively, and 
	to a decrease in the total time baseline (equal to 6154 days) of 675, 797, and 1059 days. The posterior distributions of the semi-amplitude $K_b$, orbital period $P_b$, and eccentricity $e_b$ of the candidate planet Gl~514\,b derived for each cropped dataset are shown in Fig. \ref{fig:teststabilityQPC}, and are compared with the posteriors obtained for the whole sample of HARPS$_{\rm TERRA}$+CARMENES RVs. The posteriors of $K_b$ are all in agreement within the uncertainties, and those of $P_b$ become narrower and move closer to that of the whole dataset with the increasing number of RVs. The posteriors for $e_b$ are all in agreement within the uncertainties, and the eccentricity moves to higher values and becomes more significant with the increasing number of data. The model including a Keplerian for planet b becomes more significant over the model without the planetary signal with the increasing number of RVs, as indicated by the values of the Bayesian evidence differences $\Delta\ln\mathcal{Z}$ in the plot legend. These check demonstrates the persistence of the 140-d signal over time. In Appendix \ref{app:check140d} we present an independent cross-check analysis to test the nature of this signal. Even those results support the planetary hypothesis.

	\subsubsection{Testing the 2-planet model}
	\label{sec:2planetfit}
	
	Given the large number of HARPS$_{\rm TERRA}$+CARMENES RVs, and their dense sampling, we investigated the existence of an additional planetary signal, focusing on orbits internal to that of Gl~514\,b. The RMS of the RV residuals of the 1-planet GP models is in the range 1.2--1.4 \ms, and Fig. \ref{fig:GLSresid} shows their GLS periodograms calculated for each GP model. The periodograms in blue, which correspond to residuals with the activity signal not removed from the original dataset, show power at periods longer than $P_{\star,\,rot}$. Since signals with periods longer than $P_{\star,\,rot}$ are usually strongly suppressed in the residuals of a GP model, the periodograms shown in red, corresponding to residuals with the activity signal removed from the original dataset, are not particularly informative to drive the search for signals with $P>P_{\star,\,rot}$. From this analysis, we expect to get at best hints for a possible sub \ms signal worthy of future follow-up with extreme precision RV. The set-up is the same that we adopted for the previous analysis, and we tested all the GP kernels used for the case with one Keplerian. To keep the analysis unbiased, we adopted uniform and large priors to sample the orbital periods, setting them to $\mathcal{U}$(0,100) and $\mathcal{U}$(100,200) days. In the following, the second planetary signal is identified with the subscript \textit{c}, while the subscript \textit{b} is still referred to the 140-d planet.\\
	\newline
	\textit{Test 1: QP kernel.} For planet Gl~514~b we found $K_b=1.28^{+0.21}_{-0.18}$ \ms, $P_b=140.24^{+0.44}_{-0.56}$ days, and $e_b=0.46^{+0.11}_{-0.12}$, and those of the possible innermost companion are $K_c=0.74\pm0.31$ \ms, $P_c=63.67^{+0.29}_{-2.11}$ days, and $e_c=0.53^{+0.18}_{-0.31}$. The full set of posteriors are shown in Fig. \ref{fig:corner2plecc}. The Bayesian evidence is $\ln\mathcal{Z}=-960.4$, which is equal to the Bayesian evidence of the QP model that includes only one Keplerian.\\
	\newline
	\textit{Test 2: QPC kernel.} The retrieved parameters for planet b are $K_b=1.21\pm0.17$ \ms, $P_b=140.40^{+0.27}_{-0.39}$ days, and $e_b=0.47^{+0.12}_{-0.08}$, while those of the possible innermost companion are $K_c=0.92^{+0.19}_{-0.30}$ \ms, $P_c=63.64^{+0.20}_{-1.93}$ days, and $e_c=0.58^{+0.13}_{-0.20}$. The Bayesian evidence is $\ln\mathcal{Z}=-953.9$, which is slightly higher than the Bayesian evidence of the model that includes only one Keplerian ($\Delta\ln\mathcal{Z}$=+1.3). This is not enough to claim that this model is significantly favoured over the one with only the Keplerian for planet b included, but it is suggestive that a second Keplerian signal at shorter period could be present. The posteriors for this model are shown in the second panel of Fig. \ref{fig:corner2plecc}.\\
	\newline
	\textit{Test 3: dSHO kernel.} For the candidate planet b we found $K_b=1.22^{+0.22}_{-0.20}$ \ms, $P_b=140.55^{+0.44}_{-0.54}$ days, and $e_b=0.50^{+0.12}_{-0.13}$, while for the possible innermost companion we got $K_c=0.88\pm0.30$ \ms, $P_c=63.76^{+0.15}_{-0.25}$ days, and $e_c=0.51^{+0.18}_{-0.26}$. We note that $P_c$ is well constrained, and in agreement with the results of the QP and QPC kernel. However, this model is not statistically favored over the simpler 1-planet model. The Bayesian evidence is $\ln\mathcal{Z}=-956.2$, which is only slightly higher than the Bayesian evidence of the model that includes only one Keplerian ($\Delta\ln\mathcal{Z}=+0.5$). The posteriors are shown in the third panel of Fig. \ref{fig:corner2plecc}.
	
	We summarise in Table \ref{tab:lnZtwoplanets} the Bayesian evidences for the two-planet models. 
	
	We did not find strong evidence for an additional companion orbiting at a closer distance from Gl~514 than planet b, whose main parameters remain unchanged with respect to the model with one Keplerian. However, two of three models are characterised by a slightly higher Bayesian evidence and are not disfavoured, at least suggesting the presence of a signal with period close to 64 days and sub \ms semi-amplitude $\sim$2.5-3$\sigma$ significant. This signal comes with an eccentricity around 0.5-0.6, and this naturally raises concerns about the dynamical stability of such a 2-planet system, especially against orbit crossing. To assess whether there are stable orbital configurations which are compatible with our solutions, we repeated the analysis for the QPC and dSHO cases using a more complete dynamical model (Almenara et al. in prep.; \citealt{rein12,rein15}) with a set-up unchanged, but including the orbit inclination angles as free parameters. In our models, we investigated the scenarios corresponding to co-planar and non-co-planar orbits, to assess the dynamical effects due to a non-zero relative orbital inclination angle. After sampling from the posterior \citep{danforeman13}, we got 100 000 posterior samples/orbital configurations for each of the four scenarios. To these, we applied a ``filter'' in order to select only the configurations which guarantee the dynamical stability of the system over 10$^5$ orbits of Gl~514~b. We adopted the following stability criteria: i) avoiding orbit crossing, and ii) ensuring a MEGNO chaos indicator \citep{cincotta00,cincotta03} in the range between 1.99 and 2.01. Both in the cases of co-planar/non-co-planar orbits, we found that there are thousands of possible stable configurations that survive our stability filtering, 11.46$\%$/3.76$\%$ and 3.6$\%$/1.63$\%$ of the whole posterior samples for the QPC and the dSHO kernels, respectively. The outcome of this analysis is that, in principle, a model with two Keplerians cannot be ruled out with our data, based on dynamical stability criteria, despite it is weakly favoured at best (i.e., in the case of the QPC model). Given the complexity of the problem, different analysis techniques and approaches could be used to further investigate the significance of a two-planet model, as well as collecting additional RVs with higher precision could be worthy, noticing that the signal we found would be compatible with a second planet moving through the HZ. 
	
	\begin{table}[]
		\centering
		\caption{Logarithmic Bayesian evidences $\ln\mathcal{Z}_{\rm 2pl}$ for the GP models with two Keplerians that we tested on HARPS$_{\rm TERRA}$+CARMENES VIS RVs. $\ln\mathcal{Z}_{\rm 1pl}$ is the Bayesian evidence of the model with the same GP kernel and a single Keplerian. }
		\begin{tabular}{ccc}
			\hline
			\noalign{\smallskip}
			GP kernel   & \textbf{$\ln\mathcal{Z}_{\rm 2pl}$} & \textbf{$\Delta\ln\mathcal{Z}_{\rm 2pl-1pl}$ } \\
			\noalign{\smallskip}
			\hline
			\noalign{\smallskip}
			QP & -960.4 & 0 \\
			\noalign{\smallskip}
			QPC & -953.9 & +1.3 \\
			\noalign{\smallskip}
			dSHO & -956.2 & +0.5 \\
			\noalign{\smallskip}
			\hline
		\end{tabular}
		\label{tab:lnZtwoplanets}
	\end{table}   
	
	
	\subsection{HIRES+HARPS+CARMENES dataset: Testing the existence of a longer period companion}
	\label{sec:harpshiresrvanalysis}
\begin{figure}
		\centering
		\includegraphics[width=\linewidth]{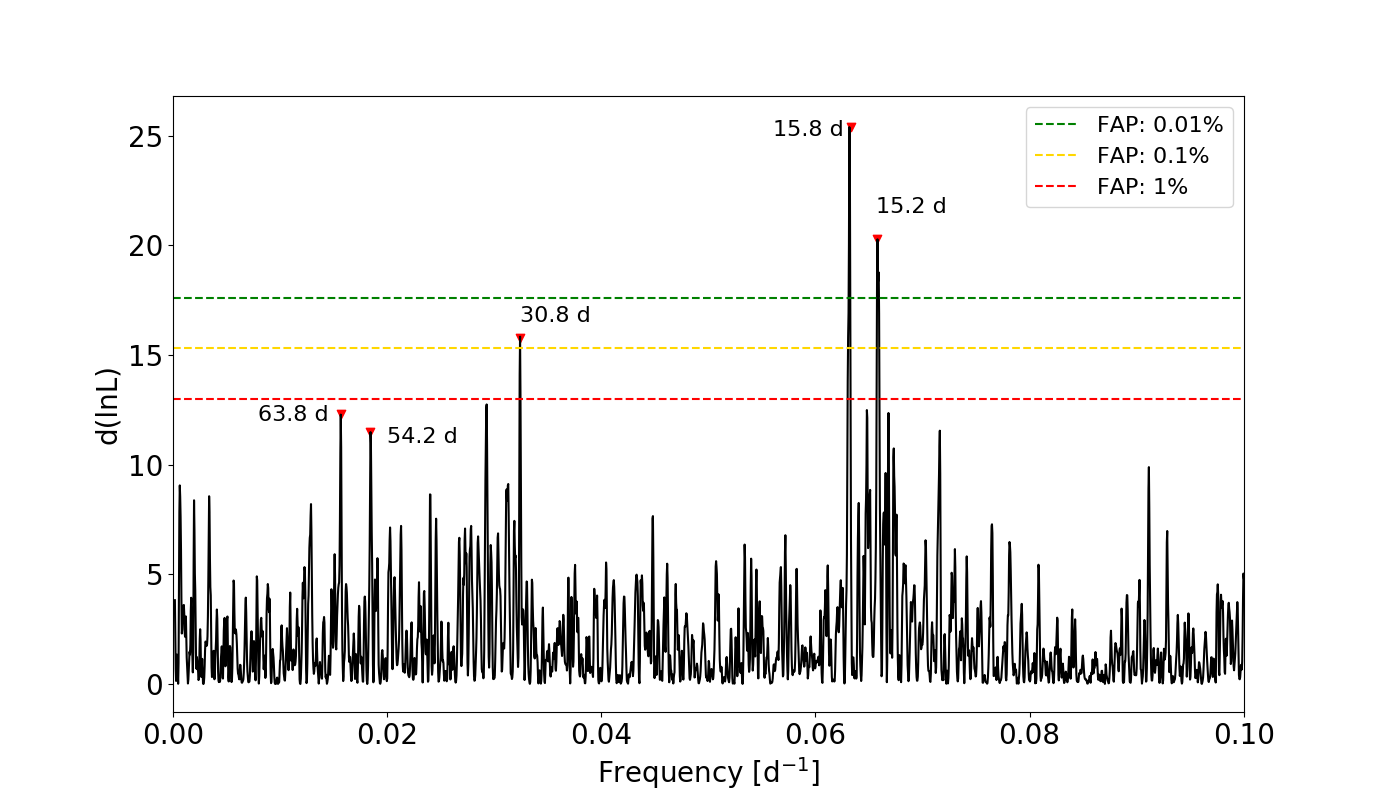}
		\caption{Maximum likelihood periodogram of the HIRES, HARPS, and CARMENES RV residuals, after removing only the best-fit signal of the candidate planet with orbital period $\sim$140 days from the original data (second column of Table \ref{tab:harpscarmeQPC}). The best-fit sinusoid with period P=357.76 days has been subtracted from the CARMENES-VIS data. Stellar activity has not been filtered out, because the best-fit model for the activity signal is linked only on RVs measured with HARPS and CARMENES. }
		\label{fig:MLPresQPC}
	\end{figure}

	Up to this point of our investigation, the existence of Gl~514\,b is the more likely hypothesis supported by our analysis. It is interesting to search for an external companion that could be responsible for the observed high eccentricity. Here we examine this possibility by first including the HIRES measurements to the RV dataset, then searching for astrometric anomalies in {\sl Gaia} and Hipparcos data (Sect. \ref{sec:astrometry}).
	Since the HIRES data are not sensitive to the 140-d signal, to speed up the analysis without affecting it we examined the RV residuals, obtained by removing our adopted best-fit solution for the Keplerian of planet b from all the datasets. The MLP of these data is shown in \ref{fig:MLPresQPC}. The periodogram does not show significant signals at long periods, as already visible in Fig. \ref{fig:MLPHIRESHARPSCARME}. Based on these premises, we considered sufficient to test only the GP model with the dSHO kernel which, as we demonstrated, provides a reliable modelling of the correlated activity signal and is computationally less demanding. For the Keplerian parameters we got $K=0.4^{+0.3}_{-0.2}$ $\ms$, $P=1579^{+1872}_{-631}$ days (with an over-density region around 1300 days), and an unconstrained eccentricity. The Bayesian evidence of this model ($\ln\mathcal{Z}=-1214.8$) is lower than that of the model without the Keplerian included ($\Delta\ln\mathcal{Z}=-2.1$), therefore it is not statistically significant over a pure correlated noise model.
	
	To explore the presence of a longer period companion, we fit the RV residuals including an acceleration term $\dot\gamma$ in place of a Keplerian. We got $\dot\gamma=-0.00026\pm0.00017$ $m s^{-1} day^{-1}$, which differs from zero only within $1.5\sigma$. The lower Bayesian evidence ($\ln\mathcal{Z}=-1222$) confirms that this model is statistically less significant than that including a Keplerian or the reference model (only GP). 
	
	In conclusion, based on our RV data and analysis framework, we do not find statistical evidence for the presence of an external companion to Gl~514\,b. 
	
	\section{\textit{TESS} photometry analysis}
	\label{sec:LCanalysis}
	
\begin{figure}
		\centering
		\includegraphics[width=\linewidth]{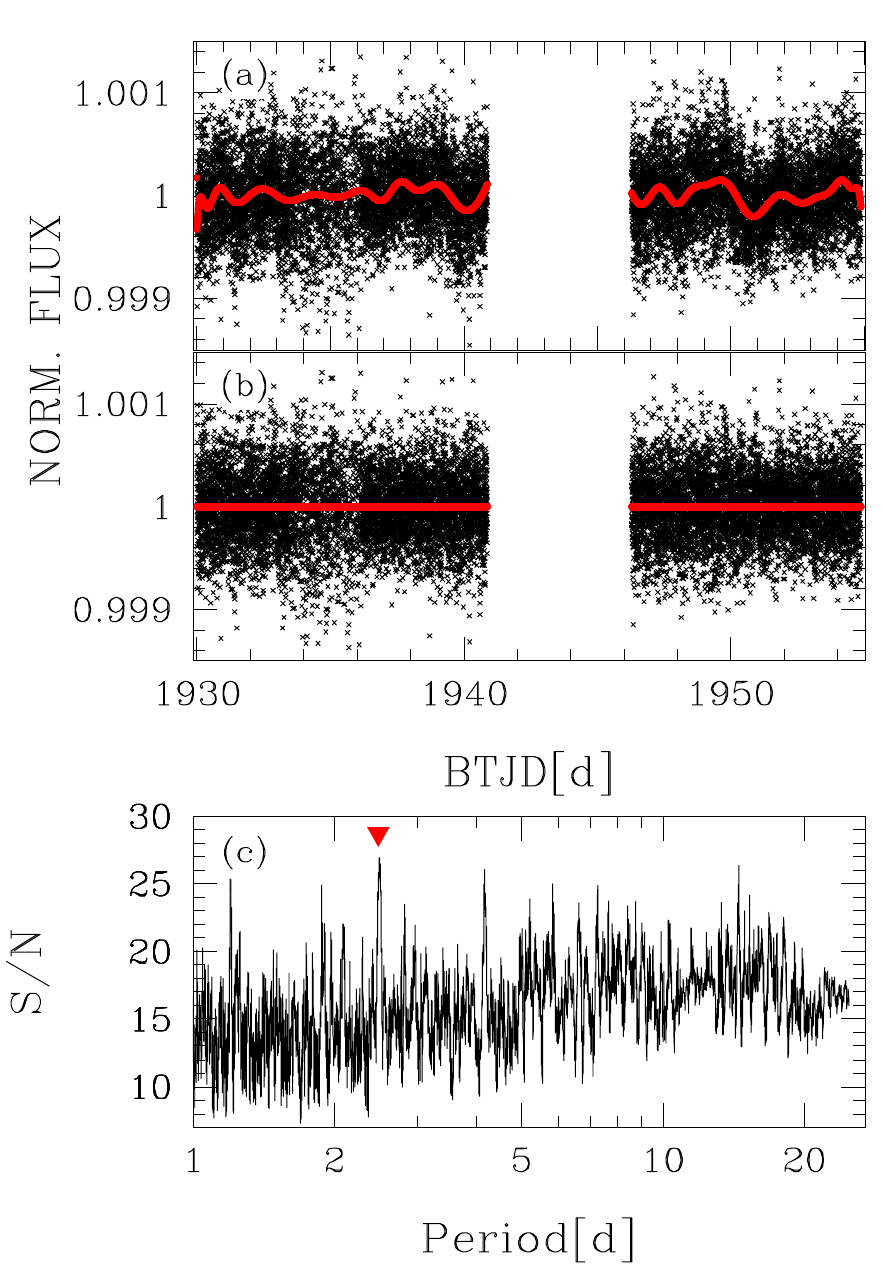}
		\caption{\textit{TESS} short-cadence light curve of Gl~514. Panel (a) and (b) show the normalised light curves before and after the subtraction of the model, respectively; the model, represented in red in panel (a), was calculated on a grid of knots spaced every 24h. Time is expressed as TESS Barycenter corrected Julian Day (BTJD = BJD - 2457000). Panel (c) shows the BLS periodogram. The red triangle indicates the period associated with the highest S/N.}
		\label{fig:tesslc}
	\end{figure}

	We searched for transit-like signals in the short-cadence \textit{TESS} light curve from sector 23, following the procedure described in \citet{Nardiello2020b}. To remove the imprint of stellar variability and other trends, we modelled the light curve with a 5th-order spline defined over a grid of $N_{\rm K}$ knots spaced 24-hours; we also removed all the points with the \textit{TESS} quality parameter \texttt{DQUALITY}$>$0, high values of the background ($>$5$\sigma_{\rm sky}$ above the mean sky value) and all the points with flux more than $4\sigma$ above the mean of the flattened light curve.
	The light curve is shown in Fig. \ref{fig:tesslc}. We calculated the Box Least Squares (BLS) periodograms \citep{Kovacs2002} of the flattened light curve searching for transit-like signals with period in the range between 1 day and the time span of the light curve. We found a peak in the BLS
	periodogram at $\sim 2.5$~d (Fig \ref{fig:tesslc}) associated to a signal detection efficiency (SDE) of $\sim 3.9$, and signal-to-noise ratio S/N of $\sim 27$. The depth of the transit model fitted to this signal is found to be $\sim 73$~ppm, which would correspond to a planet with radius $R\sim0.5 \rearth$. Assuming an empirical threshold SDE$\geq$9 to claim for a significant detection, we conclude that no transits are detected in the light curve of sector 23. 
	
	In our work, we are also interested in searching for a single transit event of Gl~514\,b, which has a geometrical transit probability of $0.5\%$ to occur. An analysis based on BLS can be useful to detect events even when only one transit falls within the time span of the observations but, in such a case, using SDE to define a detection threshold appears meaningless, as we will show hereafter. 
	We investigated what is the likelihood that the signal we detected in the \textit{TESS} data corresponds to a single transit produced by a planet with the orbital period of Gl~514\,b, To this purpose, we devised injection/retrieval simulations. Each simulated dataset is built by injecting in the original light curve a transit signal produced by a planet with an orbital period of 140 days and a radius selected from a grid of values (0.5, 0.75, 0.85, 1.0, and 2.0~$\rearth$). We constrained the phase of the orbit in such a way that the transit falls within the time span covered by the light curve. For each radius we performed 10 simulations, moving the time of central transit $T_{\rm conj}$ recursively two days ahead the $T_{\rm conj}$ generated in the previous simulation. In this way we checked how the photometric systematic errors that locally characterise the light curve affect the detection efficiency of the transit. Each simulated dataset was analysed with BLS in the same way as the original light curve. We flagged as recovered the transits for which $T_{\rm conj,\,injected}-T_{\rm conj,\, recovered}<0.1$ d. We recovered all the injected transits with $R \geq 1.0~\rearth$, which have the transit depth in agreement with that of the injected model: for $R=1 \rearth$ we recover depths in the range $310\pm40$ ppm (the expected value is 350 ppm), and for $R=2 \rearth$ we got depths in the interval $1450\pm170$ ppm (the expected value is 1350 ppm). For all the recovered transits, the main peak in the BLS periodogram has $S/N>100$, and the SDE values are very low (between 1 and 4), likely due to the presence of only a single transit. For this reason, we conclude that, in the case of Gl~514 data, the SDE is not a useful figure of merit to discriminate between recovered and missed single transits. For $R=0.85~\rearth$ we recovered 85\% of the injected transits with a $S/N>85$ and transit depths in the range $250\pm35$ ppm (the expected value is 240 ppm). The detection efficiency falls down for $R=0.75~\rearth$, and we are able to recover only 30~\% of the injected transits with $S/N>80$ and depths in the interval $210\pm50$ ppm (the expected value is 190 ppm). Finally, we could not detect transits for $R=0.50~\rearth$.
	
\begin{figure}
		\centering
		\includegraphics[width=\linewidth]{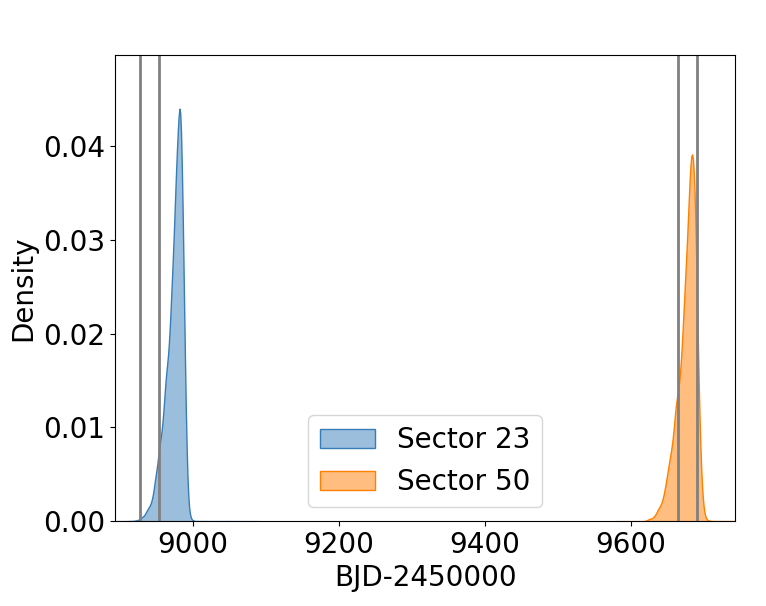}
		\caption{Posterior distributions for the conjuction times $T_{\rm conj,\,b}$ of Gl~514\,b that are close to \textit{TESS} Sectors 23 (March-April 2020) and 50 (March-April 2022). The posteriors are derived from that obtained for our best-fit QPC model. The vertical lines mark the time span of \textit{TESS} observations.}
		\label{fig:tesssectorprediction}
	\end{figure}

	The results of the simulations allow us to conclude about the presence in the \textit{TESS} light curve of a transit signal ascribable to Gl~514\,b. Assuming an inclination angle of 90$^{\circ}$ for the orbital plane (i.e. forcing the planet to transit), the measured mass of $5.6\pm0.9$ $\mearth$ would correspond to a radius $R_b=2.1^{+0.9}_{-0.6}$ $\rearth$, following the probabilistic mass-radius relation of \citep{Chen2017}. We showed that we are able to detect a single transit due to a planet with a radius as large as that predicted for Gl~514\,b within $1\sigma$, but we found no evidence for such a transit in the TESS light curve of sector~23. The signal found by BLS corresponds to a radius for which the simulations provided a null result. The non-detection may be due to three causes: (\textit{i}) the planet does not
	transit because of its geometrical configuration; (\textit{ii}) the planet transits, but the planet radius is smaller than 1 $\rearth$, and we are not able to detect it with high confidence; (\textit{iii}) the planet transits, but the transit does not fall within the time span covered by the light curve of sector~23. Looking at Fig. \ref{fig:tesssectorprediction}, the latter option appears possible. The figure shows that the time of inferior conjunction $T_{\rm conj,\,b}$ has a small likelihood to fall within the \textit{TESS} observing window during sector 23. On the contrary, the same plot shows that there will be a high probability for the time of centre transit to occur during sector 50, scheduled between 26 March and 22 April 2022. In the even more favourable case that the light curve from sector 50 will have better quality than sector 23, we expect that the detection of the transit will be a realistic possibility.   
	
	\section{Astrometric sensitivity to wide-separation companions}
	\label{sec:astrometry}
\begin{figure}
		\centering
		\includegraphics[width=\linewidth]{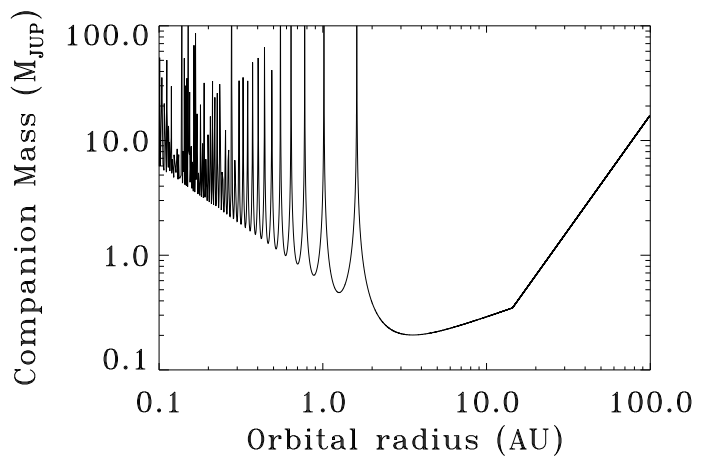}
		\caption{Detectability curve for companions of given mass and orbital distance around Gl~514 based on the proper motion anomaly technique described in \cite{Kervella19}. }
		\label{fig:astrometry}
	\end{figure}

	Gl~514 is astrometrically 'quiet'. Its entry in the {\sl Gaia} EDR3 archive reports values of astrometric excess noise and reduced unit weight error (RUWE) of 0.143 mas and 1.09, respectively. These numbers are typical of sources whose motion is well described by the standard 5-parameter astrometric model (positions, proper motions, and parallax) in {\sl Gaia} astrometry.\footnote{For reference, a threshold of RUWE $\geq1.4$ \citep{Lindegren2018} is typically used to identify astrometrically variable stars.} No statistically significant difference in the stellar proper motion at the Hipparcos and {\sl Gaia} mean epochs is reported in the \cite{Kervella22} and \cite{Brandt21} {\sl Hipparcos-Gaia} catalogues of astrometric accelerations. Given that Gl~514 is really in the Sun's backyard (its distance is 7.6 pc), the proper motion anomaly technique allows to place rather interesting limits on the presence of wide-separation companions in the planetary and sub-stellar mass regime. Using the formalism presented in \cite{Kervella19} (Eqs. (13)–(15)) we show in Fig. \ref{fig:astrometry} the mass-orbital separation sensitivity diagram from the {\sl Hipparcos-Gaia} absolute astrometry. In the approximate range $3-10$ au an object with $\sim0.2$ M$_J$ can be ruled out at the 1-$\sigma$ level. A companion with the same value of $m_p\sin i_p$ at 6.5 au (the approximate limit in orbital period sampled by the combined RV dataset) would induce an RV modulation in Gl~514 with a semi-amplitude of $\sim 3$ m s$^{-1}$, which would have likely been picked up in the RV analysis. The absolute astrometry limits however apply to companions with any orbital inclination, implying that the presence of such a low-mass gas giant can be ruled out even for quasi-face-on configurations (effectively producing too small RV signals). The proper motion anomaly sensitivity diagram of Fig. \ref{fig:astrometry} also indicates that massive giant planets and brown dwarf companions out to several tens of au would have been detected, if present. Jupiter-mass companions at Neptune-like separations would remain undetected. At 0.42 au, the upper limits on the true mass of the RV-detected planet in this work are not very illuminating ($\lesssim 1.5$ M$_J$, with a constraint on the true inclination angle $\gtrsim 1^\circ$).
	
	\section{Conclusions and perspectives}
	\label{sec:conclusions}
	We presented an analysis of nearly 25 years of RVs of the nearby M dwarf Gl~514 collected with the HIRES, HARPS, and CARMENES spectrographs. The data look dominated by signals related to stellar magnetic activity, which we corrected testing three different GP kernels of proven efficacy on quasi-periodic modulations. In all the cases, we found strong evidence for a signal which we attribute to the presence of a super-Earth with minimum mass $m_b\sin i_{\rm b}=5.2\pm0.9$ $\mearth$ (with an 87$\%$ probability that the true mass is within a factor of two larger) moving on an eccentric orbit ($e_b=0.45^{+0.15}_{-0.14}$) with orbital period $P_b=140.43\pm0.41$ days and semi-major axis $a_b=0.422^{+0.014}_{-0.015}$ au. The parameters of our adopted best-fit model are highlighted in bold face in Table \ref{tab:harpscarmeQPC}. This result was corroborated by using RVs extracted from the HARPS spectra with three different pipelines, all based on a template matching technique. We also investigated the possibility that a signal induced by an additional companion to Gl~514\,b is present in the RVs. Exploring orbital periods $P<100$ days, we did not find evidence in favour of a 2-planet model, even though our solution is suggestive of the existence of a sub \ms signal around 64 days which we deem worthy of future follow-up using very high precision RVs. No evidence was found for a longer-period planet ($P>200$ days). Available astrometric data rule out a $\sim0.2$ $M_Jup$ planet at a distance of $\sim3-10$ au, and massive giant planets/brown dwarfs out to several tens of au, while Jupiter-mass companions at Neptune-like separations would remain undetected. 
	
	One point which emerges from our analysis and deserves further attention is represented by the 3$\sigma$ significant orbital eccentricity of Gl~514\,b. In the first place, the eccentricity is a relevant parameter which can provide crucial information about the evolutionary history of a planetary system, therefore it is important to get an accurate and precise measurement of it, when possible. However, in our case it is necessary to proceed with caution, because Gl~514\,b could be actually an ``eccentric impostor''. This is a well-known problem, extensively discussed in literature (e.g. \citealt{rodigas09,trifonov17,wittenmyer13,boisvert18,wittenmyer19}). Works like those of \cite{Anglada2009} and \cite{kuster15} have discussed in detail the issue that a system of two planets on 2:1 resonant and nearly circular orbits can be confused with a single planet on an eccentric orbit. However, their results apply to systems with very different properties than Gl~514: generally, they have sparse data sampling, and the ``eccentric'' planet has a semi-amplitude much greater than that of Gl~514\,b, and significantly greater that the RMS of the residuals of the 1-Keplerian model. For Gl~514, the RMS of the residuals (1.35 \ms) is greater than the semi-amplitude of the Keplerian signal, and it looks unfeasible to distinguish between the two models. Nonetheless, we have a large number of RVs with a dense sampling, and we used GPs to model efficiently the stellar activity contribution. Therefore, we deemed interesting anyway to perform an analysis of the HARPS$_{\rm TERRA}$ and CARMENES data and check if Gl~514\,b could be an \textit{eccentric impostor}, and the system is actually be composed of two planets on 2:1 resonant and nearly circular orbits. Our analysis shows that, independently from the choice of the GP kernel, the two-planet solution is statistically never favoured over that with a single eccentric planet. That is in agreement with the more general results we presented in Sect. \ref{sec:2planetfit}, where we showed that Gl~514\,b is still fitted as an eccentric planet even when using a model with two Keplerians. 
	On the same subject, more recent works (e.g. \citealt{Hara19} and \citealt{faria22}), discuss the issue that orbital eccentricities fitted using only RVs could be spurious, as a consequence of an inappropriate modelling or not optimal data quality relative to the low semi-amplitude of the signal. \cite{Hara19} showed how one can get a wrong inference of the eccentricity without including an uncorrelated jitter term in the model and, more importantly, an incorrect result without a proper modelling of correlated signals, such as stellar activity. \cite{faria22} showed how the value of the eccentricity of Proxima d depends on the method used to extract the RVs from ESPRESSO spectra, with a better constraint obtained using a technique based on template matching. In our analysis, we included uncorrelated jitter terms, we showed that the result is independent from the model used to fit the correlated stellar activity signal, and we tested our finding against different RV extraction methods applied to HARPS spectra.  
	We conclude that the eccentric solution for the orbit of Gl~514\,b is likely non-spurious\footnote{We note that \cite{wittenmyer19} concluded that planet candidates with eccentricity $e\geq0.5$ are unlikely to be impostors}, even though the RVs analysed in this work, and the analysis framework we have selected, do not allow us to reach a strong statistical evidence in favour of this model. Given the low $\sim1 \ms$ semi-amplitude of the planetary signal, we acknowledge that further investigation of the signal and additional data are necessary to get a more accurate and precise measurement of $e_b$. Any future follow-up of Gl~514 with high-resolution spectrographs such as ESPRESSO, given a sufficiently dense sampling, could help in this regard. With this caveat in mind, for the rest of our discussion we will give credit to our result concerning $e_b$.  
	
	\begin{figure}
		\centering
		\includegraphics[width=\linewidth]{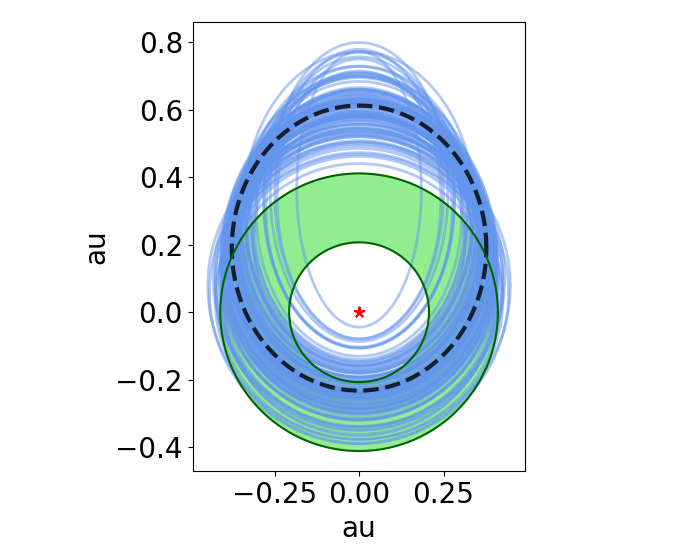}
		\caption{Schematic representation of a few possible orbits for planet Gl~514\,b (blue ellipses), randomly selected taking into account the parameter values and their errors of our adopted solution (Table \ref{tab:harpscarmeQPC}, second column). The best-fit orbit is shown with a dashed black line. The red star identify one of the foci of the ellipses occupied by the host star Gl~514\,b, and the green annulus corresponds to the conservative habitable zone. }
		\label{fig:orbitsQPC}
	\end{figure}
	
	\subsection{Gl~514\,b in the context of planets orbiting in the habitable zone of M dwarfs}
	According to the theoretical calculations by \cite{kopparapu13,kopparapu14}, the conservative HZ for Gl~514 and for a planet with mass $m_p=5\mearth$ extends between 0.207 and 0.411 au. Gl~514\,b spends nearly 34$\%$ of its orbital period within the conservative HZ of its host star, as shown in the sketch of Fig.\ref{fig:orbitsQPC}. 
	Among the known low-mass exoplanets orbiting in the HZ of nearby M dwarfs\footnote{We have used a list of planets with mass or minimum mass $<10$ $\mearth$ that we compiled querying the NASA Exoplanet Archive, and taking the missing data from the catalogue maintained by the Planetary Habitability Laboratory or from the most recent references. For the mass and radius of some planets, we used updated values: Trappist-1~d, Trappist-1~e, Trappist-1~f, and Trappist-1~g: \citet{agol21}; Proxima~b: \cite{mascareno2020}; K2-18~b: \cite{benneke2019}. For the effective temperature and luminosity of Gl~229~A~c we used values from \cite{Schweitzer19}. For Gl~163\,c we used the effective temperature determined by \citealt{tuomi13}. Gl~832~c is likely an artefact of the stellar activity (\citealt{mascareno17}, Gorrini et al. subm.).}, none has an eccentricity as high and significant as Gl~514~b. 
	
	Even though we cannot measure the radius of Gl~514\,b so far, and constrain its average composition and physical structure, nonetheless it will be interesting to use Gl~514\,b as a case study to investigate the habitability of an eccentric super-Earth orbiting a low-luminosity star using climate models. The question of the habitability of planets that experience insolation variations along their orbits, and spend a considerable fraction of time outside the HZ, is a complex problem (see, e.g., \citealt{williams2002,Kane12,bolmont16,Mendez2017}). While \cite{williams2002} concluded that long-term climate stability depends primarily on the average stellar flux received over an entire orbit, \cite{bolmont16} found that for water worlds of higher eccentricities (or host stars with higher luminosity) the mean flux approximation becomes less reliable to assess their ability to sustain a liquid water ocean at their surface. When discussing about the equilibrium temperature of an exoplanet with orbital properties similar to Gl~514\,b (with implications about climate and habitability), one fundamental parameter to be taken into account is the thermal time scale, which is defined as the time scale on which the planetary temperature fluctuations adjusts around the flux-averaged equilibrium value to the changing stellar irradiation \citep{Quirrenbach_2022}. Any temperature fluctuation is expected to be damped following an exponential decay, and the thermal time scale depends on the heat capacity per unit surface area of the planet. General considerations suggest that substantial liquid surface water reservoirs or atmospheres of several tens of bars can work as an efficient climate buffering, avoiding temperature fluctuations on short time scales. If these properties characterise Gl~514\,b, its surface equilibrium temperature should thus be damped around the flux-averaged value, which to a first approximation we estimate to be $202\pm11$ K (assuming zero Albedo). It is clear that, lacking any detailed knowledge of the physical and chemical properties, any consideration about the habitability of Gl~514\,b is presently only speculative, and further discussing this topic is out of the scope of this paper. Nonetheless, we support Gl~514\,b as a benchmark system for investigating the habitability of a super-Earth using sophisticated climate models, using tools such as \textsc{VPLanet} \citep{barnes2020}.
	
\begin{figure}
		\centering
		\includegraphics[width=\linewidth]{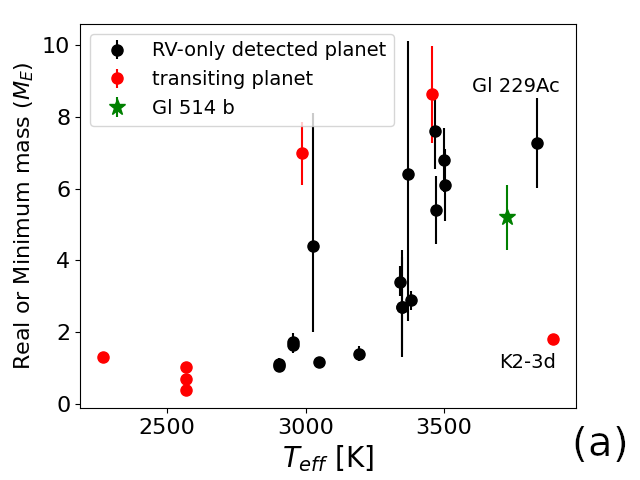}
		\includegraphics[width=\linewidth]{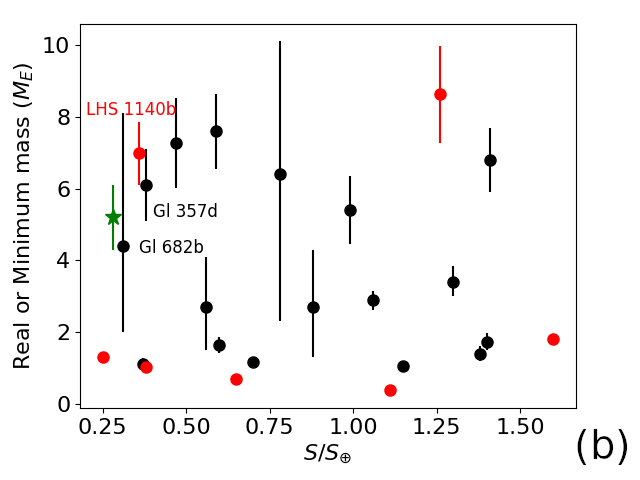}
		\includegraphics[width=\linewidth]{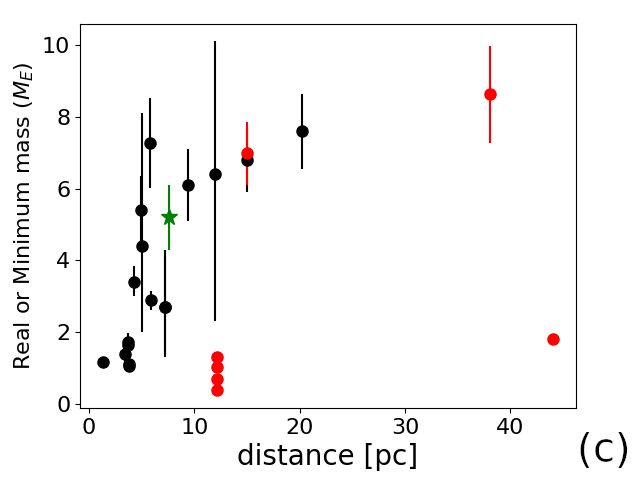}
		\caption{Gl~514~b in the context of planets with mass or minimum mass $<10$ $\mearth$ that have been detected to date within the HZ of M dwarfs. The insolation flux of Gl~514~b is the temporal average orbit value.}
		\label{fig:discussion1}
	\end{figure}

	Using the same list of potentially habitable planets, that we compiled in a way described above, we put Gl~514~b in the context of other low-mass planets detected in the HZ of M dwarfs (Fig. \ref{fig:discussion1}). Planetary masses are represented vs. the effective stellar temperature (panel (a)), insolation\footnote{For Gl~514~b we plotted the orbit-averaged insolation.} (panel (b)), and distance (panel (b)). We note that in panel (a), only two super-Earths are close to Gl~514~b, namely Gl~229~A~c and K2-3~d, thus Gl~514~b enters the very small group of low-mass planets moving within the HZ of nearby stars with spectral type earlier than M2V. When looking at panel (b), Gl~514~b is located next to Gl~682~b, but this planet has a much shorter period ($\sim$17.5 d), and its minimum mass is far less precise. Gl~514~b has a minimum mass compatible, within the uncertainties, to those of LHS~1140\,b (real mass) and Gl~357~d (minimum mass), and an insolation $\sim25\%$ lower. The potential habitability of Gl~357~d has been discussed in detail by \cite{Kaltenegger2019}. Gl~357~d is considered a prime target for observations with Extremely Large telescopes as well as future space missions, and the same expectations are even more valid for Gl~514~b, which is closer than Gl~357~d and its host star is brighter. Currently, we can only speculate about realistic outcomes of high-contrast imaging of Gl~514\,b with the Extreme Large Telescope (ELT), considering the expected performance of the Planetary Camera and Spectrograph (PCS) that will be dedicated to detecting and characterising exoplanets with sizes from sub-Neptune to Earth-size in the solar neighbourhood \citep{kasper2021}. Assuming a maximum angular star-planet separation of $\sim 80$ mas ($\sim 55$ mas for a circular orbit), PCS could in principle detect the planet if the planet-to-star \textit{I}-band flux ratio is roughly greater than $\sim 2\cdot10^{-9}$, a threshold that the Gl~514 system could realistically exceed (see Fig. 1 in \citealt{kasper2021}). 
	
	\subsection{Radio emission from sub-Alfv\'enic star-planet interaction}
	
	Very recent studies based on radio-frequency observations have proposed star-planet interaction as a possible mechanism to explain the detection of radio emission coming from nearby M dwarfs (\citealt{Turnpenney2018,Vedantham2020,Callingham2021,PerezTorres2021}). This interaction is expected to yield auroral radio emission from stars and planets alike, due to the electron cyclotron maser (ECM) instability \citep{Melrose1982}, whereby plasma processes within the star (or planet) magnetosphere generate a population of unstable electrons that amplifies the emission. In some favourable cases, the emission could be detectable. The characteristic frequency of the ECM emission is given by the electron gyrofrequency,
	$\nu_G = 2.8 \, B$ MHz, where $B$ is the local magnetic field in the source region, in
	Gauss.  ECM emission is a coherent mechanism that yields broadband ($\Delta\,\nu \sim
	\nu_G/2$), highly polarized (sometimes reaching 100\%), amplified non-thermal radiation.  
	
	If the velocity $v_{\rm rel}$ of the plasma relative to the
	planetary body is less than the Alfv\'en speed, $v_A$, i.e., $M_A = v_{\rm rel}/v_A <
	1$, where $M_A$ is the Alfv\'en Mach number, then energy and momentum can be transported
	upstream of the flow along Alfv\'en wings. Jupiter’s interaction with its Galilean
	satellites is a well-known example of sub-Alfvénic interaction, producing detectable auroral radio
	emission \citep{Zarka2007}. In the case of star-planet  interaction, the radio emission
	arises from the magnetosphere of the host star, induced by the exoplanet crossing the
	star magnetosphere, and the relevant magnetic field is that of the star,  $B_\star$,
	not the exoplanet magnetic field. Since M-dwarf stars have magnetic fields ranging from
	about 100 G and up to above 2-3 kG, their auroral emission falls in the range from a few
	hundred MHz up to a few GHz, with flux densities that could be well detected by present and future radio telescope arrays.
	
\begin{figure}
		\centering
		\vspace{-10pt}
		\includegraphics[width=\linewidth]{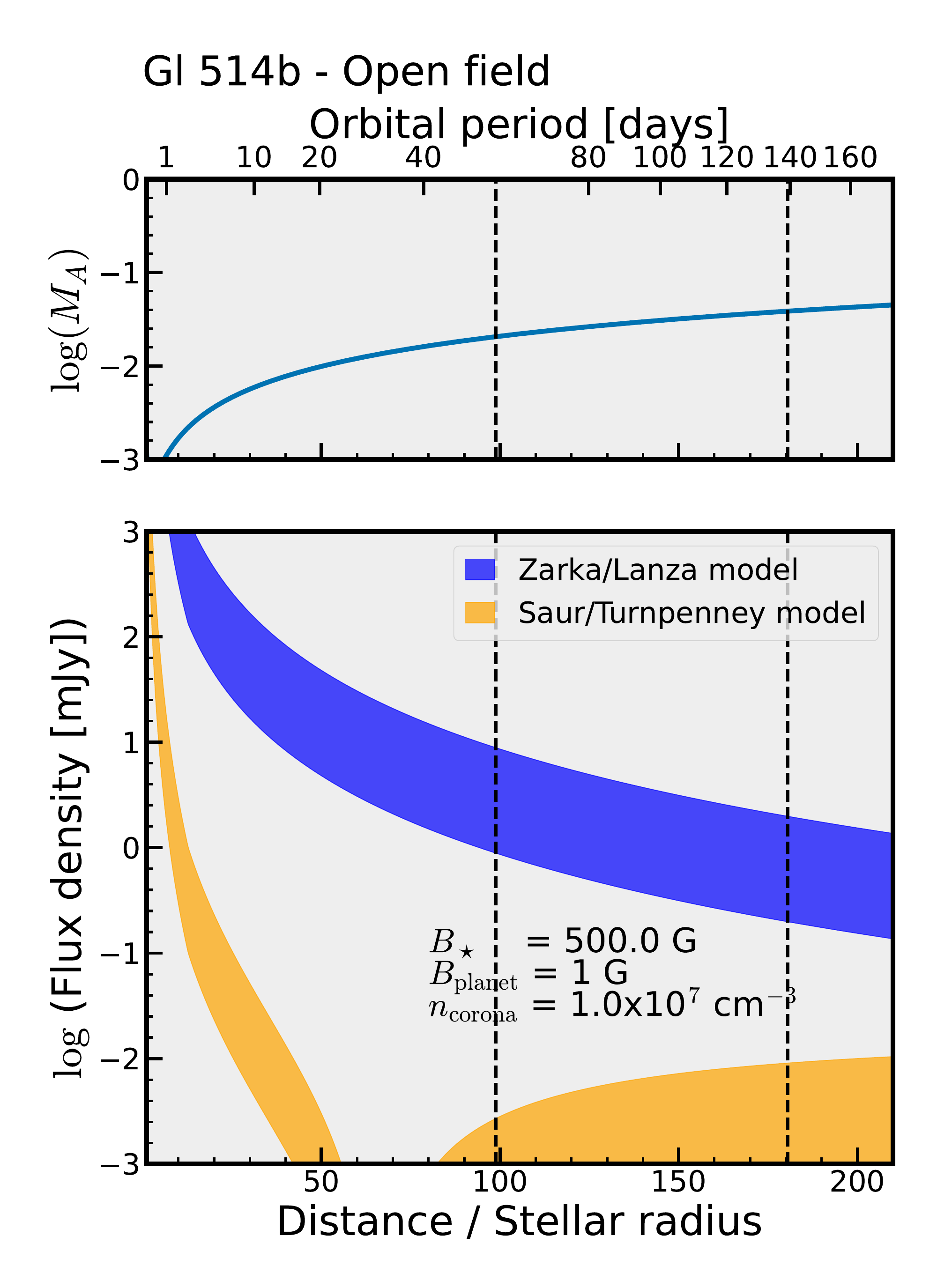}\\
		\vspace{-5pt}
		\caption{\label{fig:gl514-spi-radio} 
			Expected flux density for auroral radio emission arising 
			from star-planet interaction in the system Gl~514, as a function of orbital distance. 
			The interaction is expected to be in the sub-Alfv\'enic regime (i.e. $M_A =
			v_{\rm rel}/v_{\rm Alfv} \leq 1$; top panel)
			at distances equal to the semi-major axis and periapsis (vertical dashed line). 
		}
	\end{figure}

	Given the growing interest around the topic of star-planet interaction at low-frequencies, we find interesting to investigate this possibility for the case of Gl~514~b. We followed the prescriptions in Appendix B of \citet{PerezTorres2021} to
	estimate the flux density expected to arise from star-planet interaction at the frequency of 1.4 GHz, which corresponds to the cyclotron frequency of the
	local magnetic field of 500 G we have assumed, which is a reasonable value for a star
	with rotation period of 30 days, as indicated by the data and Fig. 5 in
	\citet{Shulyak2019}. We computed the radio emission arising from star-planet
	interaction for two different magnetic field geometries: a closed dipolar geometry, and
	an open Parker spiral geometry. For the dipolar case, the motion of the plasma relative
	to Gl~514~b happens in the supra-Alfv\'enic regime. Therefore no energy or momentum can
	be transferred to the star through Alfv\'en waves.  In the open Parker spiral case,
	however, the plasma motion proceeds in the sub-Alfv\'enic regime.  We show in Fig.
	\ref{fig:gl514-spi-radio} the predicted flux density as a function of orbital distance
	arising from the interaction of a magnetized exoplanet (1 G) with its host star.  The
	yellow and blue shaded areas correspond to the predictions of two models, and encompass
	the range of values from 0.01 to 0.1 for the efficiency factor, $\epsilon$, in
	converting Poynting flux into ECM radio emission. Note that the Zarka/Lanza model
	(blue; \citealt{Zarka2007,Lanza2009}) predicts flux densities from a few hundred $\mu$Jy up to a few mJy at the orbital
	distance of Gl~514 (dashed line). Those values are almost two orders of magnitude larger
	than expected in the Saur/Turnpenney model (yellow; \citealt{Saur2013,Turnpenney2018}), which predicts less than
	about 10$\mu$Jy in the most favourable case. If the magnetic field responsible for this
	putative cyclotron radio emission is of 500 G, as we have assumed, then observations at frequencies in the 1.0-1.8 GHz range would be useful to rule out one of the models above. However, we
	caution that the magnetic field of Gl~514 is not known, and our assumption is not more
	than an educated guess, and is likely to be uncertain within a factor of 2. As a
	consequence, observations at multiple frequencies, starting from about 150 MHz and up to
	about 2-3 GHz would be advisable to constrain better the origin of the radio emission in
	this system.

	\begin{acknowledgements}
		We thank the anonymous referee for her/his useful comments.
		This work is partly based on observations collected with the CARMENES spectrograph, which is an instrument at the Centro Astron\'omico Hispano-Alem\'an
		(CAHA) at Calar Alto (Almer\'{\i}a, Spain), operated jointly by the Junta de Andaluc\'ia and the Instituto de Astrof\'isica de Andaluc\'ia
		(CSIC). CARMENES was funded by the Max-Planck-Gesellschaft (MPG), the Consejo Superior de Investigaciones Cient\'{\i}ficas (CSIC), the Ministerio de Econom\'ia y Competitividad (MINECO) and the European Regional Development Fund (ERDF) through projects FICTS-2011-02, ICTS-2017-07-CAHA-4, and CAHA16-CE-3978, and the members of the CARMENES Consortium (Max-Planck-Institut f\"ur Astronomie, Instituto de Astrof\'{\i}sica de Andaluc\'{\i}a, Landessternwarte K\"onigstuhl, Institut de Ci\`encies de l'Espai, Institut f\"ur Astrophysik G\"ottingen, Universidad Complutense de Madrid, Th\"uringer Landessternwarte Tautenburg, Instituto de Astrof\'{\i}sica de Canarias, Hamburger Sternwarte, Centro de Astrobiolog\'{\i}a and
		Centro Astron\'omico Hispano-Alem\'an), with additional contributions by the MINECO, the Deutsche Forschungsgemeinschaft through the Major Research
		Instrumentation Programme and Research Unit FOR2544 ``Blue Planets around Red Stars'', the Klaus Tschira Stiftung, the states of Baden-W\"urttemberg and Niedersachsen, and by the Junta de Andaluc\'{\i}a. We acknowledge financial support from the Agencia Estatal de Investigaci\'on of the Ministerio de Ciencia, Innovaci\'on y Universidades through project PID2019-109522GBC5[1:4].
		
		The authors acknowledge financial support from the Agencia Estatal de Investigaci\'on of the Ministerio de Ciencia e Innovaci\'on and the ERDF
		``A way of making Europe'' through projects PID2020-120375GB-I00, PID2019-109522GB-C5[1:4], and PGC2018-098153-B-C33, and the Centre of
		Excellence ``Severo Ochoa'' and ``Mar\'ia de Maeztu'' awards to the Instituto de Astrof\'isica de Canarias (CEX2019-000920-S), Instituto de Astrof\'isica de Andaluc\'ia (SEV-2017-0709), Centro de Astrobiolog\'ia (MDM-2017-0737), Institut de Ci\`encies de l'Espai (CEX2020-001058-M), and the Generalitat de Catalunya/CERCA programme. They acknowledge financial contribution from the agreement ASI-INAF n.2018-16-HH.0.
		
		M.Damasso acknowledges financial support from the FP7-SPACE Project ETAEARTH (GA no. 313014). M.Perger and I.Ribas acknowledge the support by Spanish grant PGC2018-098153-B-C33 funded by MCIN/AEI/10.13039/501100011033 and by “ERDF A way of making Europe”, by the programme Unidad de Excelencia María de Maeztu CEX2020-001058-M, and by the Generalitat de Catalunya/CERCA programme. M.Perger also acknowledges support from Spanish grant PID2020-120375GB-I00 funded by MCIN/AEI. D. Nardiello acknowledges the support from the French Centre National d'Etudes
		Spatiales (CNES). N. Astudillo-Defru acknowledges the support of FONDECYT project 3180063. M. P\'erez-Torres acknowledges financial support from the State Agency for Research of the Spanish MCIU through the ''Center of Excellence Severo Ochoa" award to the Instituto de Astrofísica de Andalucía (SEV-2017-0709) and through the grant PID2020-117404GB-C21 (MCI/AEI/FEDER, UE). A. Su\'arez Mascare\~{n}o acknowledges financial support from the Spanish Ministry of Science and Innovation (MICINN) under 2018 Juan de la Cierva program IJC2018-035229-I. A. S. M. acknowledge financial support from the MICINN project PID2020-117493GB-I00 and from the Government of the Canary Islands project ProID2020010129. 
		This research has made use of the NASA Exoplanet Archive, which is operated by the California Institute of Technology, under contract with the National Aeronautics and Space Administration under the Exoplanet Exploration Program.
		This work is dedicated to the memory of M.T.    
	\end{acknowledgements}


	\bibliographystyle{aa} 
	\bibliography{gj514_biblio} 
	
	\begin{appendix}
		\section{Dataset extracted from the HARPS and CARMENES spectra}
		We report in Tables \ref{table:rvterradata} and \ref{table:rvnairadata} the radial velocities extracted from the HARPS spectra. The spectroscopic activity diagnostics discussed in this work are listed in Tables \ref{table:actindataharps}-\ref{table:actindatacarme}.
		The HARPS RVs extracted by \cite{trifonov20} are also available at \url{https://www2.mpia-hd.mpg.de/homes/trifonov/HARPS_RVBank.html}. Radial velocities derived from HIRES spectra are available at \url{https://cdsarc.unistra.fr/viz-bin/cat/J/MNRAS/484/L8} .
		
		\begin{table}
			\caption{Radial velocities extracted with the TERRA pipeline from HARPS spectra. An horizontal line separates pre- and post-2015 data. The complete table is made available in electronic form at the CDS.}         
			\label{table:rvterradata}  
			\begin{tabular}{ccc}   
				\hline\hline      
				\noalign{\smallskip}
				Time & RV$_{\rm TERRA}$ & $\sigma_{\rm RV_{\rm TERRA}}$ \\    
				(BJD$-2\,455\,000$) & ($\ms$) & ($\ms$) \\ 
				\noalign{\smallskip}
				\hline\hline
				\noalign{\smallskip}
				3152.621008 & -0.17 & 0.47 \\
				3516.621541 & 0.22 & 0.49 \\
				3520.663397 & 1.56 & 0.86 \\
				... & ... & ... \\
				\noalign{\smallskip}
				\hline
			\end{tabular}
		\end{table}
		
		\begin{table}
			\caption{Radial velocities extracted with the NAIRA pipeline from HARPS spectra. An horizontal line separates pre- and post-2015 data. The complete table is made available in electronic form at the CDS.}          
			\label{table:rvnairadata}      
			\begin{tabular}{ccc}
				\hline\hline      
				\noalign{\smallskip}
				Time & RV$_{\rm NAIRA}$ & $\sigma_{\rm RV_{\rm NAIRA}}$ \\    
				(BJD$-2\,455\,000$) & ($\ms$) & ($\ms$) \\ 
				\noalign{\smallskip}
				\hline\hline
				\noalign{\smallskip}
				3152.621008 & 14751.79 & 0.46 \\
				3516.621541 & 14752.1 & 0.51 \\
				3520.663397 & 14753.46 & 0.78 \\
				... & ... & ... \\
				\hline     
			\end{tabular}
		\end{table}
		
		\begin{table}
			\caption{Activity diagnostics extracted from the HARPS spectra and analysed in this work. The complete table is made available in electronic form at the CDS.}          
			\label{table:actindataharps}    
			\begin{tabular}{ccccccccc}
				\hline\hline      
				\noalign{\smallskip}
				Time & CCF FWHM & $\sigma_{\rm FWHM}$ & H$\alpha$ & $\sigma_{\rm H\alpha}$ & Ca\textsc{II}\,H$\&$K & $\sigma_{\rm Ca\textsc{II}\,H\&K}$ & Na\textsc{I} & $\sigma_{\rm NaI}$ \\    
				(BJD$-2\,455\,000$) & (\kms) & (\kms) & \\ 
				\noalign{\smallskip}
				\hline\hline
				\noalign{\smallskip}
				3152.621008 & 3.3579 & 0.0015 & 0.24642 & 0.0004 & 0.8352 & 0.0037 & 0.19078 & 0.00053 \\
				3516.621541 & 3.3542 & 0.0016 & 0.244 & 0.00044 & 0.7561 & 0.0042 & 0.17995 & 0.00056 \\
				3520.663397 & 3.3537 & 0.0025 & 0.25076 & 0.00066 & 0.7743 & 0.0064 & 0.18475 & 0.00086 \\
				... & ... & ... & ... & ... & ... & ... & ... & ...\\
				\hline       
			\end{tabular}
		\end{table}
		
		\begin{table}
			\caption{CARMENES-VIS radial velocities. The complete table is made available in electronic form at the CDS.}          
			\label{table:rvcarmedata}
			\begin{tabular}{ccc}    
				\hline\hline      
				\noalign{\smallskip}
				Time & RV$_{\rm CARMENES}$ & $\sigma_{\rm RV_{\rm CARMENES}}$ \\    
				(BJD$-2\,455\,000$) & ($\ms$) & ($\ms$) \\ 
				\noalign{\smallskip}
				\hline\hline
				\noalign{\smallskip}
				7397.755084 &   2.27 &   1.79 \\ 
				7398.717731 &   2.96 &   2.30 \\ 
				7401.766930 &   3.38 &   1.07 \\ 
				... &  ... &   ... \\ 
				\hline
			\end{tabular}
		\end{table}
		
		\begin{table*}
			\caption{Activity diagnostics extracted from the CARMENES-VIS spectra and analysed in this work. The complete table is made available in electronic form at the CDS.}       
			\label{table:actindatacarme}   
			\begin{tabular}{ccccccccc}
				\hline\hline      
				\noalign{\smallskip}
				Time & Ca-IRT & $\sigma_{\rm Ca-IRT}$ & NaD & $\sigma_{\rm NaD}$ & dLW & $\sigma_{\rm dLW}$ & CRX & $\sigma_{\rm CRX}$ \\    
				(BJD$-2\,455\,000$) & & & & & & & \\ 
				\noalign{\smallskip}
				\hline\hline
				\noalign{\smallskip}
				7397.755084 & 0.46608 & 0.00083 & 0.21765 & 0.00189 & -5.55 & 1.3 & 26.57 & 17.15 \\
				7398.717731 & 0.46992 & 0.00112 & 0.22233 & 0.00291 & -5.62 & 2.66 & 17.01 & 23.29 \\
				7401.76693 & 0.47274 & 0.00048 & 0.22578 & 0.0009 & 10.08 & 1.72 & -9.84 & 7.97 \\
				... & ... & ... & ... & ... & ... & ... & ... & ... \\
				\hline
			\end{tabular}
		\end{table*}

		\clearpage
		
		\section{Description of the GP kernels tested in this study}
		\label{app:GPkernels}
		\subsection{Quasi-periodic (QP)}
		The QP covariance matrix (e.g. \citealt{haywood2014}) has now become a standard tool to model the activity term in RV time series showing variability modulated over the stellar rotation period. In our work, an element of the matrix is defined as follows:
		
		\begin{gather} 
			\label{eq:eqgpqpkernel}
			k_{QP}(t, t^{\prime}) = h^2\cdot\exp\Bigg[-\frac{(t-t^{\prime})^2}{2\lambda_{\rm QP}^2} - \frac{sin^{2}\Bigg(\pi(t-t^{\prime})/\theta\Bigg)}{2w^2}\Bigg] + \nonumber \\
			+\, (\sigma^{2}_{\rm RV}(t)+\sigma^{2}_{\rm jit})\cdot\delta_{t, t^{\prime}}
		\end{gather}
		
		Here, $t$ and $t^{\prime}$ represent two different epochs of observations, $\sigma_{\rm RV}$ is the radial velocity uncertainty, and $\delta_{t, t^{\prime}}$ is the Kronecker delta. Our analysis takes into account other sources of uncorrelated noise -- instrumental and/or astrophysical -- by including a constant jitter term $\sigma_{\rm jit}$ which is added in quadrature to the formal uncertainties $\sigma_{\rm RV}$. The GP hyper-parameters are $h$, which denotes the scale amplitude of the correlated signal; $\theta$, which represents the periodic time-scale of the correlated signal, and corresponds to the stellar rotation period; $w$, which describes the "weight" of the rotation period harmonic content within a complete stellar rotation (i.e. a low value of $w$ indicates that the periodic variations contain a significant contribution from the harmonics of the rotation periods); and $\lambda_{\rm QP}$, which represents the decay timescale of the correlations, and is related to the temporal evolution of the magnetically active regions responsible for the correlated signal observed in the RVs.
		
		\subsection{Quasi-periodic with cosine (QPC)}
		The QPC kernel has been recently introduced by \cite{perger2021}, who found that in general it guarantees a better performance over of the QP kernel. The covariance matrix element of the QPC kernel, as implemented in our work, is 
		
		\begin{gather}
			\label{eq:eqgpqpckernel}
			k_{QPC}(t, t^{\prime}) = \exp\Big(-2\frac{(t-t^{\prime})^2}{\lambda_{QPC}^2}\Big)\cdot \Bigg[h_1^2\exp\Big(-\frac{1}{2w^2}\sin^2\Big(\frac{\pi(t-t^{\prime})}{\theta}\Big)\Big)+ \nonumber \\ 
			+h_2^2\cos\Big(\frac{4\pi(t-t^{\prime})}{\theta}\Big)\Bigg]+(\sigma^{2}_{\rm RV}(t)+\sigma^{2}_{\rm jit})\cdot\delta_{t, t^{\prime}}\,
		\end{gather}
		
		Again, $t$ and $t^{\prime}$ represent two different epochs of observations; $h_1$ and $h_2$ are scale amplitudes; $\theta$ still represents the periodic time-scale of the modelled signal, and corresponds to the stellar rotation period; $w$ still describes the "weight" of the rotation period harmonic content within a complete stellar rotation; $\lambda_{\rm QPC}$ is defined as $2\cdot \lambda_{QP}$, better representing the average lifetime of the activity-related features responsible for the stellar correlated signal in the RVs \citep{perger2021}; $\sigma_{\rm RV}$ and $\sigma_{\rm jit}$ are the radial velocity uncertainty and instrument-dependent jitter, respectively, and $\delta_{t, t^{\prime}}$ is the Kronecker delta.
		
		\subsection{Rotational (dSHO)}
		The ``rotational'' kernel is defined by a mixture of two stochastically driven, damped simple harmonic oscillators (SHOs) with undamped periods of $P_{\star,\, rot}$ and $P_{\star,\, rot}$/2. This can be obtained by combining two \texttt{SHOTerm} kernels included in the package \texttt{celerite} \citep{celerite}\footnote{\url{https://github.com/dfm/celerite/blob/main/celerite/terms.py}}. 
		The power spectral density corresponding to this kernel is 
		\begin{eqnarray}
			S(\omega) = \sqrt{\frac{2}{\pi}} \frac{S_1\omega_1^4}{(\omega^2-\omega_1^2)^2 +  2\omega_1^2\omega^2} + \nonumber\\
			+\sqrt{\frac{2}{\pi}} \frac{S_2\omega_2^4}{(\omega^2-\omega_2^2)^2 + 2\omega_2^2\omega^2/Q^2},
		\end{eqnarray}
		
		where
		\begin{gather} 
			\label{eqn:sho1}
			S_{\rm 1}=\frac{A^2}{\omega_{\rm 1}Q_{\rm 1}(1+f)}, \\ 
			S_{\rm 2}=\frac{A^2}{\omega_{\rm 2}Q_{\rm 2}(1+f)}\cdot f, \\
			\omega_{\rm 1}=\frac{4 \pi Q_{\rm 1}}{P_{\rm rot}\sqrt{4Q_{\rm 1}^{2}-1}}, \\
			\omega_{\rm 2}=\frac{8 \pi Q_{\rm 2}}{P_{\rm rot}\sqrt{4Q_{\rm 2}^{2}-1}}, \\
			Q_{\rm 1}= \frac{1}{2}+Q_{\rm 0}+\Delta Q,\\
			\label{eqn:sho2}
			Q_{\rm 2}= \frac{1}{2}+Q_{\rm 0}. 
		\end{gather}
		
		The parameters in [\ref{eqn:sho1}-\ref{eqn:sho2}], where the subscripts 1 and 2 refer to the primary ($P_{\star,\, rot}$) and secondary ($P_{\star,\, rot}$/2) modes, represent the inputs to the \texttt{SHOTerm} kernels. However, instead of using them directly, we adopt a different parametrization using the following variables as free hyper-parameters in the MC analysis, from which $S_{\rm i}$, $Q_{\rm i}$, and $\omega_{\rm i}$ are derived through Eq. [\ref{eqn:sho1}-\ref{eqn:sho2}]: the variability amplitude $A$, the stellar rotation period $P_{\star,\, rot}$, the quality factor $Q_{0}$, the difference $\Delta$Q between the quality factors of the first and second terms, and the fractional amplitude $f$ of the secondary mode relative to the primary.

		\section{Additional material relative to 1-planet models fitted to the HARPS$_{\rm TERRA}$+CARMENES RV time series}
\begin{figure*}
			\centering
			\includegraphics[width=\textwidth]{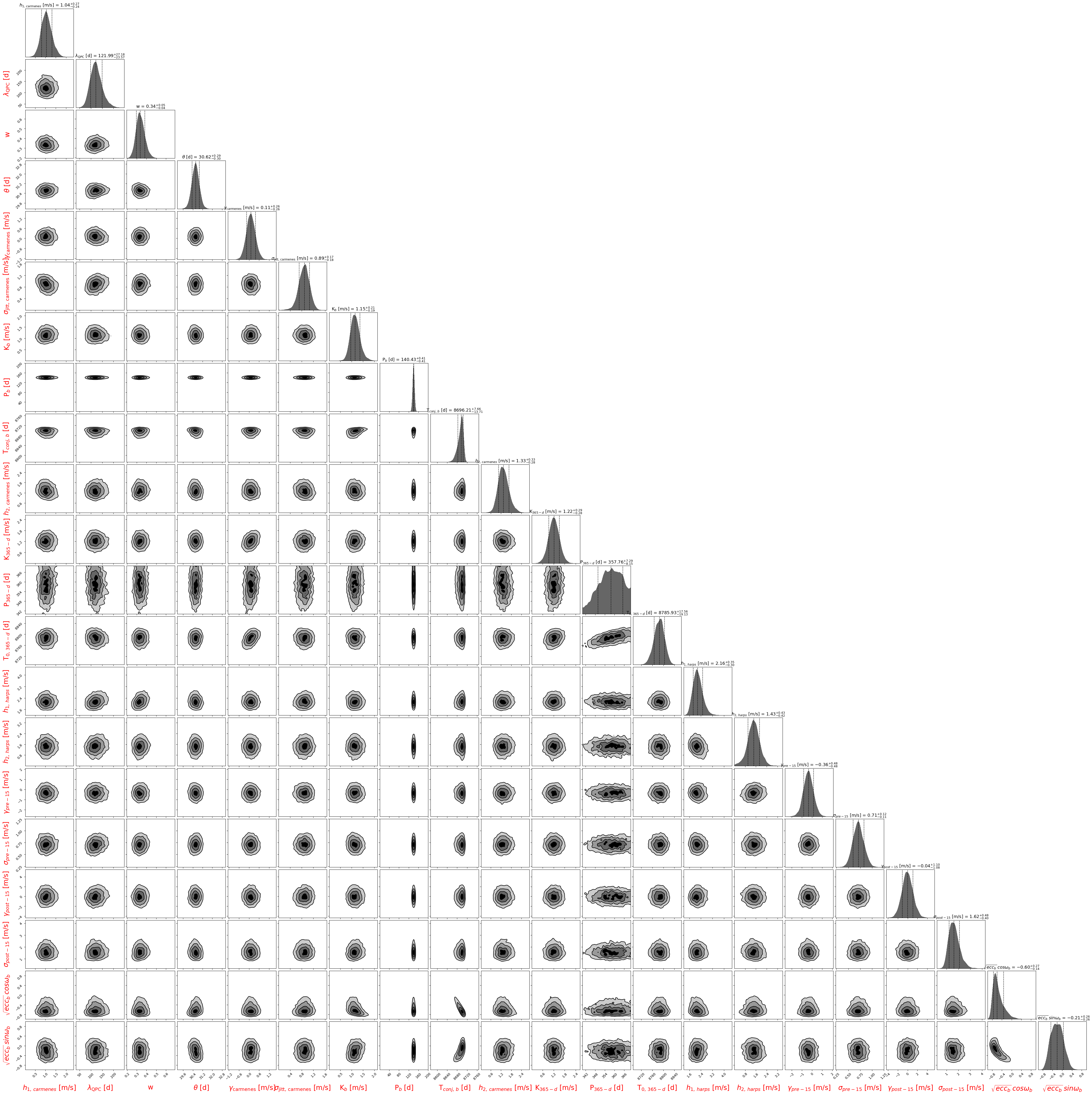}
			\caption{Posterior distributions of the free (hyper)parameters of our assumed best-fit model, including a Keplerian for planet b and a GP QPC correlated activity signal fitted to the RVs obtained from HARPS (TERRA dataset) and CARMENES VIS spectra.}
			\label{fig:corner_harpscarmenesQPC}
		\end{figure*}

		In Fig. \ref{fig:corner_harpscarmenesQPC} we show the posterior distributions, and their two by two correlations (corner plot), for all the free (hyper-)parameters of our assumed best-fit solution for the Gl~514 system, based on the analysis of the combined HARPS$_{\rm TERRA}$+CARMENES RVs. 
		
\begin{figure}
			\centering
			\includegraphics[width=\linewidth]{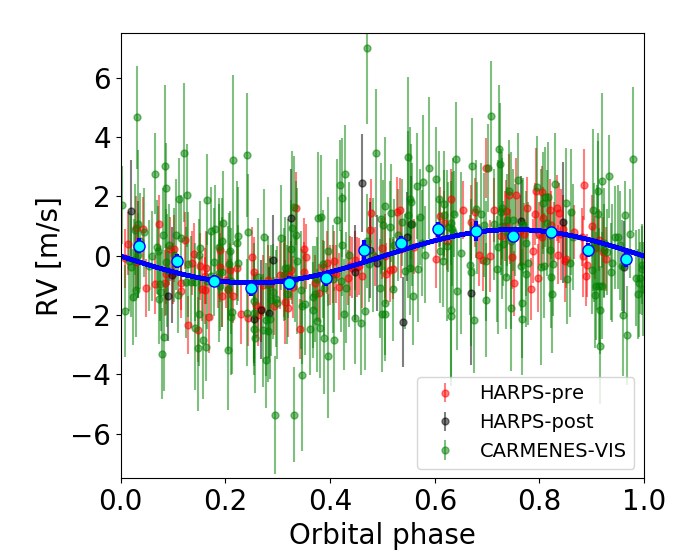}
			\caption{Phase-folded RVs of HARPS$_{\rm TERRA}$ and CARMENES-VIS showing the spectroscopic orbit of Gl~514\,b for the circular case. The blue solid line is the best-fit solution using a QPC kernel to model the activity. The error bars include the uncorrelated jitter terms added in quadrature to the formal RV uncertainties. Cyan dots correspond to RV data averaged over 15 phase bins.}
			\label{fig:fold_circ_harpscarmenesQPC}
		\end{figure}

		We show in Fig. \ref{fig:fold_circ_harpscarmenesQPC} the spectroscopic orbit of Gl~514\,b for the circular case, obtained using the QPC kernel to model the activity component.
		
\begin{figure*}
			\centering
			\includegraphics[width=0.45\textwidth]{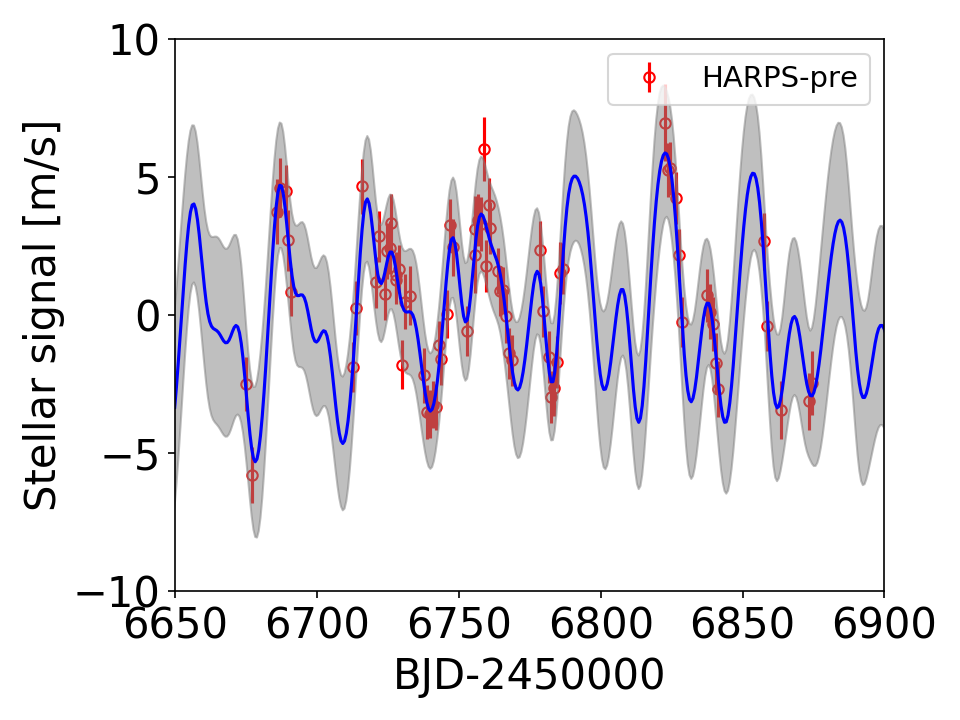}
			\includegraphics[width=0.45\textwidth]{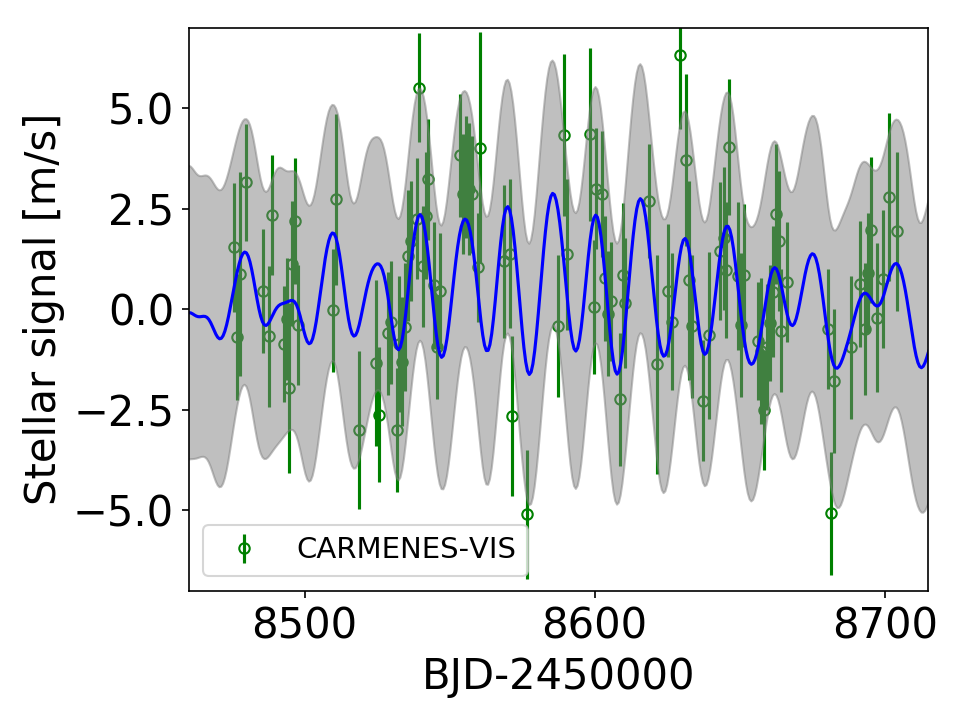}
			\caption{Sections of the residuals of the whole set of HARPS$_{\rm TERRA}$+CARMENES RVs (corresponding to the model shown in the second column of Table \ref{tab:harpscarmeQPC}, showing the QPC correlated signal mostly related to variations in stellar activity. The error bars include the uncorrelated jitter terms added in quadrature to the formal RV uncertainties.  }
			\label{fig:activityQPC}
		\end{figure*}

		In Fig. \ref{fig:activityQPC} we show two subsets of the HARPS$_{\rm TERRA}$+CARMENES RV residuals (i.e. after subtracting the Keplerian for planet b, and the additional sinusoid for the CARMENES data only) together with the GP QPC best-fit solution for the correlated term. These data correspond to the stellar activity component present in the RV time series, possibly with other not modelled small-amplitude signals included (astrophysical or instrumental), according to our adopted best model for the Gl~514 system (see the second column of Table \ref{tab:harpscarmeQPC} for the list of all the free parameters).   
		
		In Tables \ref{tab:harpscarmeQP} and \ref{tab:harpscarmeSHO} we summarise the best-fit values of the parameters for the models including the QP and dSHO kernels, which are characterised by a Bayesian evidence lower than that for the QPC model. 
		
\begin{figure*}
			\centering
			\includegraphics[width=0.6\textwidth]{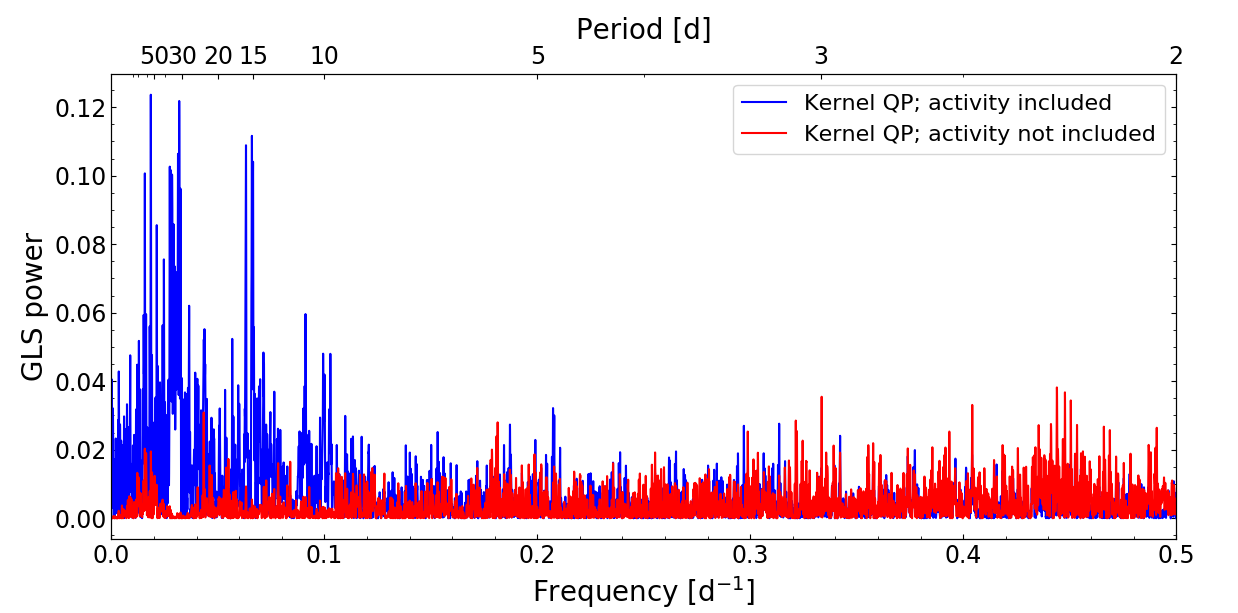}
			\includegraphics[width=0.6\textwidth]{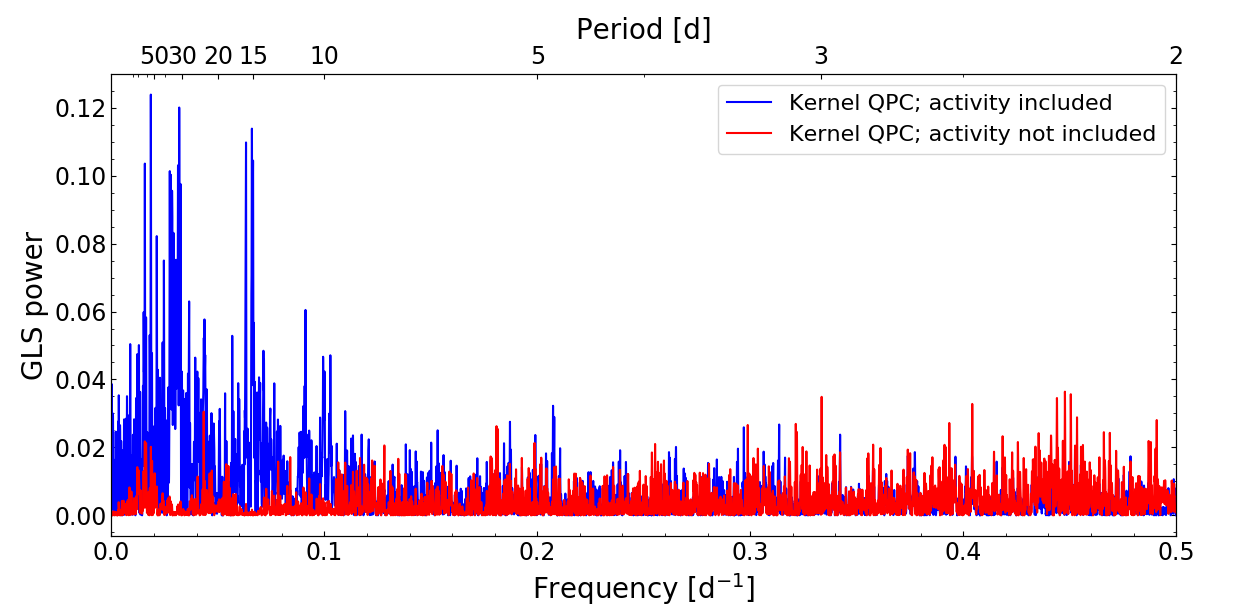}
			\includegraphics[width=0.6\textwidth]{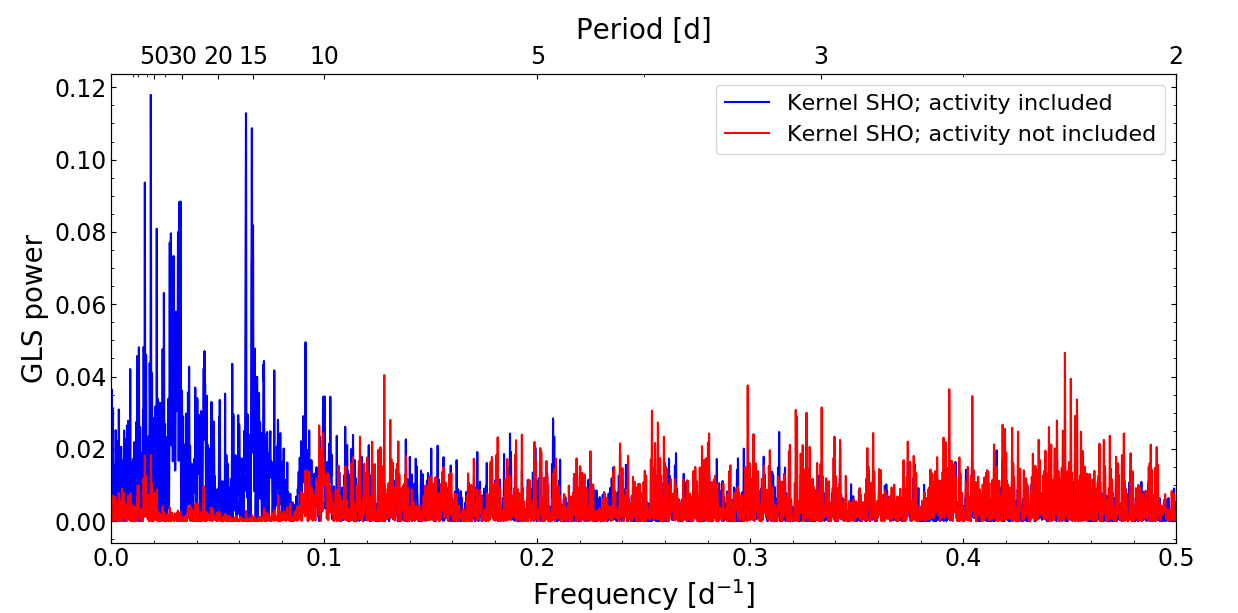}
			\caption{GLS periodograms of the HARPS$_{\rm TERRA}$+CARMENES RV residuals, after removing the best-fit models shown in the second columns of Table \ref{tab:harpscarmeQP}-\ref{tab:harpscarmeSHO}. For each kernel, periodograms of data with and without the correlated activity signal included are shown. The periodograms were calculated with the uncorrelated jitter terms added in quadrature to the formal RV uncertainties.}
			\label{fig:GLSresid}
		\end{figure*}

		In Fig. \ref{fig:GLSresid} we compare the GLS periodogram of the HARPS+CARMENES RVs residuals for all the GP kernels, after removing the best-fit models shown in the second columns of Tables \ref{tab:harpscarmeQP}-\ref{tab:harpscarmeSHO}. We show both the cases with and without the correlated GP-fitted signal included in the residuals. We note that, when the activity signal is not removed from the dataset, all the periodograms peak at 54 days, with semi-amplitude of the best-fit sinsuoid equal to 1.2$\pm$0.2 \ms. Nothing significant is seen at frequencies larger than 0.1 $d^{-1}$. When the activity signal is removed, all the periodograms do not show significant peaks, especially at low frequencies. This does not come as a surprise, because the suppression of low frequencies it has been commonly observed in residuals of GP-filtered RVs, especially at frequencies larger than the stellar rotational frequency. Nonetheless, we note that, despite their low power, peaks at around period of 50-60 days are not completely suppressed. Their nature is discussed in Sect. \ref{sec:2planetfit}.      
		
		\begin{table}
			\small
			\caption{Best-fit parameters obtained for the model with a quasi-periodic (QP) kernel applied to HARPS$_{\rm TERRA}$ and CARMENES-VIS RVs. The uncertainties are calculated from the $16^{\rm th}$ and $84^{\rm th}$ percentiles of the posterior distributions. }          
			\label{tab:harpscarmeQP} 
			\centering                 
			\begin{tabular}{lcc}       
				\hline\hline              
				\textbf{Fitted parameter} & \multicolumn{2}{c}{\textbf{Best-fit value}} \\    
				\hline              
				& $e_b$=0 & $e_b\neq$0 \\
				\hline
				\noalign{\smallskip}
				\noalign{\smallskip}
				$h_{\rm HARPS}$ [\ms] & $2.5\pm0.3$ & $2.6\pm0.3$ \\
				\noalign{\smallskip}
				$h_{\rm CARMENES}$ [\ms] & $1.7\pm0.2$ & $1.7\pm0.2$\\
				\noalign{\smallskip}
				$\theta$ [d] & $30.7\pm0.3$ & $30.7\pm0.3$ \\
				\noalign{\smallskip}
				$\lambda_{\rm QP}$ [d] & $63\pm15$ & $63^{+15}_{-12}$ \\
				\noalign{\smallskip}
				$w$ & $0.31\pm0.04$ & $0.31\pm0.03$ \\
				\noalign{\smallskip}
				$\gamma_{HARPS\,pre-2015}$ [\ms] & $-0.5\pm0.5$ & $-0.5\pm0.5$ \\
				\noalign{\smallskip}
				$\gamma_{HARPS\,post-2015}$ [\ms] & $0.1\pm1.2$ & $0.1\pm1.2$ \\
				\noalign{\smallskip}
				$\gamma_{CARMENES}$ [\ms] & $0.1\pm0.4$ & $0.2\pm0.4$ \\
				\noalign{\smallskip}
				$\sigma_{jit,\,HARPS\,pre-2015}$ [\ms] & $0.8\pm0.1$ & $0.7\pm0.1$ \\
				\noalign{\smallskip}
				$\sigma_{jit,\,HARPS\,post-2015}$ [\ms] & $1.6^{+0.5}_{-0.4}$ & $1.7^{+0.5}_{-0.4}$  \\
				\noalign{\smallskip}
				$\sigma_{jit,\,CARMENES}$ [\ms] & $1.0\pm0.2$ & $0.9\pm0.2$ \\
				\noalign{\smallskip}
				$K_b$ [\ms] & $1.01\pm0.18$ & $1.23\pm0.20$ \\
				\noalign{\smallskip}
				$P_b$ [d] & $139.75^{+0.65}_{-0.61}$ & $140.37^{+0.44}_{-0.46}$ \\
				\noalign{\smallskip}
				$T_{conj,\,b}$ [BJD-2450000] & $8668.68^{+6.67}_{-6.29}$ & $8692.00^{+10.38}_{-15.44}$ \\
				\noalign{\smallskip}
				$\sqrt{e_{\rm b}}\cos\omega_{\rm \star,\:b}$ & - & $-0.527^{+0.306}_{-0.180}$ \\
				\noalign{\smallskip}
				$\sqrt{e_{\rm b}}\sin\omega_{\rm \star,\:b}$ & - & $-0.306^{+0.268}_{-0.221}$ \\
				\noalign{\smallskip}
				$K_{365-d,\,CARMENES}$ [\ms] & $1.2\pm0.4$ & $1.3\pm0.4$ \\
				\noalign{\smallskip}
				$P_{365-d,\,CARMENES}$ [d] & $359.77^{+6.94}_{-8.59}$ & $359.81^{+6.65}_{-8.69}$  \\
				\noalign{\smallskip}
				$T_{0,365-d \,CARMENES}$ [BJD-2450000] & $8784.5\pm22.6$ & $8786.22^{+20.34}_{-21.93}$ \\
				\noalign{\smallskip}
				\hline
				\noalign{\smallskip}
				\textbf{Derived parameter} \\ 
				\noalign{\smallskip}
				\hline
				\noalign{\smallskip}
				eccentricity, $e_{\rm b}$ & - & $0.42^{+0.14}_{-0.13}$ \\
				\noalign{\smallskip}
				arg. of periapsis, $\omega_{\rm \star,\:b}$ & - &  $-2.30^{+5.27}_{-0.66}$  \\
				\noalign{\smallskip}
				\hline
				\noalign{\smallskip}
				$\ln\mathcal{Z}$ & -962.9 & -960.4 \\
				\noalign{\smallskip}
				$\ln\mathcal{Z}_{1p}$-$\ln\mathcal{Z}_{0p}$ & +2.7 & +5.2 \\
				\noalign{\smallskip}
				\hline
			\end{tabular}
		\end{table}
		
		\begin{table}
			\small
			\caption{Best-fit parameters obtained for the model with a rotational (dSHO) kernel applied to HARPS$_{\rm TERRA}$ and CARMENES-VIS RVs. The uncertainties are calculated from the $16^{\rm th}$ and $84^{\rm th}$ percentiles of the posterior distributions.}          
			\label{tab:harpscarmeSHO} 
			\centering                 
			\begin{tabular}{lcc}       
				\hline\hline              
				\textbf{Fitted parameter} & \multicolumn{2}{c}{\textbf{Best-fit value}} \\    
				\hline              
				& $e_b$=0 & $e_b\neq$0  \\
				\hline
				\noalign{\smallskip}
				$\log A_{\rm HARPS}$ & $0.9\pm0.1$ & $0.9\pm0.1$  \\
				\noalign{\smallskip}
				$\log A_{\rm CARMENES}$ & $0.6\pm0.1$ & $0.6\pm0.1$ \\
				\noalign{\smallskip}
				$\theta$ [d] & $31.0\pm0.5$ & $30.9^{+0.4}_{-0.5}$\\
				\noalign{\smallskip}
				$\log Q_0$ & $2.11^{+0.39}_{-0.42}$ & $2.25^{+0.39}_{-0.42}$  \\
				\noalign{\smallskip}
				$\log \Delta Q$ & $-3.80^{+4.41}_{-4.16}$ & $-3.77^{+4.47}_{-4.23}$ \\
				\noalign{\smallskip}
				$f$ & $3.70^{+2.57}_{-1.42}$  & $3.79^{+2.54}_{-1.46}$ \\
				\noalign{\smallskip}
				$\gamma_{HARPS\,pre-2015}$ [\ms] & $-0.2\pm0.2$ & $-0.1\pm0.2$ \\
				\noalign{\smallskip}
				$\gamma_{HARPS\,post-2015}$ [\ms] & $-0.2\pm0.6$ & $-0.3\pm0.6$ \\
				\noalign{\smallskip}
				$\gamma_{CARMENES}$ [\ms] & $0.2\pm0.2$ & $0.2\pm0.2$ \\
				\noalign{\smallskip}
				$\sigma_{jit,\,HARPS\,pre-2015}$ [\ms] & $0.8\pm0.1$ & $0.8\pm0.1$ \\
				\noalign{\smallskip}
				$\sigma_{jit,\,HARPS\,post-2015}$ [\ms] & $1.0^{+0.4}_{-0.3}$ & $1.1^{+0.5}_{-0.4}$ \\
				\noalign{\smallskip}
				$\sigma_{jit,\,CARMENES}$ [\ms] & $0.6\pm0.3$ & $0.6^{+0.2}_{-0.3}$ \\
				\noalign{\smallskip}
				$K_b$ [\ms] & $0.88\pm0.16$ &  $1.16^{+0.25}_{-0.21}$ \\
				\noalign{\smallskip}
				$P_b$ [d] & $140.23^{+0.67}_{-0.64}$ & $140.72^{+0.44}_{-0.64}$ \\
				\noalign{\smallskip}
				$T_{conj,\,b}$ [BJD-2450000] & $8674.28^{+5.72}_{-6.24}$ & $8698.24^{+6.49}_{-11.29}$ \\
				\noalign{\smallskip}
				$\sqrt{e_{\rm b}}\cos\omega_{\rm \star,\:b}$ & - & $-0.638^{+0.207}_{-0.114}$  \\
				\noalign{\smallskip}
				$\sqrt{e_{\rm b}}\sin\omega_{\rm \star,\:b}$ & - & $-0.078^{+0.267}_{-0.300}$  \\
				\noalign{\smallskip}
				$K_{365-d,\,CARMENES}$ [\ms] & $1.2\pm0.2$ & $1.2\pm0.2$ \\
				\noalign{\smallskip}
				$P_{365-d,\,CARMENES}$ [d] & $354.58^{+7.68}_{-7.10}$ & $355.21^{+7.03}_{-7.12}$ \\
				\noalign{\smallskip}
				$T_{0,365-d \,CARMENES}$ [BJD-2450000] & $8782.83^{+14.95}_{-14.81}$ & $8782.51^{+14.59}_{-14.82}$ \\
				\noalign{\smallskip}
				\hline
				\noalign{\smallskip}
				\textbf{Derived parameter} \\ 
				\noalign{\smallskip}
				\hline
				\noalign{\smallskip}
				eccentricity, $e_{\rm b}$ & - & $0.48^{+0.14}_{-0.16}$  \\
				\noalign{\smallskip}
				arg. of periapsis, $\omega_{\rm \star,\:b}$ & - & $-2.30^{+5.27}_{-0.66}$  \\
				\noalign{\smallskip}
				\hline
				\noalign{\smallskip}
				$\ln\mathcal{Z}$ & -958.2 & -956.7\\
				\noalign{\smallskip}
				$\ln\mathcal{Z}_{1p}$-$\ln\mathcal{Z}_{0p}$ & +11.9 & +13.4 \\
				\noalign{\smallskip}
				\hline
			\end{tabular}
		\end{table}
		
		\section{Results using alternative RV extraction pipelines applied to HARPS spectra}
		\label{app:nairatrifon}
		As discussed in Sect. \ref{sec:overviewdata}, in this study we also examined two alternative RV dataset extracted from HARPS spectra, namely the NAIRA and \cite{trifonov20} nightly-zero point corrected (NZP) dataset, all coming from techniques based on template matching as for the case of TERRA. We repeated all the analysis described in Sect. \ref{sec:harpscarmervanalysis} as a sanity cross-check for confirming the existence of the candidate planetary signal at $\sim$140 days. The analysis were performed using the same MC sampler, set-up, and priors of those adopted for the RVs extracted with TERRA. We summarise the results in Table \ref{tab:nairatrifonfitresults}, by reporting \textit{i}) the best-fit values of some of the signal parameters (semi-amplitude $K_b$, period $P_b$, and eccentricity $e_b$) obtained for each tested GP kernel, \textit{ii}) the corresponding logarithm of the Bayesian evidence, $\ln\mathcal{Z}$, and \textit{iii}) the difference $\Delta\ln\mathcal{Z}$ with respect to the model with no planetary signal included.    
		Independently from the algorithm used to extract the RVs, the model with the eccentricity $e_b$ treated as a free parameter is generally strongly favoured ($\Delta\ln\mathcal{Z}>5$ in five cases over six) over the model with no planet included. It is generally moderately-to-strongly favoured over the model with $e_b$ fixed to zero ($\ln\mathcal{Z}_{e_{b}\neq0}-\ln\mathcal{Z}_{e_{b}=0}\geq2.4$ in five cases over six). The eccentricity is retrieved with significance within $\sim$3-4$\sigma$. These results are in agreement with those obtained for the HARPS TERRA dataset, and strengthen our conclusion that a planetary-like signal is present in the data. Moreover, the parameters $K_b$, $P_b$, and $e_b$ are all consistent with those derived using the TERRA RVs. Based on the results shown in Table \ref{tab:nairatrifonfitresults}, we conclude that the GP QPC plus a Keplerian is the best fit model to the RVs of the two alternative pipelines, as in the case of TERRA, because the corresponding $\ln\mathcal{Z}$ is significantly the highest among all the values of the Bayesian evidence. We note that the dSHO model provides the highest significance in favour of the $\sim$140-d signal over the model with no planet included, as we found for the case of the TERRA dataset.
		
		\begin{table*}[]
			\centering
			\small
			\caption{Results for all the different GP-based fits performed on the combined HARPS+CARMENES dataset using the alternative NAIRA and \cite{trifonov20} nightly-zero point corrected (NZP) radial velocity time series. $\Delta\ln\mathcal{Z}$ = $\ln\mathcal{Z}_{\rm 1pl}$-$\ln\mathcal{Z}_{\rm 0pl}$, i.e. the difference between the Bayesian evidences $\ln\mathcal{Z}$ of the models with and without a planetary signal included. }
			\begin{tabular}{lll}
				\hline
				\noalign{\smallskip}
				\textbf{GP kernel} & \textbf{NAIRA RVs} & \textbf{Trifonov et al. (2020) RVs} \\
				\noalign{\smallskip}
				\hline
				\noalign{\smallskip}
				Quasi-periodic (QP)  & Circular orbit:  &  Circular orbit: \\
				\noalign{\smallskip}
				& $K_b$ [$\ms$] = 0.94$^{+0.19}_{-0.22}$ & $K_b$ [$\ms$] = 0.91$^{+0.17}_{-0.18}$ \\
				& $P_b$ [d] = 139.69$^{+0.72}_{-0.80}$ & $P_b$ [d] = 139.66$^{+0.70}_{-0.68}$\\
				& $T_{conj,\,b}$ [BJD-2450000] = $8669.18^{+7.02}_{-6.81}$ & $T_{conj,\,b}$ [BJD-2450000] = $8668.56^{+6.56}_{-6.31}$ \\
				& $\ln\mathcal{Z}=-966.2$; $\Delta\ln\mathcal{Z}=+1.3$ & $\ln\mathcal{Z}=-970.0$; $\Delta\ln\mathcal{Z}=+2.5$     \\
				\noalign{\smallskip}
				& Eccentric orbit:  & Eccentric orbit:  \\
				\noalign{\smallskip}
				& $K_b$ [$\ms$] =$1.24^{+0.23}_{-0.21}$ & $K_b$ [$\ms$] =$1.23^{+0.26}_{-0.22}$ \\
				& $P_b$ [d] =$140.38^{+0.46}_{-0.43}$ & $P_b$ [d] =$140.35^{+0.42}_{-0.34}$ \\
				& $T_{conj,\,b}$ [BJD-2450000] = $8697.15^{+7.54}_{-13.15}$ & $T_{conj,\,b}$ [BJD-2450000] = $8699.09^{+6.47}_{-11.74}$ \\
				& $e_b=0.48^{+0.14}_{-0.13}$ & $e_b=0.52\pm0.14$ \\
				& $\ln\mathcal{Z}=-962.9$; $\Delta\ln\mathcal{Z}=+4.6$ & $\ln\mathcal{Z}=-966.8$; $\Delta\ln\mathcal{Z}=+5.7$ \\
				\noalign{\smallskip}
				\hline
				\noalign{\smallskip}
				Quasi-periodic cosine (QPC) & Circular orbit:  &  Circular orbit: \\
				\noalign{\smallskip}
				& $K_b$ [$\ms$] = 0.84$^{+0.17}_{-0.19}$ & $K_b$ [$\ms$] = 0.81$^{+0.17}_{-0.18}$ \\
				& $P_b$ [d] = 139.91$^{+0.72}_{-0.77}$ & $P_b$ [d] = 139.87$^{+0.74}_{-0.76}$ \\
				& $T_{conj,\,b}$ [BJD-2450000] = $8671.24^{+6.85}_{-7.35}$ & $T_{conj,\,b}$ [BJD-2450000] = $8670.79^{+6.98}_{-7.16}$ \\
				& $\ln\mathcal{Z}=-958.9$; $\Delta\ln\mathcal{Z}=+2.0$ & $\ln\mathcal{Z}=-965.2$; $\Delta\ln\mathcal{Z}=+3.6$ \\
				\noalign{\smallskip}
				& Eccentric orbit: & Eccentric orbit: \\
				\noalign{\smallskip}
				& $K_b$ [$\ms$] =$1.15^{+0.23}_{-0.20}$ & $K_b$ [$\ms$] =$1.14^{+0.27}_{-0.21}$\\
				& $P_b$ [d] =$140.40^{+0.46}_{-0.40}$ & $P_b$ [d] =$140.37^{+0.38}_{-0.33}$ \\
				& $T_{conj,\,b}$ [BJD-2450000] = $8699.02^{+6.04}_{-10.78}$ & $T_{conj,\,b}$ [BJD-2450000] = $8700.17^{+5.73}_{-11.12}$ \\
				& $e_b=0.50^{+0.13}_{-0.14}$ & $e_b=0.53^{+0.14}_{-0.15}$ \\
				& $\ln\mathcal{Z}=-955.5$; $\Delta\ln\mathcal{Z}=+5.4$ & $\ln\mathcal{Z}=-962.8$; $\Delta\ln\mathcal{Z}=+6.0$ \\
				\noalign{\smallskip}
				\hline
				\noalign{\smallskip}
				dSHO (rotational) & Circular orbit:   &  Circular orbit: \\
				\noalign{\smallskip}
				& $K_b$ [$\ms$] =0.77$\pm0.17$ & $K_b$ [$\ms$] =0.88$^{+0.17}_{-0.16}$ \\
				& $P_b$ [d] =140.43$\pm0.88$ & $P_b$ [d] =140.42$\pm0.68$ \\
				& $T_{conj,\,b}$ [BJD-2450000] = $8675.34^{+6.13}_{-6.69}$ & $T_{conj,\,b}$ [BJD-2450000] = $8674.48\pm5.89$\\
				& $\ln\mathcal{Z}=-960.7$; $\Delta\ln\mathcal{Z}=+9.7$ & $\ln\mathcal{Z}=-967.5$; $\Delta\ln\mathcal{Z}=+10.7$ \\
				\noalign{\smallskip}                       
				& Eccentric orbit:   & Eccentric orbit:  \\
				\noalign{\smallskip}
				& $K_b$ [$\ms$] =1.19$^{+0.33}_{-0.26}$  & $K_b$ [$\ms$] =1.15$^{+0.27}_{-0.22}$ \\
				& $P_b$ [d] =141.11$^{+0.41}_{-0.71}$  &  $P_b$ [d] =140.91$^{+0.53}_{-0.70}$ \\
				& $T_{conj,\,b}$ [BJD-2450000] = $8702.57^{+5.08}_{-8.99}$ & $T_{conj,\,b}$ [BJD-2450000] = $8698.38^{+6.81}_{-11.39}$ \\
				& $e_b=0.56^{+0.14}_{-0.16}$ & $e_b=0.47^{+0.15}_{-0.16}$ \\
				& $\ln\mathcal{Z}=-958.4$; $\Delta\ln\mathcal{Z}=+13.0$ &  $\ln\mathcal{Z}=-966.0$; $\Delta\ln\mathcal{Z}=+12.2$  \\
				\noalign{\smallskip}
				\hline
				\noalign{\smallskip}
			\end{tabular}
			\label{tab:nairatrifonfitresults}
		\end{table*}
		
		\section{Additional tests to cross-check the existence and temporal stability of the 140-d signal}
		\label{app:check140d}
		
		To independently cross-check the results of section~\ref{sec:rvmodelling}, we analysed HARPS$_{\rm TERRA}$ and CARMENES-VIS RVs with other methods. We first use the $\ell_1$ periodogram \citep{hara2017}, a tool based on a sparse recovery technique called the basis pursuit algorithm \citep{chen1998}. The $\ell_1$ periodogram takes in a frequency grid and a covariance matrix to model noise sources as inputs. It aims to find a representation of the RV time series as a sum of a small number of sinusoids whose frequencies are in the input grid. It outputs a figure which has a similar aspect as a GLS periodogram, but with fewer peaks due to aliasing. FAPs can be calculated for each peak, and their interpretation is equivalent to that of common periodograms. 
		
		To determine the influence of the noise model, we followed \citep{hara2020} and considered a grid of covariance models and rank the alternatives with cross validation. We defined the covariance matrix $V$ so that its element at index $(k,l)$ is
		
		\begin{align}
			V_{k,\,l} & =  \delta_{k,l} \cdot (\sigma_{RV}^2 + \sigma_{W}^2 + \sigma_{C}^2 )~+~\sigma_{R}^2 \,exp\Big[-\frac{(t_k-t_l)^2}{2\tau_R^2}\Big]~+~  \nonumber \\  &+   
			\sigma_{act}^2\,exp\Big[-\frac{(t_k-t_l)^2}{2\tau_{act}^2} -\frac{1}{2} \sin^2\left(  \frac{ \pi (t_k - t_l)}{P_{act} } \right)\Big] 
			\label{eq:kernel_l1}
		\end{align}
		
		where $\sigma_{RV}$ is the nominal measurement uncertainty; $\sigma_W$ is an additional white noise jitter term; $\sigma_{C}$ is a calibration noise term; $\delta_{k,l}$ equals one if measurements $k$ and $l$ are taken within the same night, and zero otherwise; $\sigma_{R}$ and $\tau_R$ define a correlated term to model contributions due to granulation \citep{cegla2019} or instrumental effects, as defined in \cite{hara2020}; $\sigma_{act}, \tau_{act} $, and $P_\mathrm{act}$ are the hyper-parameters of a quasi-periodic covariance term to model stellar activity.
		
\begin{figure*}
			\centering
			\includegraphics[width=0.7\linewidth]{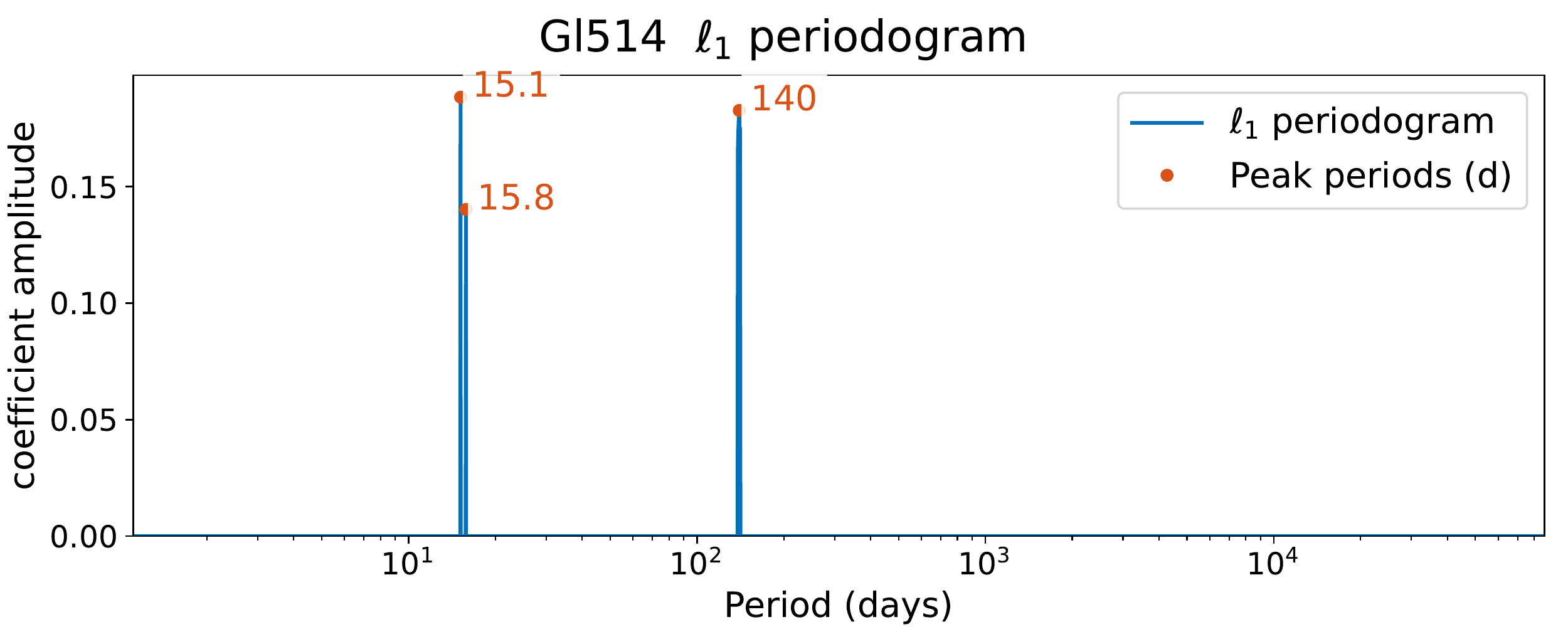}
			\caption{$\ell_1$ periodogram with free offsets and covariance models with the highest cross-validation score.}
			\label{fig:l1_perio}
		\end{figure*}

		For $\sigma_{W}, \sigma_{R}, \sigma_{QP}$, we use the grid of values $ 0.5, 1., 1.5$ m/s. $\sigma_{C}$ is fixed to 0.5 m/s. $\tau= 3$ or 6 days, $P_{act}$ is fixed to 31.4 days based on the analysis of activity indicators and $\tau_{act} = 60$ days. We try every combination of these values and find that the model with highest cross validation has $\sigma_{W} =1.5$ m/s,  $\sigma_{R} = 0.5$ m/s,  $\sigma_{QP} =1$ m/1, $\tau_R = 6$ days. We use a free offset for each of the three datasets: pre and post 2015 HARPS and CARMENES.
		The corresponding $\ell_1$ periodogram is shown in Figure~\ref{fig:l1_perio}. We find three peaks at 15.1, 140 and 15.8 days in decreasing order. Their corresponding FAP is $1.9 \cdot 10^{-5}$, $2 \cdot 10^{-3}$ and 0.97. The first two signals are statistically significant, and the second is the orbital period of the candidate planet Gl~514\,b. It is interesting to note that the peak at $P=15.8$ d has a very high FAP, contrary to the result of the MLP. The $\ell_1$ method searches for several signals simultaneously and accounts for correlated noise at the stellar rotation period. As a result, signals that could appear dominant in a regular periodogram, due either to stellar activity or aliasing, appear damped in $\ell_1$.

		To further examine their properties, we apply the method of \cite{hara2022}, which aims at testing whether the signals have consistent amplitude, phase and frequencies over time. We compute the apodized sine periodogram (ASP), defined as follows. We compare the $\chi^2$ of a linear base model $H$, and the $\chi^2$ of a model $K(\omega, t_0,\tau)$ defined as the linear model of $H$ plus an apodized sinusoid $e^{-\frac{(t-t_0)^2}{2\tau^2}} \cdot (A \cos \omega t + B \sin \omega t )$.
		The ASP is defined as the $\chi^2$ difference between the two models
		
		\begin{align}
			z(\omega, t_0,\tau) &= \chi^2_H - \chi^2_{K(\omega, t_0,\tau)}.\label{eq:z}
		\end{align}
		
\begin{figure*}
			\centering
			\includegraphics[width=16cm]{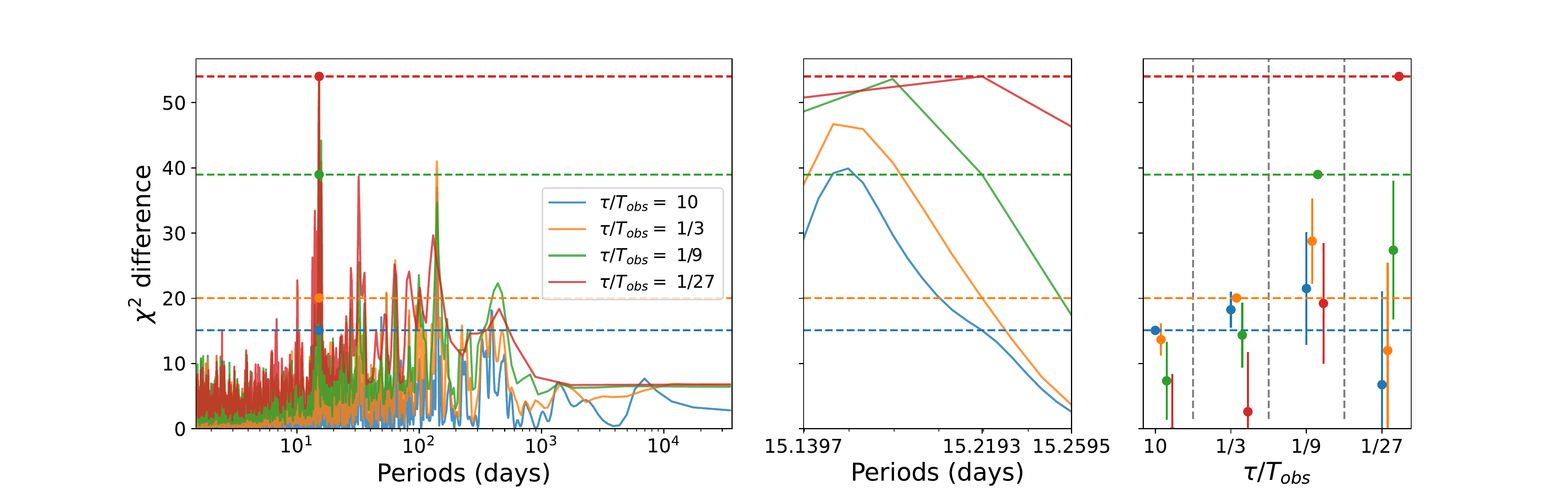}
			\caption{\textit{Left panel}: ASP corresponding to different apodization time-scales: $\tau = 10\cdot T_{obs}, T_{obs}/3, T_{obs}/9$ and $T_{obs}/27$, respectively in blue, orange, green and red. \textit{Middle panel}: zoom on the maximum peak. \textit{Right panel}: statistical tests as defined in \cite{hara2022}. }
			\label{fig:asp_iter0}
		\end{figure*}

		Like in the $\ell_1$ periodogram, our base model includes the offsets of the three datasets. For the apodized sinusoid, we consider a grid of timescales $\tau = 10\cdot T_{obs}, T_{obs}/3, T_{obs}/9$ and $T_{obs}/27$, where $T_{obs}$ is the total observation timespan of the data. In Fig. \ref{fig:asp_iter0}, we represent $z(\omega, t_0,\tau)$  as defined in Eq.~\eqref{eq:z}, maximised over $t_0$. The dominating peak has a period of 15.2 days. We want to test whether the signal is statistically compatible with a constant one (i.e. with $\tau = 10\cdot T_{obs}$). Denoting by $t_{(\tau,\omega)}$ the value of $t_0$ maximising the value of the periodogram~\eqref{eq:z} for a given $\omega$ and $\tau$,  we compute the distribution 
		\begin{align}
			D_z = z(\omega, t_{(\tau,\omega)}, \tau) - z(\omega, t_{(\tau^\prime,\omega)}, \tau^\prime)
			\label{eq:dz}
		\end{align} 
		with the hypothesis that model $K(\omega, t_{(\tau,\omega)}, \tau, A^\star, B^\star )$ is correct, where the fitted cosine and sine amplitudes $A^\star, B^\star$ are obtained by fitting model $K$ to the data. $D_z$ can easily be expressed as a generalised $\chi^2$ distribution, with mean and variance given by an analytical expression \citep{hara2022}. In Fig.~\ref{fig:asp_iter0}, horizontal dashed lines correspond to the values of $z$ at the frequency where the maximum is attained. In the right panel, the four timescale corresponds to a value of $\tau$, reported in abscissa. For each, we assume that the data contains a signal at frequency $15.2$ days and with timescale $\tau$. The points with error bars correspond to the expected value of the peaks in the ASP and their standard deviation, assuming that the timescale $\tau$ in abscissa is the correct one. These are computed as $z(\omega, t_{(\tau,\omega)}, \tau) - E(D_z)$, where $E(D_z)$ is the expectancy of $D_z$ as defined in Eq. \eqref{eq:dz}, and as the variance of $D_z$. The more each expected value (with error bars) is closer to the level defined by the dashed horizontal line of the same color, the more the value of $\tau$ in the abscissa is the best representation for the timescale associated to the signal of period $P$. From Fig.~\ref{fig:asp_iter0}, we see that assuming $\tau = 10\cdot T_{obs}$ as the correct timescale, the expected position of the ASP peaks of the other timescales is more than three sigmas away from where they are observed. Their expected positions are in agreement with the observed levels for $\tau = T_{obs}/27$, meaning that the period $P=15.2$ days is not stable over time.
		
\begin{figure*}
			\centering
			\includegraphics[width=16cm]{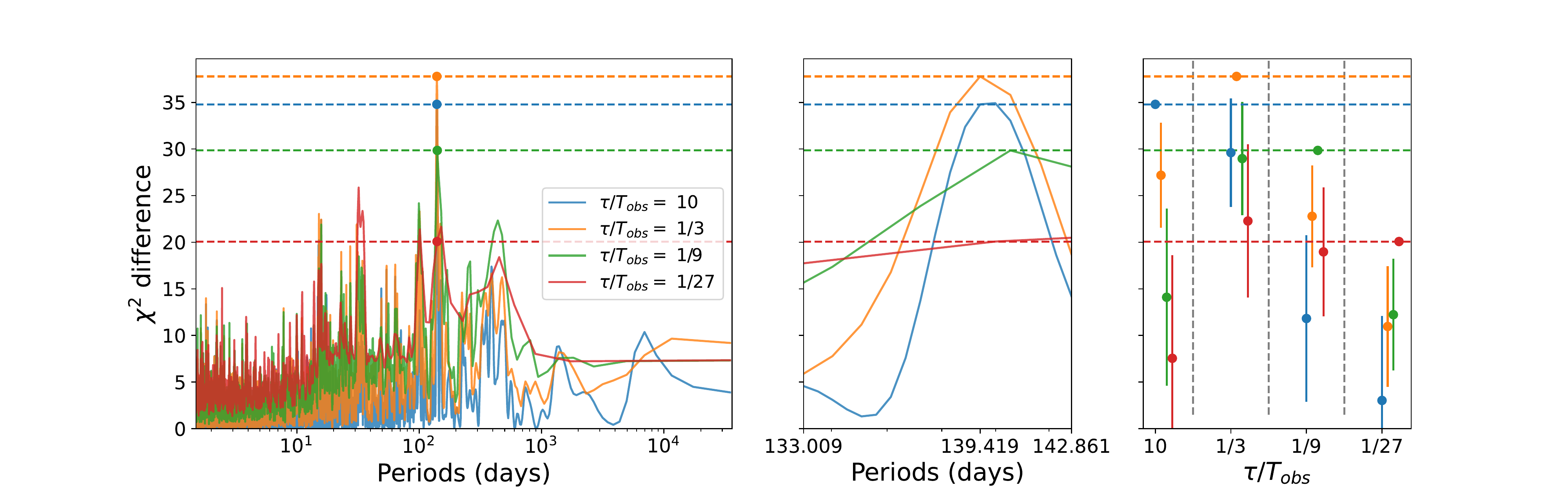}
			\caption{As in Fig. \ref{fig:asp_iter0}. The base model now includes the signal corresponding to the highest peak of the first ASP (Fig.~\ref{fig:asp_iter0}). }
			\label{fig:asp_iter1}
		\end{figure*}

		We pre-whitened the original dataset by including the 15.2-day signal in the null hypothesis/base model $H$, and we repeated the calculation of the ASP. We obtain the periodogram in Fig.~\ref{fig:asp_iter1}, which a shows a peak occurring at a period of 140 days. The statistical test in the right panel of the figure shows that this signal is compatible with having a timescale $\tau = T_{obs}/3$ or greater. This means that the signal can be credibly approximated as a sinusoid (we remark that this method does not take the orbital eccentricity into account), denoting that it compatible with a signal induced by a planet. 
		
		\section{Results relative to RV models including two Keplerians}
		\label{app:plots2plan}
		In this Section, we show plots related to the tests performed on the HARPS$_{\rm TERRA}$+CARMENES RVs discussed in Sect. \ref{sec:2planetfit}, where we examined the hypothesis of a two-planet system. 
		
		\begin{figure*}
			\centering
			\includegraphics[width=\textwidth]{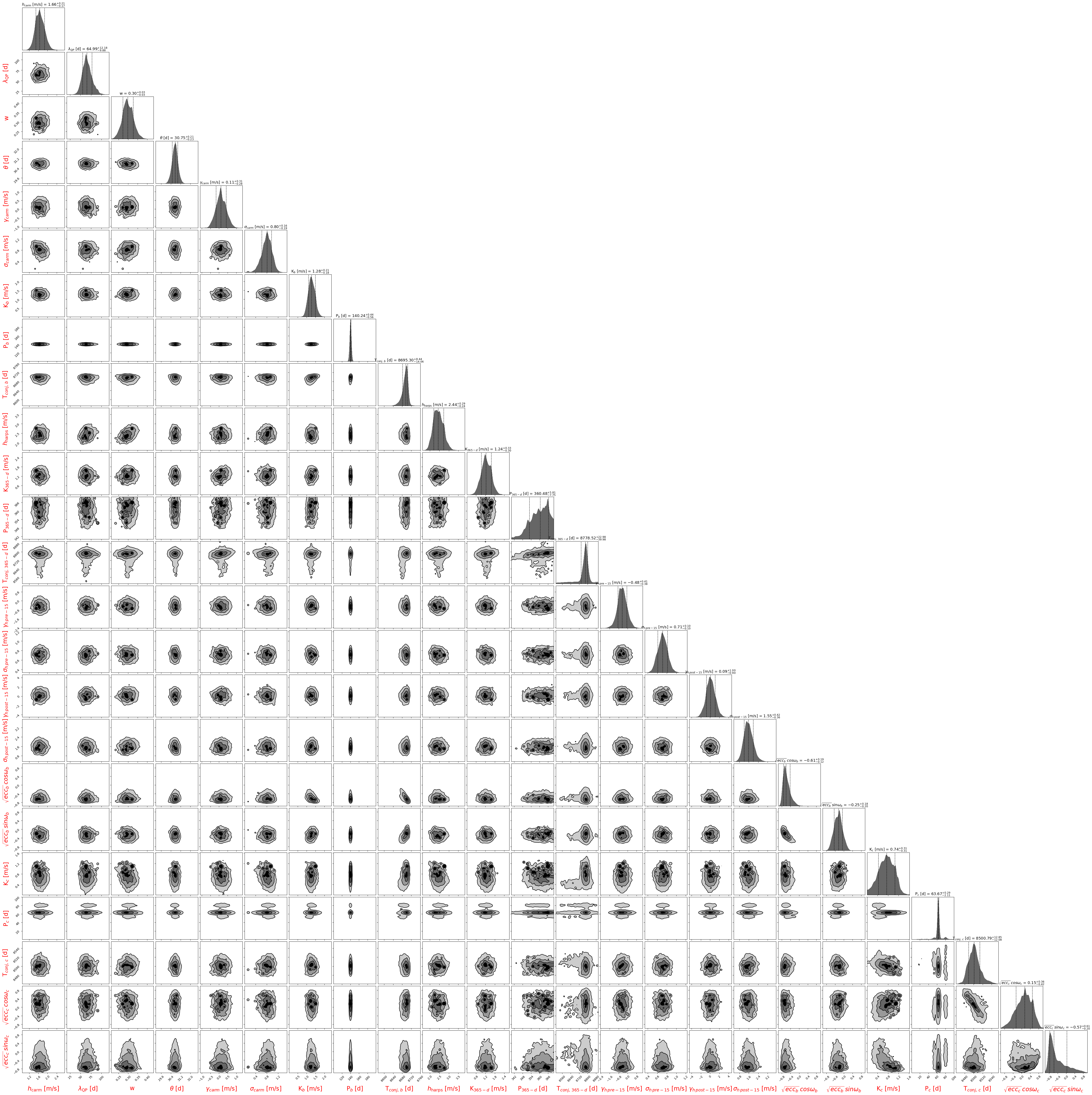}  
		\end{figure*}
		\newpage
		\begin{figure*}
			\centering
			\includegraphics[width=\textwidth]{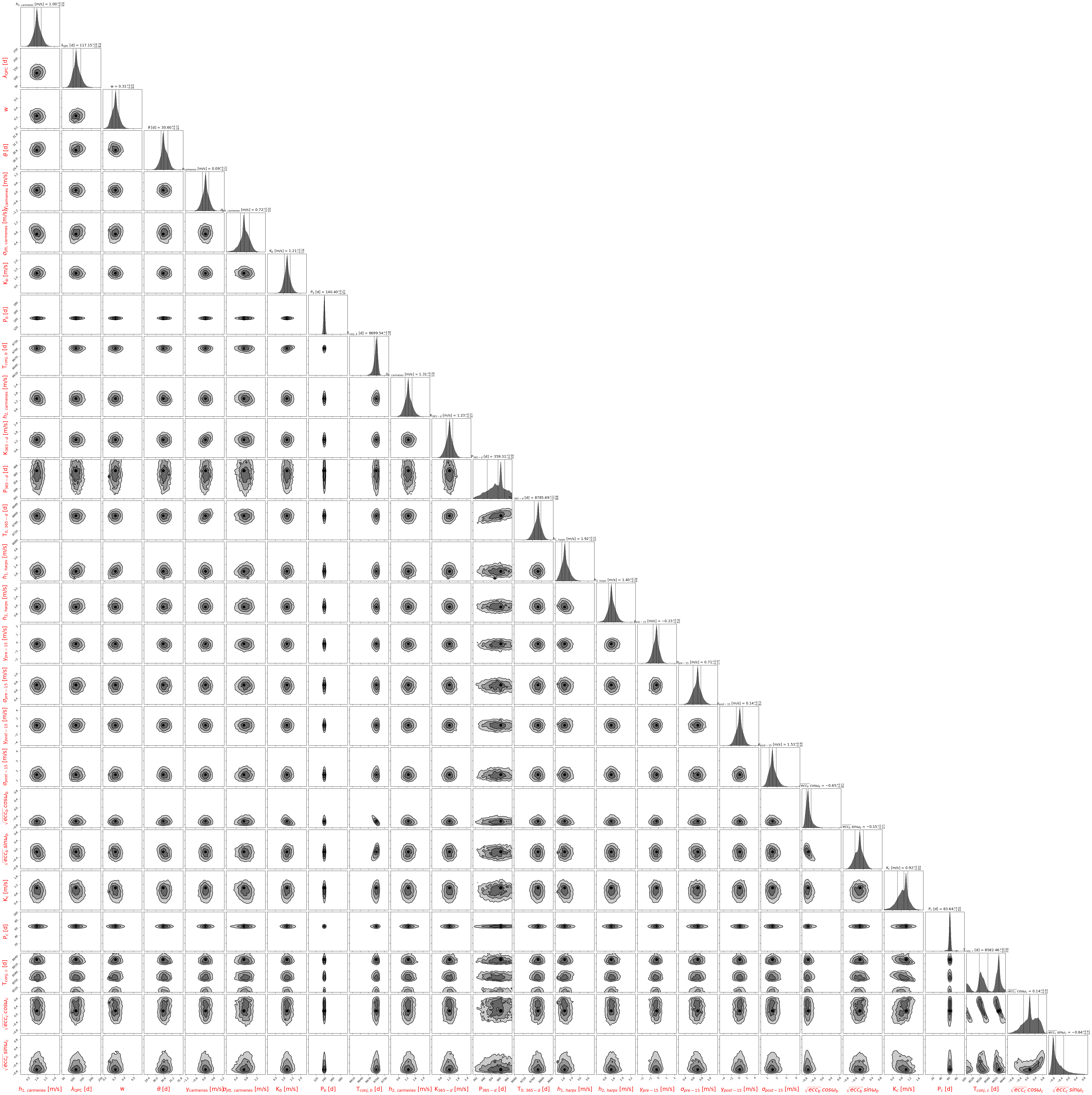}  
		\end{figure*}
		\newpage
		\begin{figure*}
			\includegraphics[width=\textwidth]{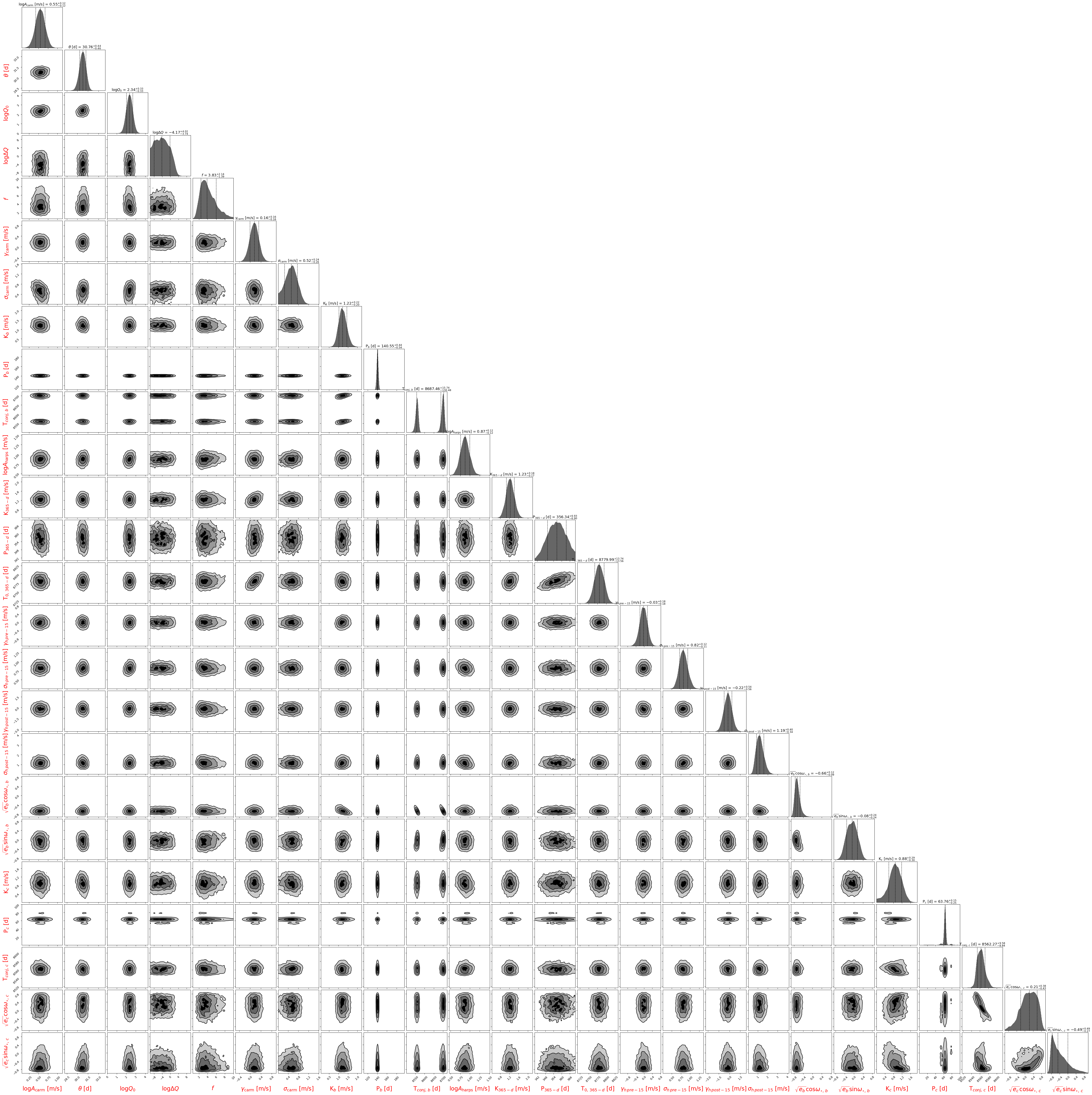}
			\caption{Corner plots showing the posterior distributions of all the free (hyper-)parameters of the two-planet models described in Sect. \ref{sec:2planetfit}. \textit{Upper panel:} QP kernel; \textit{Middle panel:} QPC kernel; \textit{Lower panel:} dSHO/rotational kernel. } 
			\label{fig:corner2plecc}
		\end{figure*}

	\end{appendix}
	
\end{document}